\documentclass{ws-ijmpa}
\usepackage[super,compress]{cite}

\usepackage{amsmath}
\usepackage{epsfig,epsf}
\usepackage{graphicx}
\usepackage{verbatim}
\usepackage{epstopdf}
\usepackage{bm}  
\usepackage{epsfig,epsf}
\def\beq{\begin{equation}}
\def\eeq{\end{equation}}
\def\bea{\begin{eqnarray}}
\def\eea{\end{eqnarray}}
\def\eq#1{{Eq.~(\ref{#1})}}
\def\fig#1{{Fig.~\ref{#1}}}
\newcommand{\bas}{\bar{\alpha}_S}
\newcommand{\as}{\alpha_S}

\newcommand{\Lb}{\left(}
\newcommand{\Rb}{\right)}
\def\pom{{I\!\!P}}
\def\reg{{I\!\!R}}
\parskip=\itemsep              

\newcommand{\nn}{\nonumber}

\newcommand{\h}{\frac{1}{2}}

\newcommand{\ga}{\gamma}

\renewcommand{\theequation}{\thesection.\arabic{equation}}

\def\pom{{I\!\!P}}
\def\reg{{I\!\!R}}

\begin{document}
\markboth{E. Gotsman, E. Levin and U. Maor}
{A comprehensive model of soft interactions  
 in the LHC  era}

%
\catchline{}{}{}{}{}
%

\title{A comprehensive  model of soft interactions  
 in the LHC era }

\author{Errol Gotsman} \address{ Department of Particle Physics, School of 
Physics and Astronomy, Raymond and Beverly Sackler Faculty of Exact 
Science, Tel Aviv University, Tel Aviv, 69978, Israel\\ 
gotsman@post.tau.ac.il} \author{Eugene Levin} \address{ Department of 
Particle Physics, School of Physics and Astronomy, Raymond and Beverly 
Sackler Faculty of Exact Science, Tel Aviv University, Tel Aviv, 69978, 
Israel\\ Departamento de F\'\i sica, Universidad T\'ecnica Federico Santa 
Mar\'\i a, Avda. Espa\~na 1680\\ and Centro 
Cientifico-Tecnol$\acute{o}$gico de Valpara\'\i so,Casilla 110-V, 
Valpara\'\i so, Chile\\ leving@post.tau.ac.il, 
eugeny.levin@usm.cl} \author{Uri Maor} \address{ Department of Particle 
Physics, School of Physics and Astronomy, Raymond and Beverly Sackler 
Faculty of Exact Science, Tel Aviv University, Tel Aviv, 69978, Israel\\ 
maor@post.tau.ac.il} %

\maketitle

~

\begin{history}
{ \it  Contribution to appear in the special issue of
the International Journal of Modern Physics A \\ on "Elastic and Diffractive
Scattering", coordinated by Christophe Royon}
\end{history}
\begin{flushright}
TAUP 2981/14
\date{today}
\end{flushright}

\begin{abstract}
 In this review we present our model which is an example of the
 self consistent approach that incorporates our theoretical understanding
 of long distance physics, based both on N=4 SYM for strong coupling  and
on the matching with the  perturbative QCD approach.
 We demonstrate how important and decisive   the LHC data  were on strong 
 interactions which led us to  a set of the phenomenological 
parameters that fully confirmed our theoretical expectations, and produced   a new
picture of the strong interaction at high energy. We also show how far we have come
 towards creating a framework for the description of  minimal bias
events  for high energy scattering without generating  Monte Carlo codes.
\end{abstract}

\ccode{PACS numbers: 13.85.-t, 13.85.-Hd, 11.55=m, 11.55 Bg}

\section{Introduction}
\par
The LHC data \cite{ALICE,ATLAS,CMS,TOTEM}  changed our
understanding of soft interactions at high energy. Regrettably, 
none of the phenomenological models based on the Reggeon 
approach\cite{DL,GLM,KAP,KMR,OST}
were 
successful in predicting this data, in spite of having a 
large number of fitted  parameters.  
This fact is not surprising since quantum chromodynamics (QCD), which 
is a microscopic theory, has a theoretical problem, i.e. accounting for 
the  
confinement of quarks and gluons. There is a wide spread belief
that the confinement of quarks and gluons is not a fundamental 
problem, this is based on the fact that lattice QCD shows that such a 
phenomenon 
exists and can be calculated. Soft interactions at high energy,  
provides us with an excellent example of processes in which the lattice 
approach fails to produce any framework which incorporates a satisfactory 
description of the experimental data. Even though confinement is a 
difficult fundamental problem, we firmly believe that 
soft interaction data, as well as its interpretation in the framework 
of a model approach, will lead to a deeper understanding of the origin of  
confinement and its properties.
\par 
To comprehend how much the LHC data contribute to our understanding 
of high energy soft processes, we compare the general characterizations 
of these processes before and after the LHC.
\begin{itemize}
\item
  Before LHC:  At the time, Donnachie and Landshoff (DL) provided a 
good description \cite{PDG} of the total and elastic cross sections 
in the ISR-Tevatron energy range. 
The DL model has a severe flaw since it does not include the 
effect
of shadowing/screening corrections on the Pomeron and the secondary 
Reggeons exchange amplitudes. 
This approximation is valid, though, in the calculation of 
the elastic hadronic amplitudes below the LHC energy\cite{DL}. 
However, it fails to describe the diffractive channels in which  
unitarity screening are significant at energies as low as the ISR.\\
The DL Pomeron trajectory is: 
\beq \label{DL}
\alpha_\pom \,=\, 1 + \Delta_\pom \,+\,\alpha'_\pom \,t\,
=\,1 + 0.08\div 0.13\,+\,0.25 \,t,
\eeq  
where $t$ is the square of the transferred momentum
of the Pomeron and the slope $\alpha'_\pom$ is measured in $GeV^{-2}$. 
Both Pomeron and secondary Reggeons stem from a natural generalization 
of QCD string theory\cite{STRING}. Therefore, this simple 
phenomenology, together with the string approach, provides a self 
consistent 
and beautiful picture of  soft interactions at high energies. 
As noted, the simple DL phenomenology of a single Pomeron exchange  
cannot reproduce the diffractive sector, which follows directly from 
the wave nature of the colliding particles. However, 
 we have a small parameter: the ratio 
$R=\h \sigma_{sd}/\sigma_{el} \approx 0.18 $ at the Tevatron energies. 
Hence, we can develop a perturbative approach with respect to this 
parameter. The partial single Pomeron exchange 
elastic amplitude at fixed impact parameters ($b$) is:
\beq \label{POMAMP}
A_\pom\Lb b \Rb \,\,=\,\,g_1\,g_2  \int d^2 b' S_1\Lb \vec{b} - 
\vec{b}^{\,'}\Rb S_2\Lb \vec{b}^{\,'} \Rb\,
e^{ \Delta_{pom}Y \,-\,\frac{b^2}{4 \alpha'_\pom   Y}}\,\,<\,\,1.
\eeq 
In \eq{POMAMP} $g_i$ is 
the vertex of the Pomeron interaction with the hadrons, 
$S_i\Lb b_i\Rb$ is the profile function at 
this vertex and $Y = \ln s$. One can see from \fig{el}, which 
illustrates \eq{POMAMP}, 
that  $A_\pom \Lb b \Rb$ reaches 1 at the Tevatron energy and, 
therefore, one expects that the DL model will need to be 
amended for the LHC range of energies. 
The second weakness of this simple model is that the small parameter 
is under estimated, as we also need to add  the double diffraction 
cross section, 
($R= \Lb \h \sigma_{sd}+ \sigma_{dd}\Rb/\sigma_{el} \approx 0.4 \div 0.5$ 
in spite of the large $\sigma_{dd}$ errors.) 
\begin{figure}
\leavevmode
\begin{tabular}{c c}
\includegraphics[width=6cm]{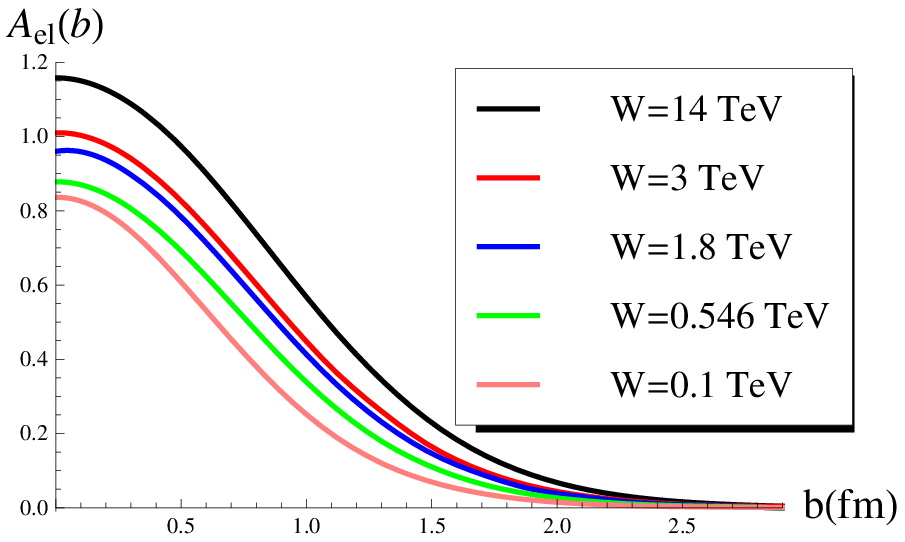}&
\includegraphics[width=6cm]{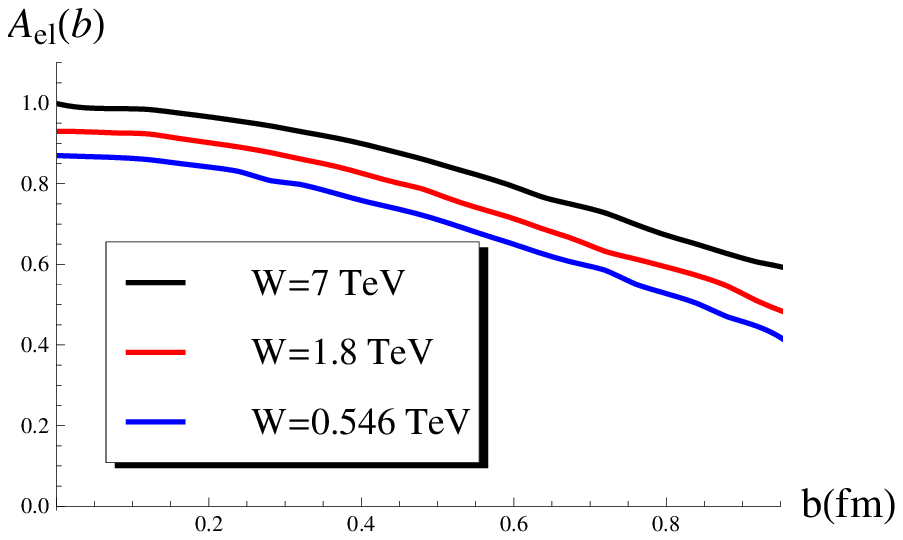}\\
\fig{el}-a &\fig{el}-b\\
\end{tabular}
\caption{The elastic amplitude calculated, using Donnachie-Landshoff Pomeronof \protect\eq{DL}
 (\fig{el}-a) and  using the data on elastic 
cross section (\fig{el}-b)
$A_{el}\Lb b \Rb = $ $\int \frac{q_\bot q_\bot}{4\pi} 
\sqrt{\frac{d \sigma_{el}\Lb s,t = -q^2_\bot\Rb}{dt} 
\frac{16\pi}{(1 + \rho)}J_0(q_\bot b)}$. For \fig{el}-b
the curves are taken from Ref.\cite{DUREL}.}
\label{el}
\end{figure}
\item Post LHC: Substantial shadowing corrections need to be taken 
into account to describe the LHC data. In models, based on 
Pomerons and their interactions, the Pomeron trajectory emerges as 
\beq \label{trajectory}
\alpha_\pom \,=\, 1 + \Delta_\pom \,+\,\alpha'_\pom \,t\,=\,1 + (0.2\div 
0.3\,) +\,\Lb \alpha'_\pom < 0.02 \Rb\,t, 
\eeq
which cannot be obtained from a string model approach. 
On the other hand, such a Pomeron provides a natural matching with 
perturbative QCD (pQCD)\cite{BFKL,LI}, 
and N=4 SYM theory\cite{AdS-CFT,POST,BFKL4,BST}.
\end{itemize}
\par
The goal of this review is to discuss two main topics: 
why and how such a Pomeron can be expected both from  theoretical 
approaches, 
and from the description of the LHC experimental data; 
and the progress we have made in our attempts to build a model 
that is able to describe the structure of the bias events without 
relying on Monte Carlo codes.
\par
As we lack a solid theoretical basis in non perturbative QCD to build a model,
we need to make   an educated guess as to which theoretical approach we 
should employ. 
Our choice will be discussed in the 
next section which, on its own, is a review of useful theoretical ideas 
regarding the strong interactions. 
In the third section we  present the key assumptions and the 
main formulae of our model. 
In section 4 we  discuss the qualitative features of the model. 
\section{Theoretical background}
\subsection{General theorems}
\par
As the methods of non-perturbative QCD are in an
embryonic stage, we rely on the consequences of 
the general features of the scattering amplitude, i.e. 
analyticity, crossing symmetry and unitarity.
\subsubsection{Unitarity}
\par
To formulate the unitarity constraints, we introduce a  
complete set of orthogonal
functions $\{ \psi_i \}$ which diagonalize the
interaction matrix ${\bf T}$
\beq \label{GT1}
A^{i'k'}_{i,k}=<\psi_i\,\psi_k|\mathbf{T}|\psi_{i'}\,\psi_{k'}>=
A_{i,k}\,\delta_{i,i'}\,\delta_{k,k'}.
\eeq
In this representation the hadron wave function can be written as
\beq \label{GT2}
\psi_h\,=\,\sum_{i} c^h_i \psi_i
\eeq
The unitarity constraints have the form
\beq \label{UNIT}
2\,\mbox{Im}\,A_{i,k}\left(s,b\right)=|A_{i,k}\left(s,b\right)|^2
+G^{in}_{i,k}(s,b),
\eeq
where $G^{in}_{i,k}$ denote the contribution of all non 
diffractive inelastic processes,
i.e. it is the summed probability for these final states to be
produced in the scattering of a state $i$ off state $k$. In \eq{UNIT} 
$\sqrt{s}=W$ is the energy of the colliding hadrons and $b$ denotes the 
impact  parameter.
A simple solution to \eq{UNIT} at high energies has the eikonal form 
with an arbitrary opacity $\Omega_{ik}$, where the real 
part of the amplitude is much smaller than the imaginary part.
\beq \label{GT3}
A_{i,k}(s,b)=i \Lb 1 -\exp\Lb - \frac{\Omega_{i,k}(s,b)}{2}\Rb\Rb,
\eeq
\beq \label{GT4}
G^{in}_{i,k}(s,b)=1-\exp\Lb - \Omega_{i,k}(s,b)\Rb.
\eeq
\eq{GT4} implies that the probability that the initial projectiles
$(i,k)$ will reach the final state interaction unchanged, regardless of 
the initial state re-scatterings, is given by
$P^S_{i,k}=\exp \Lb - \Omega_{i,k}(s,b) \Rb$.
\par
Integrating  \eq{UNIT}  over $b$ we have 
\beq \label{OPTTEOR}
2 \,\mbox{Im} A_{i,k}(s,t=0)\,= \,2 \int d^2 b\, \mbox{Im} A_{i,k}(s,b)\,
=\,\sigma_{el} + \sigma_{in} \,=\,\sigma_{tot}.
\eeq
Which is the optical theorem.
\subsubsection{The unitarity bound}
\par
Using two inputs: the unitarity constraints and the behaviour of the 
amplitude at large $b$, we derive a bound on the total cross 
section\cite{FROI}. Indeed, from \eq{UNIT} $A_{i k} \leq 1$. 
The fact that we have the lightest hadron (pion) with mass $m_\pi$, 
implies  that the amplitude decreases as 
$\exp\Lb - 2 m_\pi b\Rb$.  We can re-write \eq{OPTTEOR} and \eq{UNIT} 
in the form 
\beq \label{UB1}
\sigma_{tot}\,=\,2 \int d^2 b\,\mbox{Im} A_{i,k}(s,b)\,
\leq \,2 \int^{b^*} d^2 b \,+\,\,2 \int_{b^*} d^2 b
A_{i,k}(s,b \to \infty),
\eeq
where $b^*$ can be determined from the amplitude at large $b$, 
in which we assume that  
$A_{i,k}(s,b \to \infty)\,=\,C s^\Delta \exp\Lb - 2 m_\pi b\Rb$. 
The equation for $b^*$ has the form
\beq \label{UB2}
A_{i,k}(s,b \to \infty)\,=\,C s^\Delta \exp\Lb - 2 m_\pi b^*\Rb\,=
\,1, ~~~\mbox{with a solution}~~
b^*\,=\,\frac{\Delta}{2 m_\pi} \,\ln s.
\eeq
Using \eq{UB2}, we obtain from \eq{UB1} that
\beq \label{UB3}
\sigma_{tot}\,\leq \,4 \pi \Lb \frac{\Delta}{2 m_\pi} \,\ln s\Rb^2\,=
\, \mbox{Const }\,\ln^2 s.
\eeq
\subsubsection{Good-Walker mechanism}
\par
We have eluded to the fact  that the processes of diffraction dissociation 
play an 
important role in the description of high energy scattering. Indeed, 
in the framework of the Pomeron approach, they 
provide a qualitative measure of the contribution of the shadowing 
correction.  
The origin of diffraction lies in the wave nature of the scattering 
particles and can be illustrated in the following way: 
In the initial state we have the wave function of two non-interacting hadrons
\beq \label{GW1}
\Psi_{in}\,=\,\psi_{h_1} \psi_{h_2}\,=\,\Lb \sum_{i} c^{h_1}_i \psi_i \Rb 
\,\Lb \sum_{k} c^{h_2}_i \psi_k \Rb, 
\eeq
while the wave function of the final state has the form
\beq \label{GW2}
\Psi_{fin}\,=\,{\bf T} \Psi_{in}\,=\,\sum_{i,k} A_{ik} 
c^{h_1}_i\,c^{h_2}_k\,\psi_i \psi_k \,\neq\,\psi_{h_1} \psi_{h_2}.
\eeq 
\par
Generally, the final sate is not the same as the initial state. Only if 
$\psi_h = \psi_k$, does the interaction lead to the case where the two 
hadrons in the   
final state  are identical to the two hadrons in the initial state.
Therefore, the interaction results in a cross section which is 
proportional to 
$\Big{|}\langle \Psi_{fin }|{\bf T}|\Psi_{in }\rangle\Big{|}^2$, 
which can be re-written in the form
\beq \label{GW3}
\Big{|}\langle \Psi_{fin}|{\bf T}|\Psi_{in} \rangle\Big{|}^2\,=
\,\sum_{i,k} A^2_{ik} \Lb c^{h_1}_i\Rb^2\,\Lb c^{h_2}_k \Rb^2\,=
\,\langle \psi_{h_1} \psi_{h_2}| {\bf T^2} | \psi_{h_1} \psi_{h_2}\rangle.
\eeq
The elastic cross section is proportional to 
$\Big{|}\langle \Psi_{in}|{\bf T}|\Psi_{in}\rangle\Big{|}^2\,=
\,\Big{|}\langle \psi_{h_1}\psi_{h_2}|{\bf T}|\psi_{h_1} \psi_{h_2}\rangle
\Big{|}^2$.
Finally, 
\beq \label{GW4}
\sigma_{diff}\,\,\propto\,\,\langle \psi_{h_1} \psi_{h_2}|{\bf T^2}| 
\psi_{h_1} \psi_{h_2}\rangle
\,-\,\Big{|}\langle \psi_{h_1} \psi_{h_2}|{\bf T}| 
\psi_{h_1} \psi_{h_2}\rangle\Big{|}^2.
\eeq 
\par
These ideas on the origin of diffraction were introduced in the early 50's 
by Landau, Pomeranchuk, Feinberg, Ahiezer, Ter-Mikaelyan and Sitenko (see 
the review by Feinberg and Pomeranchuk\cite{FP}) and were crystallized 
and put into an elegant theoretical framework by Good and Walker\cite{GW}.
\par
\subsection{Reggeon approach}
\par
For more than five decades the Reggeon approach has provided the main tool 
for high energy scattering phenomenology. 
This approach connects the existence of 
resonances with the asymptotic behaviour of high energy scattering
(see Refs.\cite{COL,SOFT,LEREG}). 
When considering the exchange of a resonance with a spin $j$, one has to
also include all excitation with spin $j + 2$, $j + 4$, ...  (keeping 
all other quantum numbers unchanged). 
These particles lie on a Regge trajectory
$\alpha_\reg(t)$ with $\alpha_\reg(t = M^2_j) = j$. 
The contribution to the
scattering amplitude initiated by the exchange
of all resonances can be described as an exchange of
the new object: the Reggeon, and its contribution to the scattering
amplitude is given by a simple function:
\bea\label{RGEX}
&&A_R (s,t)\,\,=\,\,g_p(m_1,M_1,t)\,g_t(m_2,M_2,t) \cdot R\Lb s,t\Rb\,\,\\
&&\mbox{with a Reggeon propagator}\,\,R\Lb s,t\Rb \,\,=\,\,\frac{ \Lb 
\frac{s}{s_0}\Rb^{\alpha_\reg(t)}
\,\pm \,\Lb - \frac{s}{s_0}\Rb^{\alpha_\reg (t)}}{ \sin \pi \alpha_\reg(t)}\,\,\equiv\,\,\eta\Lb t\Rb \Lb\frac{s}{s_0}\Rb^{\alpha_\reg (t)}\nn
\eea
$\alpha_\reg (t)$ is a function of the momentum transfer which we call 
the Reggeon trajectory. \eq{RGEX} reflects the factorization property of 
the Reggeon: the dependence on mass of the interacting hadrons is concentrated 
in the vertices $g_{p}(m_i,M_1, t)$ and $g_{t}(m_2,M_2, t)$, while the 
Reggeon propagator depends only on $s$ and $t$. The functions 
$g_{p,t}(m_i,M_i,t)$ are  phenomenological functions that describe 
the vertex of the Reggeon interaction with hadrons with masses 
$m_i$ and $M_i$. In the region of positive $t$ the zeros of 
$\sin \pi \alpha_\reg(t)$ generate the resonances with a mass 
$M_j$ at $\alpha_\reg(t\,=\,M^2_j)\,=\,j$.  The factor
 $\eta\Lb t\Rb$ depends only on the Reggeon trajectory and
 determines the phase of the amplitude.


     \begin{figure}
  \leavevmode
  \begin{center}
 \includegraphics[width=7cm]{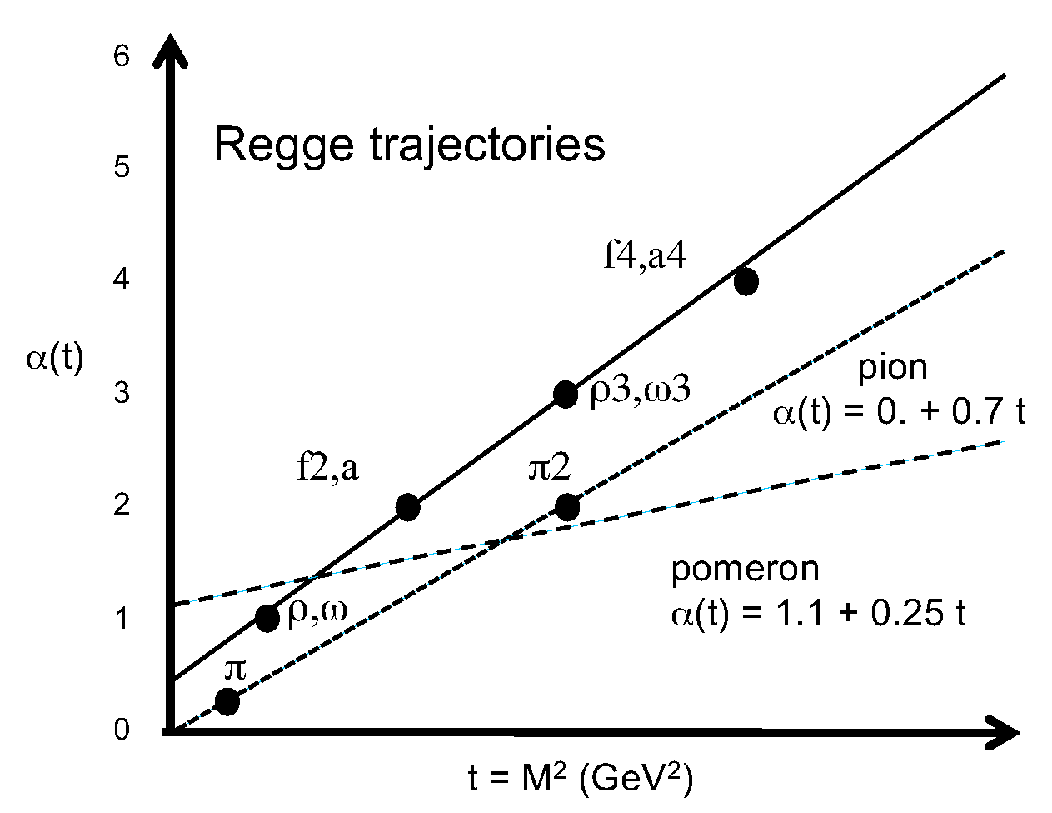}   
  \end{center}
       \caption{ The typical examples of the Reggeon trajectory 
with resonances at $t\,>\,0$ \cite{PDG} adapted from Ref. \cite{RTR}. The
 dotted line in the right figure shows the DL Pomeron. }
                      \label{traj}
        \end{figure}

The name of the new object, as well as the form of the amplitude,
came from analysing  the properties of the scattering amplitude in
the $t$ channel, using the angular momentum representation\cite{COL}. 
\par
Among the Reggeons which  correspond to the measured 
spectrum of their corresponding resonances, there is one Reggeon which 
does not have a resonance on its trajectory. Its existence was assumed 
to be able to describe the behaviour of the total cross section. 
This special Reggeon is called the Pomeron, 
and it has an intercept which is close to unity, 
as the energy behaviour of the total cross section does not display a 
decrease with increasing energy. Since $\alpha_\pom(0) \to 1$ the exchange 
of the 
Pomeron leads to a mostly imaginary amplitude, and it generates processes 
of multiparticle production.
\par
\begin{figure}
\begin{tabular}{c c}
\leavevmode
\includegraphics[width=8cm]{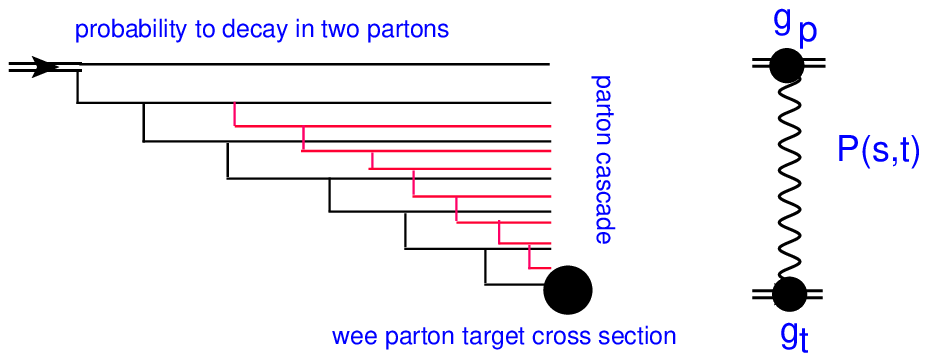} &  
\includegraphics[width=4cm]{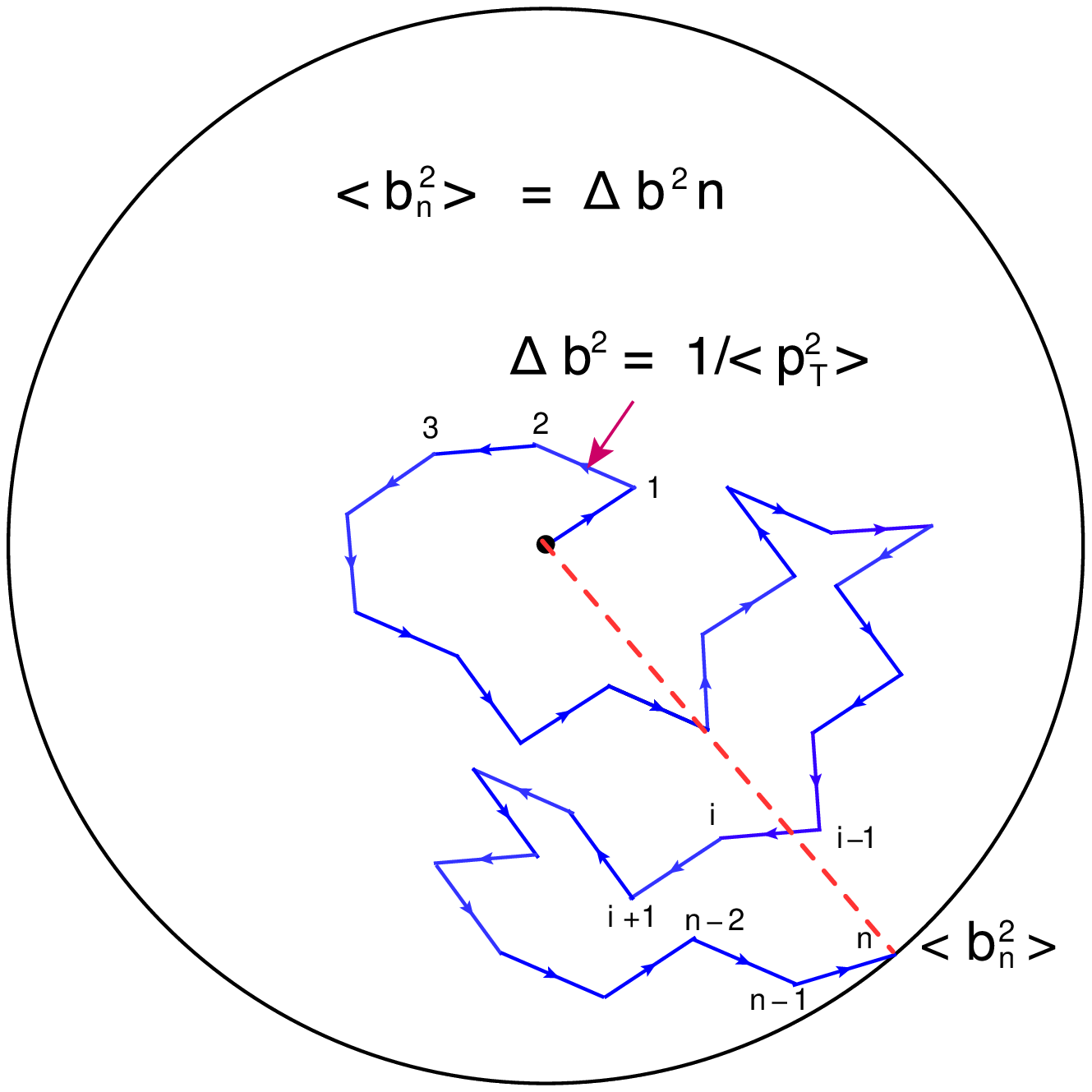}\\
\fig{par}-a &\fig{par}-b\\
\end{tabular}
\caption{ The parton approach to high energy scattering: 
longitudinal structure of the parton cascade (\fig{par}-a) 
and Gribov's diffusion (\fig{par}-b).}
\label{par}
\end{figure}

The $s$-channel structure of the Pomeron has a simple explanation in the 
framework of the parton picture\cite{FEYN,GRIB1,GRIB2} which are 
summarized in the two pictures of \fig{par}.
\par
\fig{par}-a states that the total cross section induced by a single 
Pomeron exchange is equal to
\bea\label{PART}
\hspace{-1.5cm}&&\sigma = \sum_{n-2}^\infty\,
\int^Y_0 \!\!\!\!\!\! d y_1\!\!\!  
\int ^{ y_1} _0\!\!\! \!\!\!d y_2 \dots \!\!\! 
\int^{y_{n-1}}_0\!\!\! \!\!\! \!\!\! d y _n  
\prod^{n-1}_{i=2} \,d^2 p_{i,T} \nn\\
&& \Psi^*\Lb \{ x_i, \vec{p}_{i,T}\}, y_n, \vec{p}_{n,T}\Rb
\,\Psi\Lb \{ x_i, \vec{p}_{i,T},\}, \vec{p}_{n,T}\Rb \sigma_{\mbox{parton}}
\Lb y_n, p_{n,T}\Rb,
\eea
where $\Psi$ is the wave function of the partons (point-like particles) 
which have restricted transverse momentum $p_{i,T} \leq \mu$.
$\mu$ does not depend on the total energy. We assume that the partons are 
distributed uniformly in the rapidity range $(0, Y)$ and the integral 
over $dy_n$ converges.
\par
\fig{par}-b illustrates  Gribov's diffusion picture in the 
transverse plane of the partons populating the parton cascade. 
It is based on the uncertainty principle in which 
$\Delta b \,p_{i,T}\,\sim\,1$ 
for the emission of a parton in the cascade. This figure shows that 
after $n$-emissions the partons are distributed in the area with 
a radius $b^2_n\,\,=\,(1/< p_{i,T}>^2 )\, n$. Since $n \propto Y$ we get 
$R^2\,\propto\,(1/<p_{i,T}>^2) \,Y\,\,=\,\,\alpha'_\pom\,Y$.
\par
\begin{figure}
\begin{center}
\leavevmode
\includegraphics[width=10cm]{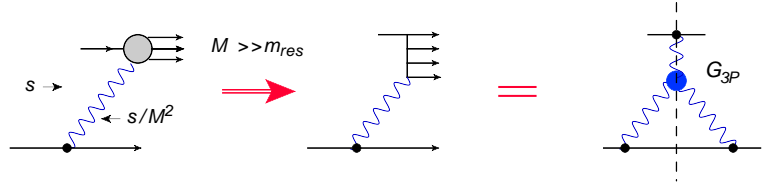} 
\end{center}
\caption{The parton approach to high energy scattering: 
processes of diffraction production of large mass. 
Wavy lines denote Pomerons. The blob shows the triple Pomeron vertex.}
\label{3pom}
\end{figure}

In the parton approach the processes of  diffraction dissociation 
at large mass ($M > m_{res}$ where $m_{res}$ is the mass of the 
resonances) is closely related to the interaction of the Pomerons, 
see \fig{3pom}.
It could be described by the following expression
\beq \label{3POM}
\frac{M^2 \,d \sigma^{3\pom}_{diff}}{d M^2}\,=
\,2\,\int d t\, g_p(t=0)\,g^2_t(t)\,G_{3 \pom}(t)\, 
\mbox{Im}P\Lb M^2,t=0\Rb \,P\Lb \frac{s}{M^2},t\Rb\, P^*\Lb 
\frac{s}{M^2},t\Rb. 
\eeq
If $\Delta_\pom\,=\,0$, this contribution corresponds to a  
diffractive channel, which is not included in the GW mechanism.  
Averaging \eq{GW3} leads to a divergent series. 
However, in the case of $\Delta_\pom \,>\,0$, 
the integral of \eq{3POM} over $M$ is convergent, and this contribution 
is a part, of the GW mechanism\cite{GUS}. 
\fig{3pom} shows an example of a triple Pomeron interaction. 
\par
Our knowledge of the Pomeron interaction was summarized in 
Gribov Pomeron calculus\cite{GRIB3}, which can be formulated by the 
following path integral:
\begin{eqnarray}
&&Z[\Phi, \Phi^+]\,=\,\int\,D \Phi\,D\Phi^+\,e^S\,\,\,
\mbox{with}\,\,S \,=\,S_0\,+\,S_I\,+\,S_E\,\nonumber\\,
&&S_0\,=\,\int d Y\,d^2 b\, \Phi^+(Y,b )\,\left\{-\,\frac{d}{dY}\,\,+
\,\,\Delta\,\,+\,\alpha'_\pom \nabla^2 \right\} \Phi(Y, b);\nonumber\\
&& S_I\,\,=\,\,G_{3\pom} \int dY\,d^2 b\,\left\{\Phi(Y,b)\,\Phi^+(Y,b)\,
\Phi^+(Y,b)\,\,+\,\,h.c.\right\}.\label{GRT}
\end{eqnarray} 
\par    
$S_E$ specifies the interaction with hadrons or nuclei. 
$\Phi\Lb Y,r\Rb$ describes the Pomeron with rapidity $Y$ and 
impact parameter $b$. It turns out that \eq{GRT} has a simple 
statistical interpretation and can be re-written as an equation 
presenting the probability $P_n$ to have $n$-Pomerons at rapidity $Y$. 
The equation has the form specified in Ref.\cite{GRPO,BOR}.
\bea  \label{EQCAS}
-\,\frac{\partial\,P_n(y, b)}{\partial \,y}\,\,+
\,\,\alpha'_\pom \nabla^2 P_n(y, b)\,\,&=&
\,\,G_{3\pom}\left\{ -\,n\,P_n(y, b)\,+\,\,(n-1)\,P_{n-1}(y,b) 
\right\}\,\,\,\label{STI}\\
& + &\,\,G_{3\pom}\,\left\{ -\,n\,(n - 1)\,P_n(y, b)\,
+\,(n+1)\,n\,P_{n+1}(y, b) \right\}.\nonumber
\eea 
\par
The problem of the Pomeron interactions has not been 
completely solved, as 
we have failed to find theoretical arguments for restricting the 
number of  Pomeron interaction vertices, 
as well as finding the term $S_E$ in 
\eq{GRT}.
\subsection{Perturbative QCD}
\par
In pQCD the high energy Pomeron behaviour,  
arises in a natural way with an intercept 
$\alpha_\pom(0)\,=\,1\,+\,\mbox{C}\bas$ \cite{LONU,BFKL,LI}, 
where $\bas$ is the QCD coupling. However, in spite of the power-like 
increase with energy, the expression for the high energy contribution 
to the scattering amplitude, differs significantly  from 
the corresponding Pomeron expression. The best way to see this is to 
compare the QCD contribution 
(which is a BFKL Pomeron) with \eq{RGEX}. 
The propagator of the BFKL Pomeron for the scattering of two dipoles with 
sizes $r$ and $R$ has the form \cite{BFKL}
\bea \label{BFKLPOM}
&&P_\pom \Lb Y, b \Rb\,\Longrightarrow \,P_{\mbox{\tiny BFKL}}\Lb Y,r,R\Rb\,
=\,\frac{r\,R}{\sqrt{Y}}\,e^{ \Delta_{\mbox{\tiny BFKL}}\,Y\,-\,\frac{\ln^2\Lb 
r^2/R^2\Rb}{D \,Y}}, \\
&&\mbox{with}~~\Delta_{\mbox{\tiny BFKL}}\,=\,2 \ln 2\,\bas~\mbox{and} ~D\,
=\,14 \zeta(3) \bas \,=\,16.828 \,\bas\nn.
\eea
\par
Note that:\\ 
(i) The BFKL Pomeron is not an angular 
momentum pole but a branch cut, since its Y-dependence  
has an additional $\ln s$ term.\\ 
(ii) It does not depend on the impact parameter.\\ 
(iii) The BFKL propagator 
depends on the sizes of dipoles, consequently, 
the BFKL Pomeron does not factorize..
\par
Despite a different structure of the Pomeron in QCD, 
the key partonic formula 
of \eq{PART} works with $\sigma_{\mbox{parton}}
\Lb y_n, p_{n,T}\Rb\,\,=\,\,\sigma^{BA}_{\mbox{\tiny dipole-dipole}},$ 
where $\sigma^{BA}$ is calculated in the Born approximation of 
pQCD. This cross section does not depend on $Y$.
The similarities between the QCD and the partonic cascades, 
leads to the same triple BFKL Pomeron mechanism for diffractive 
production of large masses shown in \fig{3pom}. 
The similarities between the Reggeon calculus and the interaction between   
BFKL Pomerons, is clearly seen in the path integral formulation of   
the BFKL Pomeron interaction, where, instead of \eq{GRT}, the
$S_0$ and $S_I$ contributions have different forms\cite{BRN}
than in \eq{GRT}:
\bea 
&&S_0\,=\,\int\,dY\,dY'\,d^2 x_1\, d^2 x_2\,d^2 x'_1\, d^2 x'_2\,
\Phi^+(x_1,x_2;Y)\,\nabla_1^2 \nabla^2_2 \Lb \frac{\partial}{\partial Y}\,-\, 
{\cal H}\Rb \,\Phi(x'_1,x'_2;Y');\nn\\
&&S_I\,=\,\frac{2\,\pi \bas^2}{N_c}\,\int \,d Y'\,\int
\,\frac{d^2 x_{1} d^2x_{2}\,d^2 x_3}{x^2_{12}\,x^2_{23}\,x^2_{13}}\,\label{S}\\
&&\cdot\{\left(x^4_{12}\nabla^2_1\nabla^2_2\Phi(x_1,x_2;Y')\,\right)\,\cdot\,
\Phi^+(x_1,x_3;Y')\,\Phi^+(x_3,x_2;Y')\,\,+\,\,h.c. \}; \nn\\
&&
{\cal H} f(x_1,x_2;Y) \,\,= \,\,\frac{\bas}
{2\pi}\,\int\,d^2 x_3\,\frac{x^2_{12}}{x^2_{23}\,x^2_{13}}\,
\left(f(x_1,x_2;Y)\,-\,f(x_1,x_3;Y)\,-\,f(x_3,x_2;Y) \right). \nn
\eea
\par
Comparing \eq{GRT} and \eq{S}, one can see that in QCD, we have the
 field depending  on the size of 
the interacting dipole. The particular expression 
for the triple Pomeron interaction depends on the strength of the 
triple Pomeron coupling, it is of the order of 
$\bas^2$ $(G_{3\pom} \propto \bas^2$). 
There is no Gribov's diffusion (the term $\alpha'_\pom \nabla^2_b$ 
in \eq{GRT}). In principle, the four Pomeron interaction should appear in 
\eq{S} but it is suppressed (see Ref.\cite{AKLL}).
\subsection{N=4 SYM}
\par

At present N=4 SYM is the only theory we know which can deal with a 
large coupling constant. 
Hence, we  use this theory as the guide to handle  physics phenomena 
in this regime. The attractive feature of this theory is that N=4 SYM
 with small coupling, leads to normal QCD-like physics (see Refs.
\cite{POST,BFKL4})  with OPE  and linear equations for DIS,
 as well as the BFKL equation for the high energy amplitude.
The high energy amplitude reaches the unitarity limit: black disc
 regime, in which half of the cross section is due to elastic
 scattering, and half is associated with processes of  multiparticle 
production.

\begin{figure}[ht]
\centerline{\epsfig{file=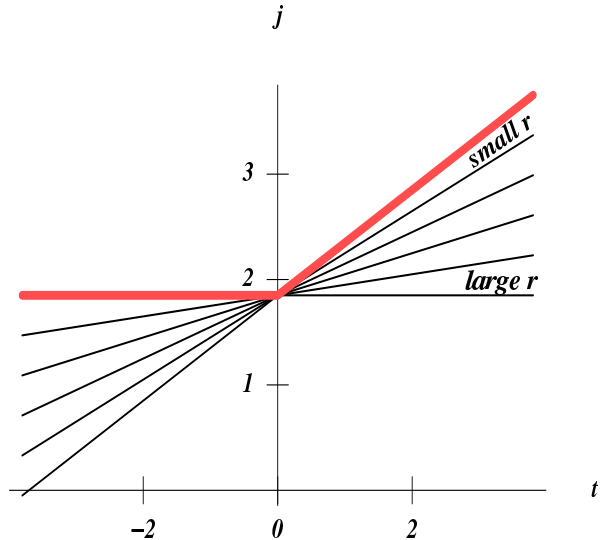,width=85mm}}
\caption{The behaviour of the Pomeron trajectory in N=4 SYM according 
to Ref.\cite{BST}. The figure is taken from Ref.\cite{BST}}
\label{pom4}
\end{figure} 
This theory has an analytical solution due to AdS/CFT 
correspondence, and can be reduced to  weak gravity 
in $AdS_5$ space.
\par
In the strong coupling limit, the following are 
the main features of this theory\cite{BST,HIM,COCO,BEPI,LMKS}:\\ 
(i) It has a soft Pomeron which, in this case, is the reggeized graviton 
with a large intercept $\alpha(0)_\pom = 2 - 2/\sqrt{\lambda}$, where, 
$\lambda = 4 \pi N_c \alpha^{YM}_S$.  $\alpha^{YM}_S $ is the QCD-like 
coupling;\\ 
(ii) The main contribution to the total cross section at high 
energy is due to the processes of elastic scattering and 
diffractive dissociation;\\  
(iii) The leading Pomeron trajectory has a form shown in 
\fig{pom4}. Namely, a Regge pole with $\alpha'_\pom = 0$ in the 
scattering region ($t \,<\, 0$), while $ \alpha'_\pom \,>\, 0$
in the resonance region of positive $t$;\\  
(iv) The Pomerons (gravitons) interact with the triple Pomeron vertex 
which is small ( $\propto 2/\sqrt{\lambda}$);\\  
(v) The only source of diffraction production is the GW mechanism 
in which the fifth coordinate $z$ plays a role of the degree of freedom. 
At first sight the small value of $\alpha'_\pom$ is not related to the 
small size of the partons in this theory. It is related to the small 
values of the fifth coordinate $r$ (see \fig{pom4}). The physical 
meaning of this coordinate is the typical size of the colliding particles.  
It should be stressed that all these features appear in the theory and 
therefore, for the first time, we have a theoretical justification for the 
Reggeon-type phenomenology for high energy scattering. 
\par
To summarize, our intent is to consider  
the Pomeron in N=4 SYM, which is based on the discovery 
that, actually, it is a BFKL Pomeron. 
The following glossary aims to translate QCD to this theory.
\par
\begin{center}
{\bf  Glossary~~~ ${ \equiv}$\,\,~~~ AdS-CFT correspondence:}
~
\begin{tabular}{  c c c}
N=4 SYM &    & { QCD}\\
&~~~~~~~~~~~ & \\
{Reggeized graviton} \,&${ \iff}$&\,{ BFKL Pomeron}\\
{$z $} \,&${ \iff}$ & $\,r $\,(dipole size)\\
$ 1 - 2/\sqrt{\lambda}$ & $ \iff$&\,$ \Delta_{BFKL}$ 
(intercept of the BFKL Pomeron)\\
$2/\sqrt{\lambda} $\,&$ \iff$&\, $D_{BFKL}$ (see \eq{BFKLPOM}) \\
\end{tabular}
\end{center}
\section{General features of our model}
\subsection{Main assumptions and parameters of our model}
\par
We have built our model using the main characteristic features 
of N=4 SYM and QCD approaches.
We assume:
\begin{itemize}
\item \quad {\it The Pomeron is a Regge pole}. This assumption is made 
so as to simplify our calculations, as we need 
an approach which is convenient to determine our parameters.
\item \quad  {\it  $\Delta_\pom$ is large ($0.2 \div 0.3$)}. 
As we have discussed, the intercept of our Pomeron turns out to be 
large, both in N=4 SYM and in pQCD. 
The range of the values for $\Delta_\pom$ is taken from a fit of 
deep inelastic HERA data in the framework of N=4 SYM \cite{LEPO}.
\item\quad
The fact that we are successful in describing the experimental data with 
an input 
 BFKL-like (``hard") Pomeron,
means that  as a result of  screening corrections our "hard"  Pomeron 
   transmutes into a "soft"
 Pomeron, with a small effective  $\Delta_{\mbox{eff} (\pom)} \approx 
0.1$.   In other words, our input Pomeron which is BFKL-like ("hard"), 
 due to screening corrections, behaves like a "soft" Pomeron.
\item \quad {\it $\alpha'_\pom \,\,=\,\,0$}. This constraint stems both 
from N=4 SYM and pQCD. We have checked that the  
data, including the LHC output, imposes a very small value 
$\alpha'_\pom \leq \,0.028 \,GeV^{-2}$ (see below).
\item \quad {\it Large GW components}. Large GW components occur
naturally  in N=4 SYM, in which this mechanism is the only source 
of the diffractive production. For the sake of simplicity we replace the 
rich structure of the produced states by one wave function, and  
develop 
a two channel model to describe the GW mechanism.
\item 
\quad {\it Only $G_{3\pom}$}. In QCD, the triple BFKL 
Pomeron vertex is small ($\propto\,\bas^2$), and the vertices for more 
than three Pomeron interactions contain an additional suppression. 
We restrict ourselves, and consider only $G_{3\pom}$ coupling, so as to 
provide a 
natural matching with the QCD approach of \eq{S}.
\item \quad {\it  $G_{3\pom}$ is small}. 
In QCD, $G_{3\pom}\,\propto\,\alpha^2_S$ while in N=4 SYM 
$G_{3\pom} \ll 2/\sqrt{\lambda}$.
\end{itemize}
\subsection{The Lagrangian of the model}
\subsubsection{$S_0$ and $S_I$}
\par
Using the above assumptions in terms of a functional integral we get:
\beq\label{FI}
Z[\Phi, \Phi^+]\,\,=\,\,\int \,\,D \Phi\,D\Phi^+\,e^S 
\,\,\,\mbox{with}\,\,\,\,S \,=\,S_0\,+\,S_I\,+\,S_E,
\eeq
where, 
\beq \label{S0}
S_0\,=\,\int d Y \Phi^+(Y)\,\Big\{-\,\frac{d }{d Y}\,
+\,\Delta_\pom\Big\}\,\Phi(Y)
\eeq
describes the  free Pomeron trajectory with an intercept $\Delta_\pom$ 
and a slope 
$\alpha'_\pom = 0$. 
These two features that occur both in N=4 SYM and QCD, have
been included in \eq{S0}.
\par
$S_I$ characterizes the interaction between Pomerons and has the form:
\beq \label{SI}
S_I\,\, =\,\,G_{3\pom}\int d Y\,\Big\{\Phi(Y)\,\Phi^+(Y)\,\Phi^+(Y)\,
+\,h.c. \Big\}.
\eeq
Note that we only take into account  the triple Pomeron interaction. 
This form provides a natural matching with the pQCD 
approach\cite{BART},
and with the BFKL Pomeron calculus (see \eq{S})\cite{BRN}.
We will specify $S_E$ which is responsible for the interaction of the
target and colliding projectile, in two processes: 
proton-proton 
and proton-nucleus interactions.
\par
Reggeon Field Theory with the action given by \eq{S0} and \eq{SI}
has been solved for arbitrary $S_E$  (see  Refs. \cite{AMATI,KOLE}). 
 This theory leads to a total cross section that 
decreases at ultra high energies. Thus it is necessary to include the 
four Pomeron 
vertex to prevent this decrease\cite{BOMU}. 
We will show below that this decrease occurs at ultra high energies, 
and we will specify the range of energies for which our model  
is trustworthy.
\subsubsection{$S_E$ for hadron-hadron collisions}
\par
We need to incorporate in our procedure a sufficiently large 
GW\cite{GW} component which is required so as to  describe    
low mass diffraction, and
which follows from the N=4 SYM approach\cite{BST}.
We develop a two channel model which takes into account  
the GW mechanism, and in which the observed physical 
hadronic and diffractive states are written in the form 
\beq \label{MF1}
\psi_h\,=\,\alpha\,\Psi_1+\beta\,\Psi_2\,;\,\,\,\,\,\,\,\,\,\,
\psi_D\,=\,-\beta\,\Psi_1+\alpha \,\Psi_2, 
\eeq
where, $\alpha^2+\beta^2\,=\,1$. Note that GW diffraction 
is presented by a single wave function $\psi_D$.
The wave functions $\Psi_1$ and $\Psi_2$ diagonalize the interaction 
matrix (see \eq{GT1} with $i,k = 1,2$).
Bearing in mind \eq{MF1}, we can write $S_E$ in the form
\beq \label{SEP}
S_E\,=\,-\,\int dY'd^2b \sum^2_{i=1}\Big\{\Phi\Lb Y'\Rb g^{(i)}\Lb 
b\Rb\,\delta\Lb Y'\,-\,0\Rb\,+\,\Phi\Lb Y'\Rb g^{(i)}\Lb b 
\Rb\,\delta\Lb Y - Y'\Rb\Big\}.
\eeq
$g^{(i)}$ denotes the vertex of the Pomeron  
interaction with the state $i$. This is described 
by either the wave functions $\Psi_1$ or $\Psi_2$. 
$Y'\,=\,0$
($Y'\,=\,Y$) 
indicate respectively, the
position of the target and projectile, in rapidity.  
$b$ denotes the impact parameter.
We parameterize $g^{(i)}\Lb b \Rb$ as 
\beq \label{SPB}
g^{(i)}\Lb b \Rb\,=\,g^{(i)}\,S\Lb b \Rb\,=
\,\frac{g^{(i)}}{4\pi}\,m^3_i\,b\,K_1\Lb m_i\,b \Rb,
\eeq
where, $S\Lb b \Rb$ is the Fourier 
transform of the dipole formula for
the form factor $1/((1 + q^2/m^2_i)^2$. $K_1(z)$ is
the modified Bessel function of a second kind , the McDonald function,
(see Ref.\cite{RY} formulae {\bf 8.4}). 
\subsubsection{$S_E$ for hadron-nucleus scattering}
\par
Using \eq{SEP}, and neglecting the correlations between nucleons in a
nucleus, the $S_E$ term can be written in the form:
\bea \label{SEA}
&&S_E\,=-\,\int d Y' d^2 b \sum^2_{i=1}\\
&&\Big\{ \Phi\Lb Y'\Rb g^{(i)}\Lb b \Rb\,\delta\Lb Y-Y'\Rb\,+\,\Phi\Lb Y'\Rb 
\int d^2 b' \,g^{(i)}\Lb \vec{b} - \vec{b}^{\,'} \Rb\,S_A\Lb b'\Rb\,\delta
\Lb Y'-0\Rb\Big\}\nn.
\eea
The last term is well known, and we refer the reader
to Refs\cite{BRN,LMKS} which, as far as we know, are the most
recent papers where this derivation is based on Feynman
diagrams. For heavy nuclei $|\vec{b} - \vec{b}^{\,'}|\,\ll\,R_A$
and the second term in \eq{SEA} can be replaced by:
\beq \label{SEA1}
\Phi\Lb Y'\Rb \int d^2 b' \,g{(i)}\Lb \vec{b}-\vec{b}^{\,'}\Rb\,S_A\Lb b'
\Rb\,\,\xrightarrow{|\vec{b}-\vec{b}^{\,'}|\,\ll\,R_A}\,\Phi\Lb Y'\Rb\,g^{(i)}
\,S_A\Lb b\Rb.
\eeq
However, in the case of not very heavy nuclei (air for example), 
the radii are not very large and we cannot neglect the $b'$ dependence. 
$S_A\Lb b\Rb\,=\,\int dz\rho\Lb z,b\Rb$, with $\int d^2\,S_A
\Lb b\Rb\,=\,A$. $\rho$ denotes the density of the nucleons in a nucleus.  
\subsubsection{Small parameters and selection of the Pomeron diagrams}
\par
Using \eq{FI},\eq{S0},\eq{SI} and \eq{SEP} 
we can find expressions for all 
experimental observables measured in 
proton-proton interactions\footnote{We need to introduce additional 
phenomenological parameters to describe the main characteristics 
of the inelastic processes, which take into account the hadronization stage, 
in terms of a microscopic approach based on QCD.}.  
We simplify the problem using the fact that in N=4 SYM, the vertices of 
the 
Pomeron interaction with hadrons turn out to be larger than the triple Pomeron 
vertex. Taking this into account we can
define  
a new small parameter, 
\beq \label{NSP}
Q\,\,=\,\,\gamma^2 \,s^{\Delta_\pom}\,\,\,\ll\,\,1\,; ~~~~~\mbox{while}
~~~~H\,\,=\,\,g^{(i)}\,G_{3 \pom}\,s^{\Delta_\pom}\,\,\geq\,\,1
\eeq
\par
\begin{figure}[t]
\centerline{\epsfig{file=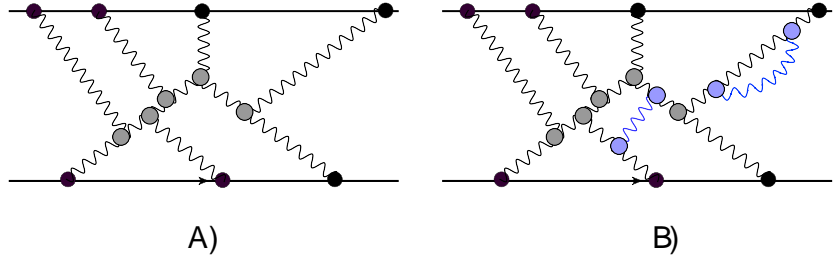,width=110mm}}
\caption{The set of diagrams that contribute to the scattering amplitude of 
proton-proton scattering in the kinematic region given by \eq{NSP}.
\fig{ppset}-A shows the net diagrams which are proportional to $H^{n}$.
\fig{ppset}-B shows two examples of the diagrams which are proportional 
to an additional power of the small parameter $Q$. 
This figure which is proportional to an extra $Q^2$ was neglected.
The wavy lines denote the soft Pomerons. The black circles denote
$g^{(i)}$, while the gray circles describe the triple Pomeron vertices.}
\label{ppset}
\end{figure}

The set of parameters that we will discuss below confirm our 
expectations. In \eq{NSP} 
$\gamma^2 \,=\,\int\,G^2_{3 \pom}\Lb k_{T,1}=0,k_T,k_T\Rb d^2 k_T,$ 
were $k_{T,i}$ are transverse momenta of the three Pomerons. 
The main contributions, which are proportional to $H^n$
stem from the `net' diagrams of \fig{ppset}-a. 
The small parameter $Q$
is obtained from the diagram of \fig{ppset}-b in which 
the interaction is between two Pomerons that
are not attached to a proton or to the Pomeron loop diagrams (see two 
examples in \fig{ppset}-b).
\subsection{Summing large Pomeron loops (Mueller-Patel-Salam-Iancu (MPSI) 
approximation)}
\par
As has been mentioned, the approach given by the functional integral of 
\eq{FI} can be solved \cite{AMATI,KOLE}. 
However, in our model we prefer to 
develop the approximate method of calculation which leads to a simpler 
set of formulae. These formulae allow us to organize the fitting 
procedure in an economical way.
The main idea of this approximation, which we call the 
Mueller-Patel-Salam-Iancu (MPSI) approximation,
is the following: At high energy, in the kinematic region
\beq \label{KR}
Y\,\leq\,\frac{\Delta^2_\pom}{g^2_{3 \pom}}\,\equiv\,\frac{1}{\gamma},
\eeq
only large Pomeron loops, with a rapidity size of the order of $Y$,
contribute to the high energy asymptotic behaviour of the scattering 
amplitudes.
\subsubsection{The simplest loop diagram.}
\par
\begin{figure}[ht]
\centerline{\epsfig{file=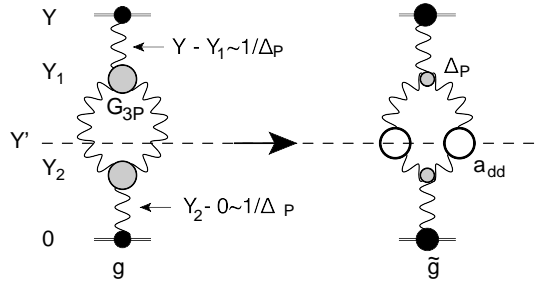,width=70mm}}
\caption{MPSI approximation: an example of the first loop diagram. 
Wavy lines denote the BFKL Pomerons,
the blob stands for the scattering amplitude ($a_{dd}$) of two partons  
and $G_{3\pom}$ denotes the triple BFKL Pomerons vertex. $\Delta_\pom$  
the Pomeron intercept while $g$($\tilde g$) denotes the Pomeron-hadron 
vertex.}
\label{mpsi1}
\end{figure}

The MPSI approximation has been discussed in detail
in Refs. \cite{MPSI,LEPR,LMP,GLMM}. 
Here we illustrate the method using the
example of the first Pomeron loop
diagram given in \fig{mpsi1}. Using \eq{FI} or the generating
function approach, we obtain the
contribution of this diagram in the form
\bea
A\Lb \fig{mpsi1}\Rb\,&=&\label{MPSI0}\\
&=&  \,-\,g_1\,g_2 G^2_{3\pom}\,
\int^Y_0\,d\,Y_1\,\int^{Y_1}_0\,d\,Y_2\,P(Y - Y_1)\,
P^2(Y_1 - Y_2)\,P(Y_2 - 0) \,\nn\\
&=&\,-\,g_1\,g_2 G^2_{3\pom}\,\int^Y_0\,d\,Y_1\,
\int^{Y_1}_0\,d\,Y_2\,\,e^{\Delta\,(Y + Y_1 - Y_2)}\,\,\nn\\
&=&\,-\,\,\,\frac{g_1\,g_2 G^2_{3\pom}}{\Delta^2_\pom}\,
\left\{e^{2\,\Delta_\pom\,Y}\,\,+\,\,\,e^{\,\Delta_\pom\,Y}\,\,
+\,\,\Delta_\pom\,Y\, e^{\,\Delta_\pom\,Y} \right\}\,\nn\\
&=& \,-\,\tilde{g}_1\,\tilde{g}_2\,\left\{a^2_{dd}\,e^{2\,\Delta_\pom\,Y}\,\,
+\,\,a^2_{dd}\,e^{\,\Delta_\pom\,Y}\,\,+\,\,\Delta_\pom\,a^2_{dd}\,Y\,
e^{\,\Delta_\pom\,Y} \right\} \label{MPSI1}.      
\eea
$G_{3\pom}\,=\,\Delta_\pom\,a^2_{dd}$, $\tilde{g}_i\,=\,g_i/\sqrt{a_{dd}}$
(see the notation in \fig{mpsi1}).
\par
The main idea of the MPSI approximation is to take into account only 
the first term in \eq{MPSI1}, while neglecting other terms, 
since they are 
suppressed by  $\exp[-\Delta_\pom Y]$. This term results from  the   
integration of 
$Y - Y_1 \approx 1/\Delta_\pom$ and $Y_2 - 0 \approx 1/\Delta_\pom$ 
(see \fig{mpsi1}).\\
\subsubsection{Dressed Pomeron.}        
\par
\begin{figure}[ht]
\centerline{\epsfig{file=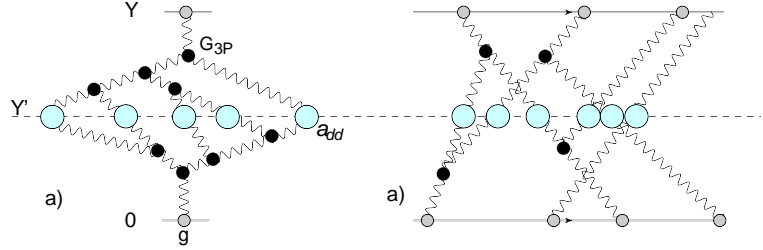,width=100mm}}
\caption{MPSI approximation: enhanced diagrams (\fig{mpsi}-a) 
and the net diagrams of \fig{ppset}-a (\fig{mpsi}-b). 
Wavy lines denote the BFKL Pomerons,
the blob stands for the scattering amplitude of two partons  
$G_{3\pom}$ is the triple BFKL Pomerons vertex.}
\label{mpsi}
\end{figure}

In the following we utilize the MPSI approximation 
so as to find the sum of enhanced diagrams (see \fig{mpsi}) 
which change the Green function of the Pomeron. 
From \fig{mpsi} it is clear that the MPSI approximation reduces the sum 
of 
the Pomeron loops to the product of two cascade of Pomerons, where only 
the 
splitting of one Pomeron into two, is taken into account. 
The Pomeron cascade satisfies \eq{EQCAS}, which can be reduced to the 
equation of the generating function \cite{MUCD,LALE,LELU}:
\beq   \label{Z}
Z(y,\,u)\,\,=\,\,\sum_n\,\,P_n(y)\,\,u^n,
\eeq
where, $P_n\Lb y\Rb$ is the probability to find $n$-Pomerons 
at rapidity $y$.
At rapidity $y=Y$, there is only one fastest parton (dipole), 
which is $P_1(y\,=\,Y)\,=\,1$, while $P_{n>1}(y\,=\,Y)\,=\,0$. 
This is the initial condition for the generating function:
\beq \label{INC}
Z(y\,=\,Y)\,=\,u\,;~~~~~~~~~~~
Z(y,\,u\,=\,1)\,\,=\,\,1.
\eeq
The condition at $u=1$, follows from the physical meaning of $P_n$ 
as a  probability.
\par 
The generating functions of the projectile $Z^p \Lb Y -Y' \Rb$ 
and the target  $Z^t \Lb Y' \Rb$ (see \fig{mpsi}) satisfy 
a very simple equation that describes the parton cascades, 
in which a parton can only decay  into two partons. 
This equation follows from \eq{EQCAS}. It has the form:
\beq \label{GFEQ}
\,\,-\frac{\partial\,Z(y,\,u)}{\partial\, y}\,\,
=\,\,-\,\Delta_\pom\,u\,(1\,-\,u)
\,\,\frac{\partial\,Z(y,\,u)}{\partial\, u}.
\eeq
The solution of the equation above is:
\beq \label{GFSOL}
Z\Lb y, u\Rb\,\,=\,\,\frac{u}{u \,\,+\,\,(1 - u)\,e^{\Delta_\pom y}}. 
\eeq
\eq{GFSOL} satisfies the initial and boundary conditions of \eq{INC}.  
The amplitude in the MPSI approximation has the following 
form\cite{MPSI,LELU}:
\bea \label{MPSI6}
&& N^{MPSI}\Lb Y\Rb\,\,=\\
&&=\,\,\sum^\infty_{n=1}\,\frac{(-a_{dd})^n}{n!}\,\Lb\frac{\partial}
{\partial  u^{(1)}}\Rb^n\, Z\Lb Y-Y'; u^{(1)}=1\Rb\,\Lb\frac{\partial}
{\partial u^{(2)}}\Rb^n Z\Lb Y' - 0; u^{(2)}=1\Rb\nonumber\\
&&=\,\,1\,\,-\,\,\exp\left\{\,-\,a_{dd} \,\frac{\partial}
{\partial u^{(1)}}\,\frac{\partial}{\partial
u^{(2)}}\,\right\}\,Z\Lb Y - Y'; u^{(1)} \Rb\,
Z\Lb Y'- 0; u^{(2)}\Rb |_{u^{(1)},=\,u^{(2)}= 1}\nonumber.
\eea

From \eq{MPSI6} we can see that the MPSI approximation is the 
$t$-channel unitarity constraint adjusted to Reggeon Calculus, 
in the form of a generating function.  
From this picture we can find the 
sum of enhanced diagrams from the knowledge of the 
cascade described by the `fan' diagrams. The physical meaning of the 
 parameters are: $a_{dd}$ is the low energy amplitude for 
two partons (dipoles) scattering at an arbitrary rapidity $Y'$, and 
$\Delta_\pom$ is the value of the vertex for the decay of one parton (dipole) 
to two parton (dipoles). 
It should be stressed that the answer does not depend on the value of 
$Y'$, which  should be chosen somewhere in the central region of the 
scattering.
\par
Using \eq{MPSI6} we obtain the Green's function of the dressed Pomeron 
in a closed form\cite{GLMM}:
\beq \label{MPSI7}
G\Lb Y\Rb\,=\,1 \,-\,\exp\Lb \frac{1}{T\Lb Y\Rb}\Rb\,\frac{1}{T\Lb Y\Rb}
\,\Gamma\Lb 0,\frac{1}{T\Lb Y\Rb} \Rb~~\mbox{with}~~
T\Lb Y \Rb \,=\,a_{dd}\,e^{\Delta Y}.
\eeq
\subsubsection{Net diagrams of \fig{ppset}-a (Ref.\cite{GLMMA})}
\par
Before starting the discussion on the summation of different processes 
in the MPSI approximation, we wish to make a general remark 
on our further presentation.
 Thus far, we have discussed the derivation of our results  
and the results of our computations, 
however, lack of space does not allow us to continue in this user 
friendly style. 
 Most of the 
theoretical calculations are new and  are  dispersed  over all our 
papers.
For the convenience of the reader 
we have included in the titles of the subsections, the references to the 
relevant papers. 
\par
Using the MPSI approach, the sum of the Pomeron diagrams, with $S_E$ given 
in \eq{SEP}, has been calculated in Ref.\cite{GLMMA}. 
An especially simple form presents the sum of the net diagrams of 
\fig{ppset}-a\cite{GLMMA}. For the amplitude that has been introduced 
in \eq{GT1},\eq{GT2} and \eq{GT3}, we obtain:
\beq \label{MPSI8}
A_{i,k}\Lb Y; b\Rb\,\,
=1 \,\,-\,\,\exp\left\{-\int d^2 b'\,
\,\frac{\Lb \tilde{g}_i\Lb\vec{b}'\Rb\,\tilde{g}_k\Lb\vec{b} - 
\vec{b}'\Rb\,T(Y)\Rb}{1\,+\,T(Y)\,\left[\tilde{g}_i\Lb\vec{b}'\Rb 
+ \tilde{g}_k\Lb\vec{b}-\vec{b}'\Rb\right]}\right\}.
\eeq 
\subsubsection{Diffractive production (Refs.\cite{LEPR,GLMM,GLMMA})} 
\par
\begin{figure}[ht]
\begin{tabular}{c c  c}
\epsfig{file=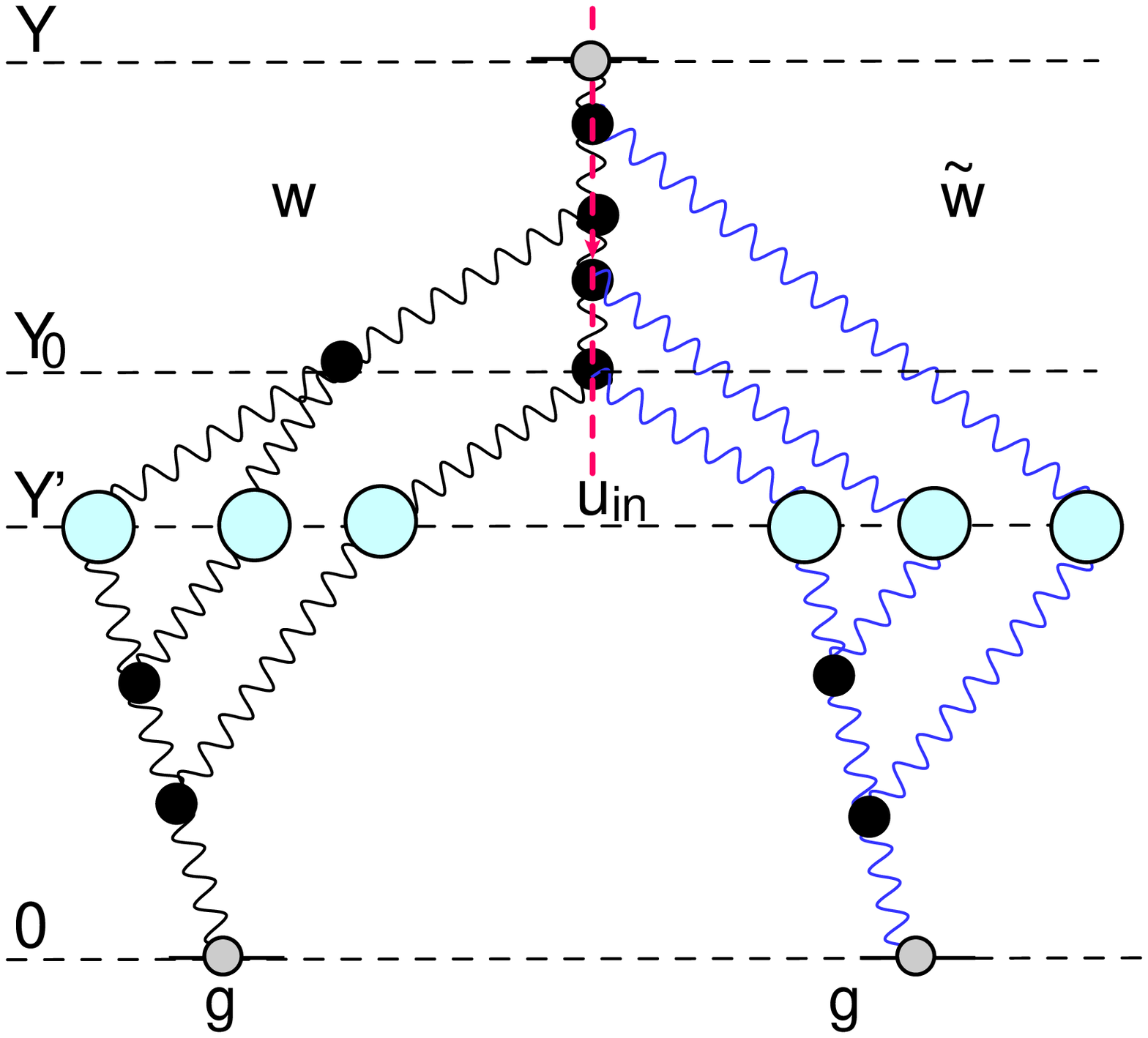,width=50mm,height=50mm}&~~~~~~~~~&
\epsfig{file=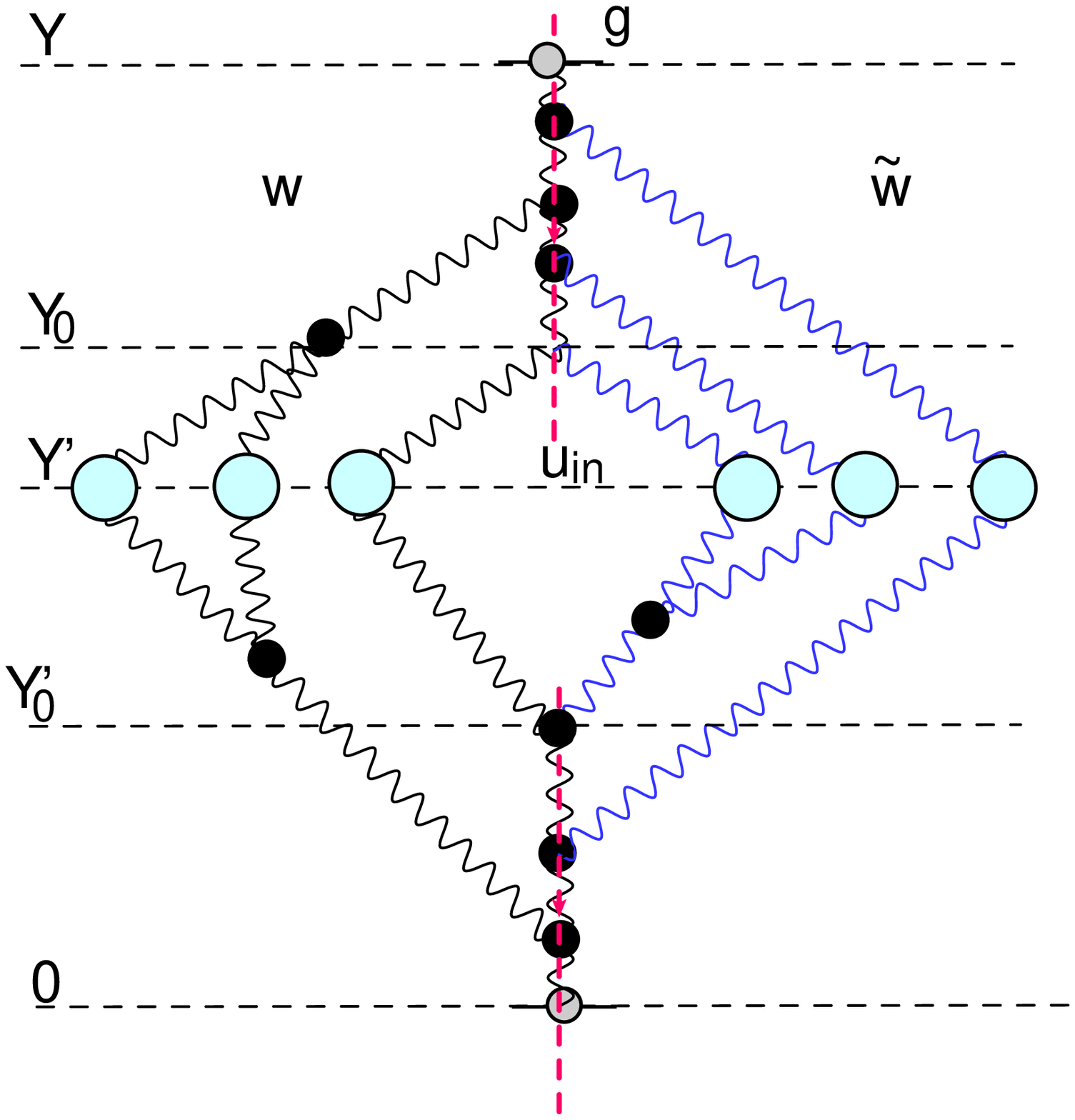,width=50mm,height=50mm}\\
\fig{mpsid}-a & &\fig{mpsid}-b\\
\end{tabular}
\caption{MPSI approximation: single (\fig{mpsid}-a) and double 
(\fig{mpsid}-b) diffractive production. Wavy lines denote the BFKL Pomerons,
the blob stands for the scattering amplitude of two partons.  
$G_{3\pom}$ is the triple BFKL Pomerons vertex. Wavy lines that are 
crossed by the dashed lines denote the cut Pomeron.}
\label{mpsid}
\end{figure}

As we have mentioned, the parton cascade generates the processes of 
diffractive production (see \fig{3pom}). 
In \fig{3pom} we see that the cross section 
of single diffractive production is closely related to the triple Pomeron 
interaction. The upper Pomeron in \fig{3pom} describes the process of 
multi-particle production by the Pomeron. 
The cross section of such processes is determined by 
$2 \mbox{Im} P\Lb s,t\Rb$ which used to be called the cut Pomeron. 
In Ref.\cite{LEPR} the technique developed, allows us 
to calculate the cross sections of single and double diffraction in 
the MPSI approximation summing Pomerons and cut Pomerons. 
The diagrams for these two processes are shown in \fig{mpsid}. 
We sum these diagram \cite{GLMM}  and the sum has the following form
\bea  \label{SD}
&&N^{MPSI}_{sd}(Y,Y_m = Y - Y_0 = \\
&&= \frac{\Delta_\pom a_{dd}^2}{6}\,\frac{e^{\Delta_\pom (2\,Y - Y_m)}}
{T_{SD}^2\Lb Y,Y_m\Rb}\,T_{SD}\,((T_{SD}-1)^2 -2) + e^{1/T_{SD}}
(1 + 3 T_{SD})\,\Gamma_0(1/T_{SD}), \nn
\eea
\beq \label{TSD1}
T_{SD} \Lb Y,Y_m \Rb\,\,=\,\, a_{dd}\,\exp\Lb \Delta_\pom (Y - Y_m)\Rb\,
\left\{\exp\Lb \Delta_\pom Y_m\Rb - 1 \right\}.
\eeq
\par
The expression for the integrated cross section for double diffraction 
(see \fig{mpsid}-b) can be obtained directly from the unitarity constraint 
of \eq{UNIT}, as the diagrams that describe the elastic and single 
diffraction cross sections, 
do not contribute to the set of Pomeron 
diagrams, that describe the exact
Pomeron Green's function (see \fig{mpsi}-a and \fig{mpsi}-b ).
The unitarity constraint is given by:
\beq \label{UNDD}
2 \,N^{MPSI}\,\,=\,\,N^{MPSI}_{DD} \,\,+\,\,N^{MPSI}_{in},
\eeq
where, $N^{MPSI}_{in}$ stands for the inelastic cross section.  
\par
It was shown that $N^{MPSI}_{in}$ is equal to 
$N^{MPSI}\Lb 2 T(Y)\Rb$ 
(see Refs.\cite{KL,BORY,LEPR}). 
Consequently, the integrated double diffraction
cross section can be written in the form:
\beq \label{TDD}
N^{MPSI}_{DD} \Lb Y\Rb\,=\,2 \,N^{MPSI} \Lb T(Y)\Rb \,-\,N^{MPSI} \Lb 
2\,T(Y)\Rb.
\eeq
\subsection{Qualitative features of our model.} 
\par
Current  $\pom$ models have  changed 
our perception of the Regge Pomeron.
Before detailing with specific features of our model, we make a few 
general
remarks resulting from  basic principles.
\begin{itemize}
\item
Scattering amplitudes are constrained by the  s-channel uninitarity bound,
where in the black disc limit  
$\sigma_{el}\,=\,\sigma_{inel}\,=\,0.5 \, \sigma_{tot}$.
\item
The  Pumplin bound\cite{P} is a
direct consequence of \eq{GW3} and \eq{GW4} and the unitarity
constraints for the S-matrix,i.e
\beq \label{SMA}
S\,S^+\,=\,1;~~~S \,=\,1 + i\,{\bf T};\,\,~~~ i \Big( {\bf T^+}\,\,-
\,\,{\bf T}\Big)\,\,=\,\,{\bf T^+\,T},
\eeq 
and can be written as:
\beq
\sigma_{el}+\sigma_{diff}^{GW} \leq \frac{1}{2}\sigma_{tot}.
\eeq
$\sigma_{diff}^{GW}$ is the sum of GW soft diffractive cross sections.
\item
 Below the black disc limit,
$\sigma_{el} \leq \frac{1}{2}\sigma_{tot}-\sigma_{diff}^{GW}$
and $\sigma_{inel} \geq \frac{1}{2} \sigma_{tot}+\sigma_{diff}^{GW}.$
\end{itemize}
 More specific comments regarding our model:
     \begin{itemize}   
     \item                                                                    
Our model is based on a fitted  
bare non screened Pomeron in a 2 channel
Good-Walker (GW) system, composed of soft elastic
and GW diffractive scatterings channels.
Our $\pom$ basic parameters are
$\Delta_{\pom}=0.2-0.3$ and a very small $\alpha^{\prime}_{\pom}$
($\approx$ 0.028).
\item
 Since $\Delta_{\pom}$ is non zero there is 
no dynamic distinction between low and high mass diffraction.
We have treated  the Pomeron interactions
separately, as their
dependence on the Pomeron parameters is different from that of
the GW components.
\item
Our approach, based on Reggeon calculus,
(see \eq{FI},\eq{S0},  \eq{SI} and \eq{SEP})
satisfies t-channel unitarity constraints. In practice
t-channel unitarity, induced by multi $\pom$ interactions,
leads to "high mass" GW diffraction and, consequently, additional
screening of the GW sector.
\item  In spite of  large screening,
 neither experimentally, nor in any of the models on the market, is the 
diffraction
 cross section   small, and it does not appear to decrease even at 
energies 
of W = 100 TeV. 
 
 \item  In our two channel model two partial amplitudes ($A_{12}(s, b)$
 and $A_{22}(s, b)$)
 reach unity at small $b$, but one amplitude ($A_{11}(s,b)$) is less
 than 1 at small b, even at $W = 100\,TeV$.
 
 \end{itemize} 
 \par
As we have discussed, the LHC data led to small value 
of the slope of the Pomeron trajectory, which is in accord with 
the theoretical expectation that has been discussed above.  
However, at first sight, this statement is in contradiction with the 
measured shrinkage of the diffraction cone as a function 
of energy, which is considerable. 
Actually, this is not true if the value of the Pomeron 
intercept is large. To see this, it is enough to discuss the simple 
eikonal formula of \eq{GT3} with $\Omega$ given by \eq{RGEX},i.e.
\beq \label{QF1}
A\Lb s, b\Rb\,\,=\,\,i\Lb 1 -  e^{-A_P (s,b)}\Big)\,\,=
\,\,i\Big( 1 -  \exp\Lb - S\Lb b\Rb\Lb\frac{s}{s_0}\Rb^{\Delta_\pom}\Rb\Rb,
\eeq
where (see \eq{POMAMP}), 
\beq \label{QT2}
S\Lb b\Rb\,\,=
\,\,\int \frac{d^2}{4 \pi^2}\,e^{ \vec{q}\cdot \vec{b}}\,g_p(m_1,M_1,t - q^2)
\,g_t(m_2,M_2,t= -q^2). 
\eeq
At large $s$, $A_P (s,b=0)\,\,\gg\,\,1$ and $A\Lb s,b=0\Rb = 1$. 
However, at large $b$ $S\Lb b \Rb\xrightarrow{m b\,\,\gg\,1}\,\exp\Lb -  
m \,b \Rb$ and $A\Lb s, b\Rb\,=\,A_P (s,b)\,\,\ll\,\,1$.
Therefore, the typical value of $b$, which contributes to the integral
\beq \label{QT3}
\langle b^2 \rangle\,\,\,=\,\,\,\int b^2 d^2 b A\Lb s, b\Rb\Big{/}  
\int d^2 b A\Lb s, b\Rb,
\eeq
stems from
\beq \label{QT4}
A_P (s,b=b^*)\,\,=\,\,g^2 \Lb\frac{ s}{s_0}\Rb^{\Delta_\pom}\, \exp\Lb - m\,b\Rb\,\,
\approx\,\,1.
\eeq
It gives
\beq \label{QT5}
b^*\,\,=\,\,\frac{\Delta_\pom}{m} \,\ln\Lb s/s_0\Rb \,\, +
\,\,R_0\approx 1/m.
\eeq
Therefore, $\langle b^2 \rangle\,=\,\Lb \frac{\Delta_\pom}{m} 
\,\ln\Lb s/s_0\Rb \,\, +\,\,R_0\Rb^2$.
One can see that if 
$\Delta_\pom = 0.2 \div 0.3$ and $m \sim 1 \,GeV$ 
\footnote{As has been mentioned in section 2.1.2 $m$ 
was expected to be approximately the pion mass. However, 
in Ref.\cite{SN} arguments have been provided that $m$ could be 
a mass of the lightest glueball. $m \sim 1 \,GeV$ 
is a reasonable estimate.}
$\langle b^2 \rangle\,\,=\,\,R_0^2 \,\,+\,\,2\,\frac{\Delta_\pom}{m} \ln\Lb 
s/s_0\Rb + \Lb\frac{\Delta_\pom}{m} \ln\Lb s/s_0\Rb\Rb^3 \to  R_0^2 \,\,+
\,\,2\,\frac{\Delta_\pom}{m} \ln\Lb s/s_0\Rb\,\,=\,\,
R_0^2\,\,+\,\,2\,(0.2 \div 0.3 )\ln\Lb s/s_0\Rb$. 
This value of the average $b^2$ is close to the experimental one.
\section{Physical observables in our model.} 
\subsection{ Classical soft Physics (Refs.\cite{GLMM,GLMMA})}
\subsubsection{Total and elastic cross sections }   
\par
We start with the formulae for the classical set of soft 
interaction data at high energy: total, elastic and diffractive 
dissociation cross sections. 
In our two channel approximation for the Good-Walker mechanism, 
the elastic amplitude has the following form:
\beq \label{SCL1}
a_{el}(s, )\,=\,i \Lb \alpha^4 A_{1,1}\,+\,2 \alpha^2\,\beta^2\,A_{1,2}\,
+\,\beta^4 A_{2,2}\Rb.
\eeq
\begin{itemize} 
\item
Note he elastic amplitude $a_{el}(s,b)=1,$ when, and only when,
$A_{1,1}(s,b)=A_{1,2}(s,b)=A_{2,2}(s,b)=1$.

\item When $a_{el}(s,b)=1,$ all diffractive amplitudes
at the same values of (s,b) vanish (see \eq{SCL3}).

\end{itemize}

$A_{i,k}$ are given by
\bea \label{SCL2}
A_{i,k}\Lb Y; b\Rb\,&=&\,\,1 \,\,-\,\,\exp\Big( - \h \Omega^\pom_{ik}\Lb s, b\Rb\Big) \\
&=&\,1 \,-\,\exp\left\{-\int d^2 b'\,
\,\frac{\Lb \tilde{g}_i\Lb\vec{b}'\Rb\,\tilde{g}_k\Lb\vec{b} - 
\vec{b}'\Rb\,G(Y)\Rb}{1\,+\,G(Y)\,\left[\tilde{g}_i\Lb\vec{b}'\Rb 
+ \tilde{g}_k\Lb\vec{b}\,-\,\vec{b}'\Rb\right]}\right\}.\nn
\eea
$G\Lb Y\Rb$ is given by \eq{MPSI7}.  
\eq{SCL2} can be obtained from \eq{MPSI7} 
by the replacement $T\Lb Y\Rb \to G\Lb Y\Rb$.
It reflects the fact that in \eq{SCL2} 
we have replaced the `bare' Pomeron by the dressed one.
The physical observables can be calculated using the following set of 
formulae which stem from the unitarity constraint of \eq{UNIT}:
\bea
&&\sigma_{tot}\Lb s \Rb\,=\,2\,\int d^2 b\,
\mbox{Im}\, a_{el}\Lb s, b\Rb;~~~~~
\sigma_{el}\Lb s \Rb\,=\,\int d^2 b\,|a_{el}\Lb s, b\Rb|^2 
\label{sigma};\\
&&d \sigma_{el}/dt=\pi|f\Lb s,t\Rb|^2;~\sigma_{tot}\,=\,4\pi 
\mbox{Im} f\Lb s,t\Rb;~a_{el}\Lb s, b\Rb\,=
\,\frac{1}{2\pi}\int d^2q\,e^{i\vec{q}\cdot\vec{b}} f\Lb s,t = 
- q^2\Rb; \nn\\
&&\sigma_{in }\Lb s \Rb = \int d^2 b\,G^{in}\Lb s, b\Rb~\mbox{with}~
G^{in}(s,b)\,=\,\alpha^4 G^{in}_{1,1}\,+
\,2 \alpha^2\,\beta^2\,G^{in}_{1,2}\,+
\,\beta^4 G^{in}_{2,2}\label{sigmain}\\
&&B_{el}\,=\,\int d^2 b\,b^2\,\mbox{Im}\, a_{el}\Lb s,b\Rb
\Big{/}\Lb 2 \int d^2 b\,\mbox{Im}\,a_{el}\Lb s,b\Rb\Rb\label{bel}.
\eea       
\subsubsection{Diffractive production cross sections }
\par
In our model we have two sources leading to 
diffractive production\cite{GLMMA}: 
the first one is due to Good-Walker mechanism in two channel model, and the second stems from 
the enhanced diagrams of \fig{mpsid}-a. As we have mentioned, 
when $\Delta_\pom \neq 0$, there is no difference between the 
origins of the two, and we shall call the second one `large mass 
diffraction'. 
The Good-Walker term has the following form:
\beq \label{SCL3}
\sigma^{GW}_{SD}\,=\,\int\,d^2 \Lb\alpha\beta\{-\alpha^2A_{1,1}
+(\alpha^2-\beta^2)A_{1,2}+\beta^2 A_{2,2}\}\,\Rb^2.
\eeq      
$A_{i,k}$ are given by \eq{SCL2} and the term which describes 
diffractive production in the region of large mass, 
\bea\label{SCL4}
&&\sigma^{\mbox{Large mass}}_{SD}\,\,=\,2 \int d Y_m \int d^2 b \,\\
&&\left\{\,\alpha^6\,A^{SD}_{1;1,1}\,\,+\,\alpha^2\beta^4 A^{SD}_{1; 2, 2} 
+ 2\,\alpha^4\,\beta^2 \,A^{SD}_{1;1,2}\,+\,\beta^2\,\alpha^4 
\,A^{SD}_{2; 1, 1}\,+\,2\,\beta^4\alpha^2
\,A^{SD}_{2; 1,2}\,+\,\beta^6\,A^{SD}_{2; 2,2} \right\}, \nonumber
\eea   
with,\bea \label{SCL5} 
&&A^{SD}_{i;k, l}\Lb Y, Y_m\Rb\,=\\
&&\int d^2 b'\, 2 \Delta\,e^{-\Delta Y_m} \,T^2_{SD}
\Lb Y,Y_m\Rb\,\tilde{g}_i\,\sqrt{\tilde{g}_k\tilde{g}_l}
\,e^{ - 2 \tilde{g}_i -\tilde{g}_k - \tilde{g}_l}\,\int\frac{d \xi_2 
\,d \bar{\xi}_2\,\sqrt{\xi_2 \bar{\xi}_2}}{\Lb
1 \,+\,(\xi_2 + \bar{\xi}_2) \cdot T_{SD}\Rb^2}e^{ - \xi_2 - \bar{\xi}_2}
\,\nn\\
&& \times\,\,\,\,J_{1}\Lb 2 \sqrt{\tilde{g}_k\,\xi_2}\Rb
J_{1}\Lb 2 \sqrt{\tilde{g}_l\,\bar{\xi}_2}\Rb\,\,
\left\{1\,-\,\exp\Lb \frac{-\tilde{g}_i}{1\,+\,\xi_2\,T_{SD}}\Rb\right\}\,\,
\left\{1\,-\,\exp\Lb \frac{-\tilde{g}_i}{1\,+\,\bar{\xi}_2\,T_{SD}}
\Rb\right\}.\nonumber
\eea
$\tilde{g}_i$, in \eq{SCL5}, denotes $\tilde{g}_i\Lb\vec{b}'\Rb$.
Similar notations apply to $\tilde{g}_k$ and $\tilde{g}_l$.

\beq \label{TSD}
T_{SD}\Lb Y; Y_m=\ln\Lb M^2/s_0\Rb\Rb\,\,\,=\,\,\,a_{dd}\,
\Lb e^{\Delta Y_m}\, - \,1\Rb\,e^{ \Delta (Y - Y_m)}.
\eeq
\par
The same two mechanisms contribute to double diffractive 
production leading to:
\beq \label{SCL6}
\sigma^{G W}_{DD}\,\,\int \,d^2 b \,\, \alpha^2\,\beta^2
\left\{-\alpha^2\,A_{1,1}\,+\,(\alpha^2 - \beta^2 )\,A^{1,2}\,
+\,\beta^2 A_{2,2} \right\}^2.
\eeq

$A_{I,k}$ are given by \eq{SCL2} and the term which is 
determined by the Pomeron interaction (see \fig{mpsid}-b),  
which contribute to the large masses:
\beq \label{SCL7}
\sigma^{\mbox{Large mass}}_{DD}\,\,=\,\,\int\,d^2 b\,
\left\{\alpha^4\,A^{DD}_{1,1}\,+ \,2 \alpha^2\,\beta^2 A^{DD}_{1,2}\,
\,\beta^4\,A^{DD}_{2,2}\,\right\}.
\eeq
\bea\label{SCL8}
A^{DD}_{i,k}\Lb Y; b\Rb \,\,&=&\,\,
2 A^{el}_{i,k}\Lb T\Lb Y\Rb; b \Rb\,\,\,-\,\,| A^{el}_{i,k}\Lb T\Lb Y
\Rb; b \Rb|^2 \,\\
&-& 2\,\int \,d Y_m A^{SD}_{i;k,k}\Lb T_{SD}\Lb Y, Y_m; b \Rb; b 
\Rb\,\,-\,\,A^{in}_{i,k}\Lb 2 T\Lb Y \Rb;b\Rb\nonumber.
\eea

\subsection{Estimates of the value  of parameters and the range of
applicability}
\par
Our model has a set of parameters, determined by fitting  the  
experimental data. The advantage of our model is that for some of these 
parameters we can utilize estimates obtained from matching with pQCD. 
\beq \label{EST1}
\Delta_\pom\,\propto\,\bas;~~~G_{3 \pom}\,\propto\,\bas^2; 
~~~ g_i\,\propto\,1,
\eeq
which leads to $ \ga\,\ll\,\Delta_\pom\,\ll\,g_i$.
\par
 To estimate the range of energy where the MPSI 
approximation is valid, we notice that the most dangerous term in 
\eq{MPSI1} 
is the last one, which is proportional to $\Delta_\pom a^2_{dd} Y$. 
Such terms lead to the renormalization of the Pomeron intercept. 
In order to neglect them, we have to restrict the range of $Y$ by
\beq \label{EST2}
\Delta_\pom a^2_{dd} Y \,<\, 1; ~~~~Y\,<\,\frac{1}{a^2_{dd}\,\Delta_\pom}\,=
\,\frac{1}{\ga}.
\eeq
In our approach we did not take into account the four Pomeron interaction. 
As shown in Ref.\cite{AKLL}, this interaction becomes  important when 
$Y \,>\,(2/\Delta_\pom)\ln\Lb 1/\as\Rb$ and its strength is 
approximately $a^2_{dd} = \ga/\Delta_\pom$. 
We will see below that our fit requires  
$\Delta_\pom \,\approx 0.23$ and $\ga=0.0045$, which leads to small 
$a^2_{dd} \approx 0.015$.  Therefore, the value of $Y$, for which we 
need to  include the four Pomeron interaction, 
is approximately $Y \sim 50\div 60$.
\subsection{Data and fitting procedure}
\subsubsection{Data prior to the LHC era}
\par
We started constructing our model in 2007, prior to the commissioning 
of the LHC. At that time the highest energy accelerator data available was 
that emanating from the Tevatron at W = 1.8 TeV. There were two 
incompatible measurements available for the total cross section at that 
energy. That of the CDF collaboration \cite{CDFt} which gave a value of 
80.03 $\pm$ 2.24 mb, and the second by the E710 collaboration \cite{E710} 
which published a value of 72.1 $\pm$ 3.3 mb. The E710  value was close to 
that predicted by the Donnachie-Landshoff formalism\cite{DL}, and in accord 
with the values cited by most models (including GLM). 
\par
Currently, with the publication of the LHC 
total cross section by the TOTEM collaboration \cite{TOTEM}, i.e.
$\sigma_{tot}$ = 98.6 $\pm$  2.2 mb  at W =  7 TeV, 
most models that reproduce this value, 
are in accord with the CDF measurement at W = 1.8 TeV. 
There is a lesson to be learnt from this, that experimental results 
should not be discriminated against, on the basis of 
 theoretical prejudice!

\subsubsection{Low energy behaviour of the scattering 
amplitude}
As we shall see  in the next section, most of our data base consists of
lower energy points from ISR and Sp${\bar p}$S/SppS
(W $\approx$ 20 -70 GeV),
where the contribution of the secondary Regge exchanges are
important.
A secondary Reggeon has an energy behaviour
$\exp\Lb \Delta_\reg  (Y - 0)\Rb$.
The sum $\pom + \reg$ describes the energy behaviour of
 the elastic scattering amplitude
 without screening corrections.
 \begin{figure}
 \centerline{\epsfig{file=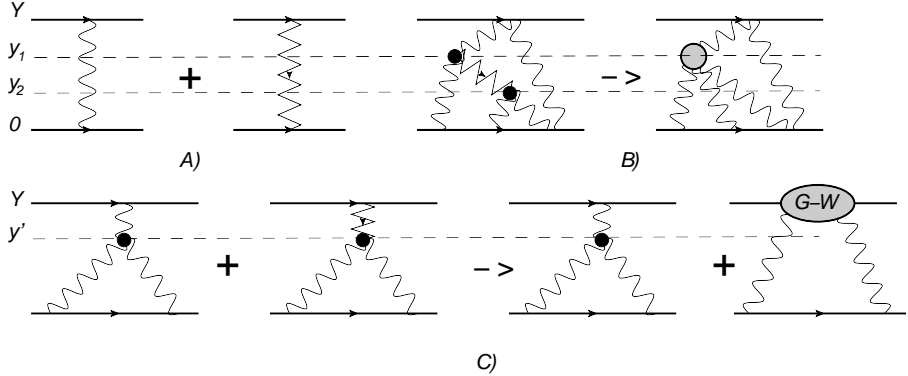,width=120mm}}
\caption{Contributions of secondary  Reggeons denoted by zigzag lines. 
Wavy lines denote the Pomeron. \fig{lowen}-A shows the contribution to
  the scattering amplitude due to exchange of Pomeron and Reggeon.
 \fig{lowen}-B and \fig{lowen}-C illustrate that the exchange of
 secondary Reggeons can be reduced to inclusion of the vertices
for the Pomeron-Pomeron interaction, or can be included into Good-Walker 
mechanism.}
\label{lowen} 
 \end{figure}
 This sum replaces the single Pomeron exchange in
the definition of $\Omega_{ik}$.
 Inserting this sum everywhere in the more complicated diagram
(see \fig{lowen}),
 one can see that the integrations over rapidities reduces
the contributions
 of the secondary Reggeons. By introducing  new vertices
 for the Pomeron-Pomeron interactions (see \fig{lowen}-B) the
integration over $y_1 - y_2$ can be replaced by a
 new $\pom \to 3 \pom$ vertex, or it
can be absorbed into G-W mechanism (see \fig{lowen}-C).
Since in our approach the vertices, other than the triple
 Pomeron vertex are considered to be small,
 we arrive at the conclusion that for lower energies
  we only need to replace the single Pomeron
 exchange by $\pom + \reg$, in the definition of $\Omega_{ik}$.
\begin{figure}[ht]
\centerline{\includegraphics[width=80mm]{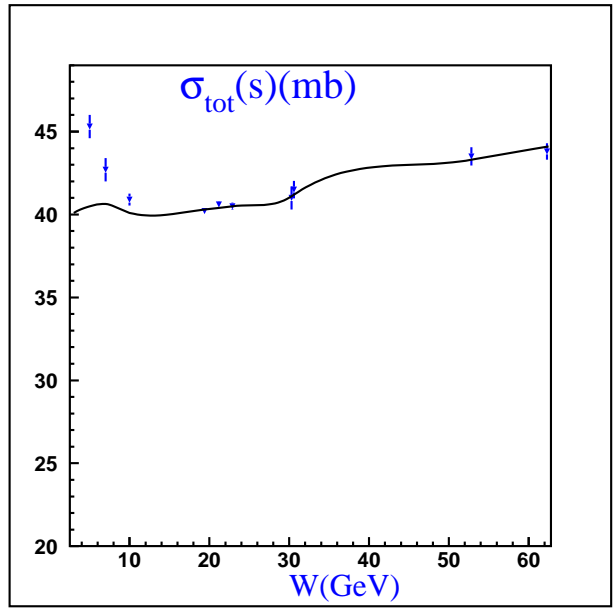}}
\caption{Behavior of the total cross section 
$\frac{1}{2}[\sigma_{\bar{p}p}\,+\,\sigma_{pp}]$ at low energies
  }
\label{sigttle}
\end{figure}

In \fig{sigttle} we  compare our prediction for lower energies
 with the experimental data and obtain
a satisfactory description to within 10\%.
 The conclusion is very simple: we do not need an
 additional source to describe the lower
energy behaviour of the amplitude.
  
\subsubsection{The fit to  the data and its phenomenology}

\paragraph{ The main formulae of the fit }
The main formulae, that we have derived from \eq{SI}, \eq{MF1},
\eq{SEP} and \eq{SPB}, are written in \eq{SCL1}.
Note that $g_i $ in \eq{SCL1} have dimension of inverse momentum (see
\eq{S})
 as well as $G_{3\pom}$, while $\gamma$ is dimensionless.
 Actually $\gamma^2 \,=\,\int d^2 k \,G^2_{3 \pom}$,
but because we do not know
 the dependence of $G_{3\pom}$ with respect to the transverse momenta of
Pomerons, we consider $\gamma$ and $G_{3\pom}$
as independent parameters of the fit.

We deal with the secondary reggeon in the same way as in
Ref.\cite{GLMM} adding
 $\Omega^{\reg}$ to $\Omega^\pom$ in \eq{SCL1}, i.e. 
 \beq \label{OMFIN}
 \Omega_{i k}\Lb s, b\Rb\,\,=\,\,\Omega^\pom_{i k}\Lb s, b \Rb\,\,+\,
\,\Omega^\reg_{i k}\Lb s, b \Rb. 
\eeq 
 
  $\Omega^{\reg}$ is taken in the following form:
 \beq \label{OMREG}
  \Omega^{\reg}_{i k}\Lb s, t\Rb\,\,=\,\,g^\reg_i g^\reg_k \eta\Lb t\Rb
 e^{- \Lb  \frac{1}{4}\Lb R^2_{0 i}\,\,+\,\,R^2_{0 k}\Rb\,\,+\,
\,\alpha'_\reg\ln \Lb s/s_0\Rb \Rb\,|t|} \,\Lb 
\frac{s}{s_0}\Rb^{\Delta_\reg}, 
  \eeq
with the signature factor $\eta\Lb t \Rb$ which is equal to
\beq \label{SIGNT}
\eta\Lb t \Rb\,\,=\,\,\frac{ 1\,\,\pm\,\,e^{ i \pi \alpha_\reg\Lb t
 \Rb}}{\sin\Lb \pi \alpha_\reg\Lb t \Rb\Rb}.
\eeq 


\paragraph{The strategy of our fitting procedure.}
In our model the Pomeron is specified by nine  parameters, the Tevatron
data (which was the highest energy data available at that time) on its own 
is not sufficient to determine the parameters.
Consequently, we have also to include ISR - Sp${\bar p}$S/SppS
lower energy (W $\approx$ 20-70 GeV) data.
This data has small errors which facilitate
a reasonably reliable fit. To reduce the number of Reggeon
parameters, we define
$\sigma_{tot} = \frac{1}{2}(\sigma_{tot}(pp) + \sigma_{tot}(p{\bar p})$.
The inclusion of the Regge sector of our fit requires five additional 
parameters. i.e. we have fourteen parameters in all.

Our data base has 58 experimental data points,
which include the average
i.e.
($\sigma_{tot} = \frac{1}{2}(\sigma_{tot}(pp) + \sigma_{tot}(p{\bar p})$),
integrated elastic cross sections, integrated single and double
 diffraction cross sections, $B_{el}$, to which we have added a
consistency check of the  CDF data
$\frac{d\sigma_{el}}{dt}$ (-t $\leq 0.5 GeV^2)$,
$\frac{d^2 \sigma_{sd}}{dt dM^2/s}$ (-t = 0.05 GeV$^2$) and $B_{sd}$.   
The data points were fitted to determine the 14 free parameters of our
model. We fit  the entire data base  simultaneously.

 \paragraph{ The results of the fit (prior the LHC)}
Our fit is based on 58 experimental data points.
The model gives a good reproduction
of the data, with a $\chi^2/d.o.f.$ = 1.56.   
However, a large contribution to the value of $\chi^2/d.o.f.$
stems from the uncertainty of the
value of two single diffraction cross sections,
and of the CDF total cross section\cite{CDFt}
at the Tevatron (W =1.8 $TeV$).
Neglecting the contribution of these
three points to the total $\chi^2$ we
obtain $\chi^2/d.o.f. = 0.86$.

\subsubsection{Comments on the parameter values of the fit}
It should be stressed that the value of our phenomenological
parameters (see Table 1 ) are in
agreement with the theoretical estimates of \eq{EST1}.
We obtain a value of $\alpha'_{\pom}\, = \,0.028$, which is sufficiently 
close to zero to justify our approximation of taking 
$\alpha'_{\pom}\,=\,0$ in our summation of diagrams.  
Choosing the typical soft scale $\mu = 1\,GeV$ we can see
that $g_i \mu \approx 1$ and
\beq \label{FI3}
G_{3\pom}\mu  \approx \gamma \,  
\approx\,\Delta^2_\pom \,\ll\,\Delta_\pom \,\ll\,g_i \mu.
\eeq
As we have discussed, we sum all diagrams in an
approximation in which $g_i G\Big(T\Lb
Y\Rb\Big) \,\geq\,1$ while $\Delta^2_\pom G\Big(T\Lb Y\Rb\Big)
\,\,\ll\,\,1$.
The values of the fitted parameters support
the use of this approximation.
\par
As mentioned previously,  
most models for soft interactions, 
which were proposed prior to the
measurements at the LHC, are only marginally compatible 
with LHC data, our GLM model has the same deficiency.
We investigated  possible causes of the problem,
by considering separate fits to the high energy ($W > 500\, GeV$), and low
energy ($W < 500\, GeV$) data. 
Our output results are moderately higher than our
previous predictions. Our results for total and elastic cross sections
are systematically lower than the recent Totem and Alice published values,   
while our
results for the inelastic and forward slope agree with the data.
If, with additional experimental data, the errors will be reduced, 
while the central cross section values remain unchanged, 
we will need to reconsider the physics on which our model is built.
Our procedure for adjusting parameters may be deficient, requiring
a more sophisticated data analysis which may yield
satisfactory results. Recall that the fitted data base \cite{GLM}
contained no LHC data. Moreover,
the low energy ($W < 500 GeV$) total, elastic
and diffractive cross sections which constitute the major portion of the
fitted data points have rather small errors.
Consequently, our fitting procedure is not well balanced as the
main contribution to our $\chi^2/d.o.f.$ stems from the low energy data.
\newline
To  check the second option,
we removed the low energy data and fitted only 
the high energy data ($W > 500 GeV$),
including the available LHC soft cross section data points,
so as to determine the Pomeron parameters.
Having adjusted these parameters,
we tuned the value of the Reggeon-proton vertex, which enabled us
to obtain a smooth cross section behaviour through the ISR-LHC energy
range.
We hoped that this exercise would
clarify to what extent our model has intrinsic deficiencies,
or do we just have a technical problem in the procedure for
adjusting our free parameters.

At this stage we were fairly pessimistic about our model reproducing all 
the "soft physics" data emanating from LHC 
and concluded our paper\cite{GLMM} 
 as follows:
"In spite of the fact that the values of the parameters,
extracted from our current fitting, are slightly different from our
previous values, the overall picture remains unchanged.
Our updated total and elastic cross sections
are slightly lower than the published TOTEM values\cite{TOTEM}, but still
within the relatively large experimental error bars. Should future LHC
measurements confirm the present TOTEM values,
we will need to revise our dynamic picture for soft scattering."
\subsubsection{Success at last}
In our last paper \cite{GLMPL} we  found
a set of parameters in our model for soft interactions
at high energy, that successfully describes all high energy experimental
data, including the LHC data.
This model is based on a single Pomeron with a large intercept
$\Delta_\pom = 0.23$, and a slope $\alpha'_\pom = 0.028 (\approx 0)$,
that describes both long and short distance processes. It also provides
a natural matching with perturbative QCD. All features of our
model are similar to the expectations of N=4 SYM, which at present is the
only theory that is able to treat strong interactions on a theoretical
basis.

\par
In this paper we retract the pessimistic concluding statement we had in 
\cite{GLMPL}
as we find that the conclusion was premature.
The set of parameters in our previous
paper was found by fitting all data with energy $W \geq 500 GeV$,
including the LHC data.
In the present version of our GLM model\cite{GLMPL} we made no changes 
other than tuning the 3 Pomeron parameters.
Our tuned $\Delta_{\pom}$ changes from 0.21 to 0.23, while $G_{3\pom}$
and $\gamma$ the Pomeron-proton vertex, are unchanged. The small change
in the value of $\Delta_{\pom}$ is sufficient to produce the desired
results in our $\sigma_{tot}$ and $\sigma_{el}$ output values
for LHC energies, while
the changes in the output values of the other observables are small enough
not to spoil the good reproduction of the data achieved in
Ref.\cite{GLMLAST}.
Taking $\alpha'_\pom=0$, allows us to sum all diagrams
having Pomeron interactions\cite{GLM}. This is the great
advantage of such an approach. 
In our model we only take into account the triple Pomeron
interaction vertices ($G_{3\pom}$), this provides a natural matching to
the hard Pomeron, since at short distances $G_{3\pom} \propto \as^2$,
while other vertices are much smaller.
\subsubsection{Results}
\begin{figure}[h]
\begin{tabular}{c c c}
\includegraphics[width=0.3\textwidth]{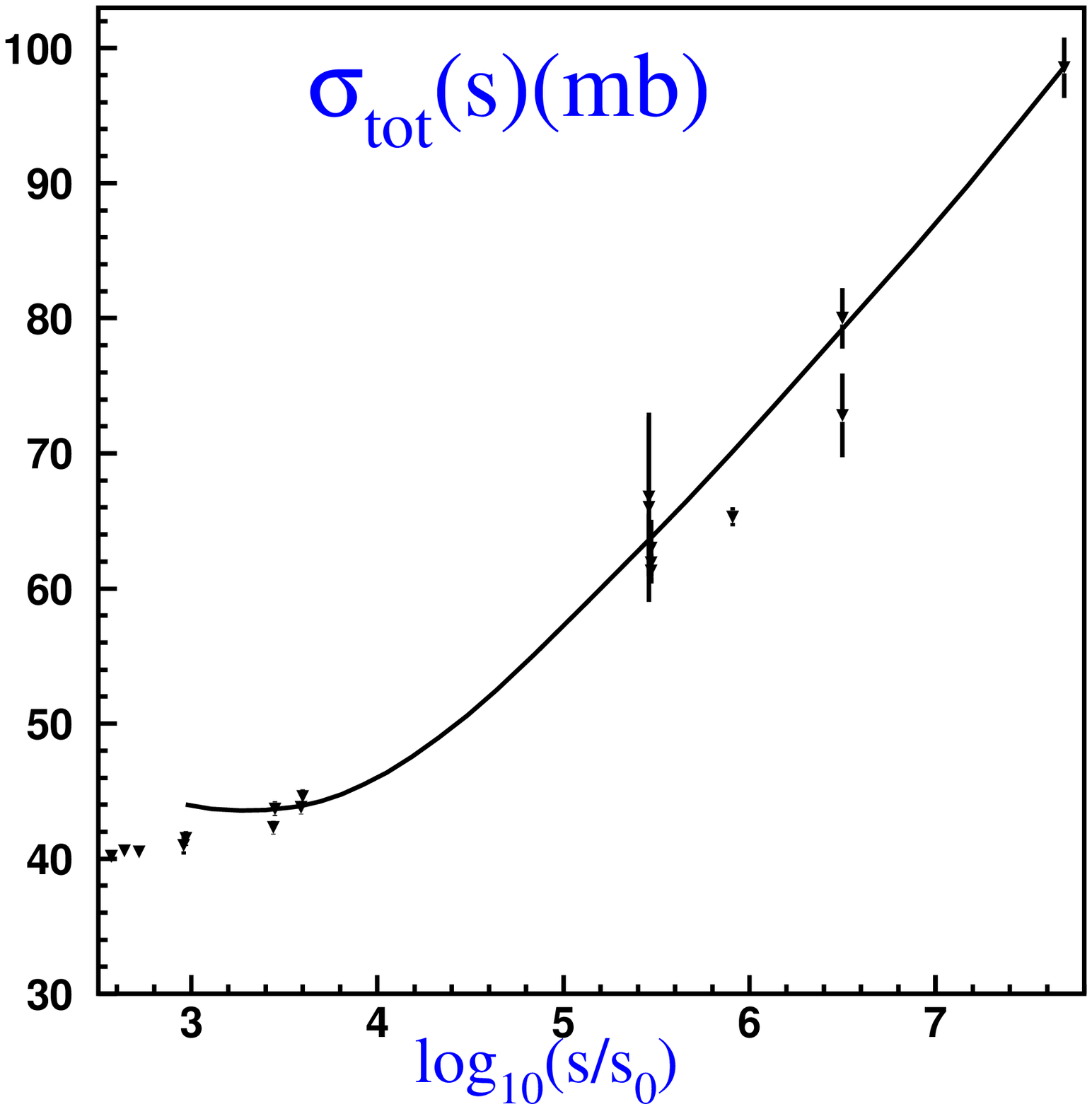}&
\includegraphics[width=0.3\textwidth]{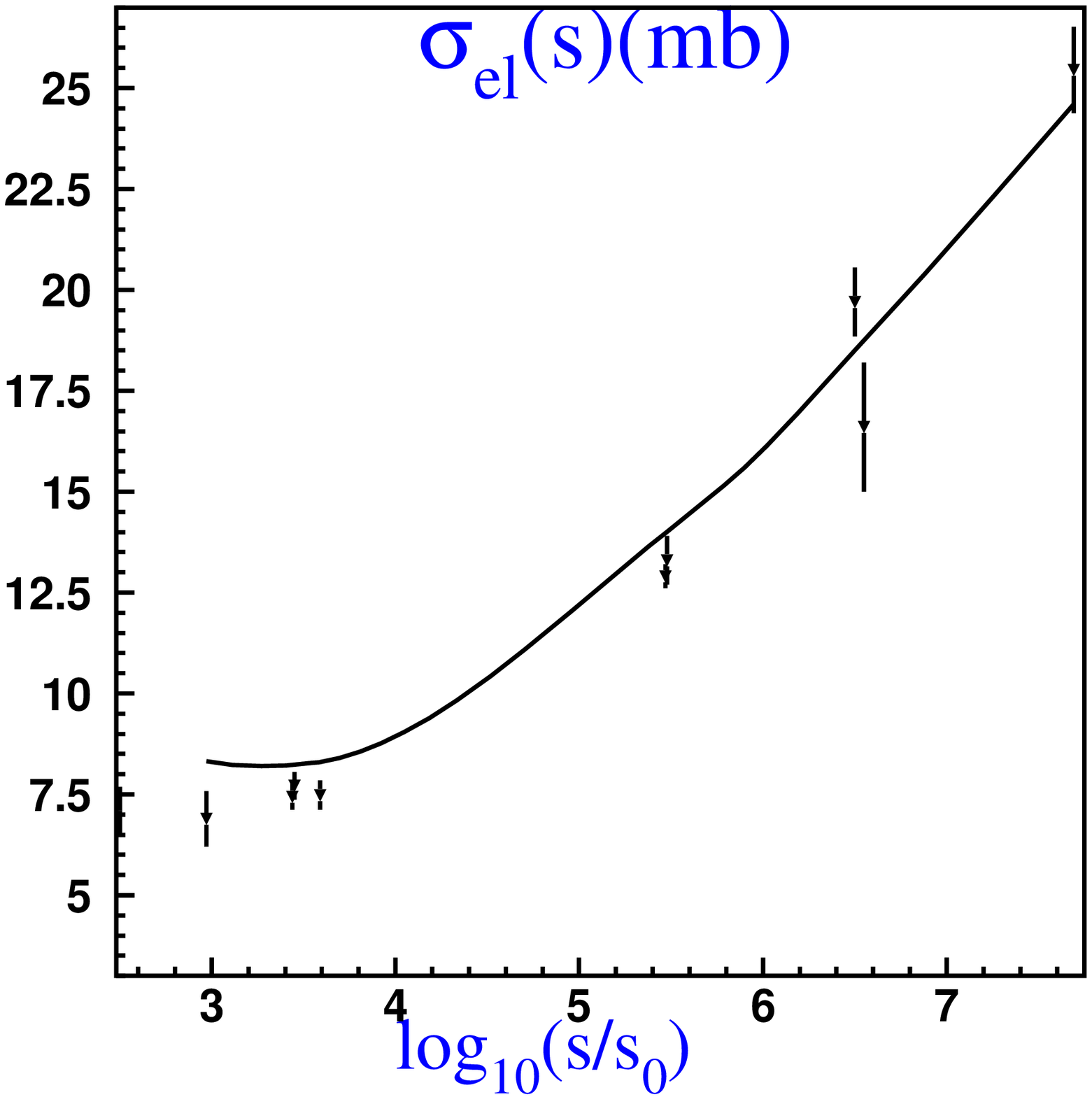}&
\includegraphics[width=0.3\textwidth]{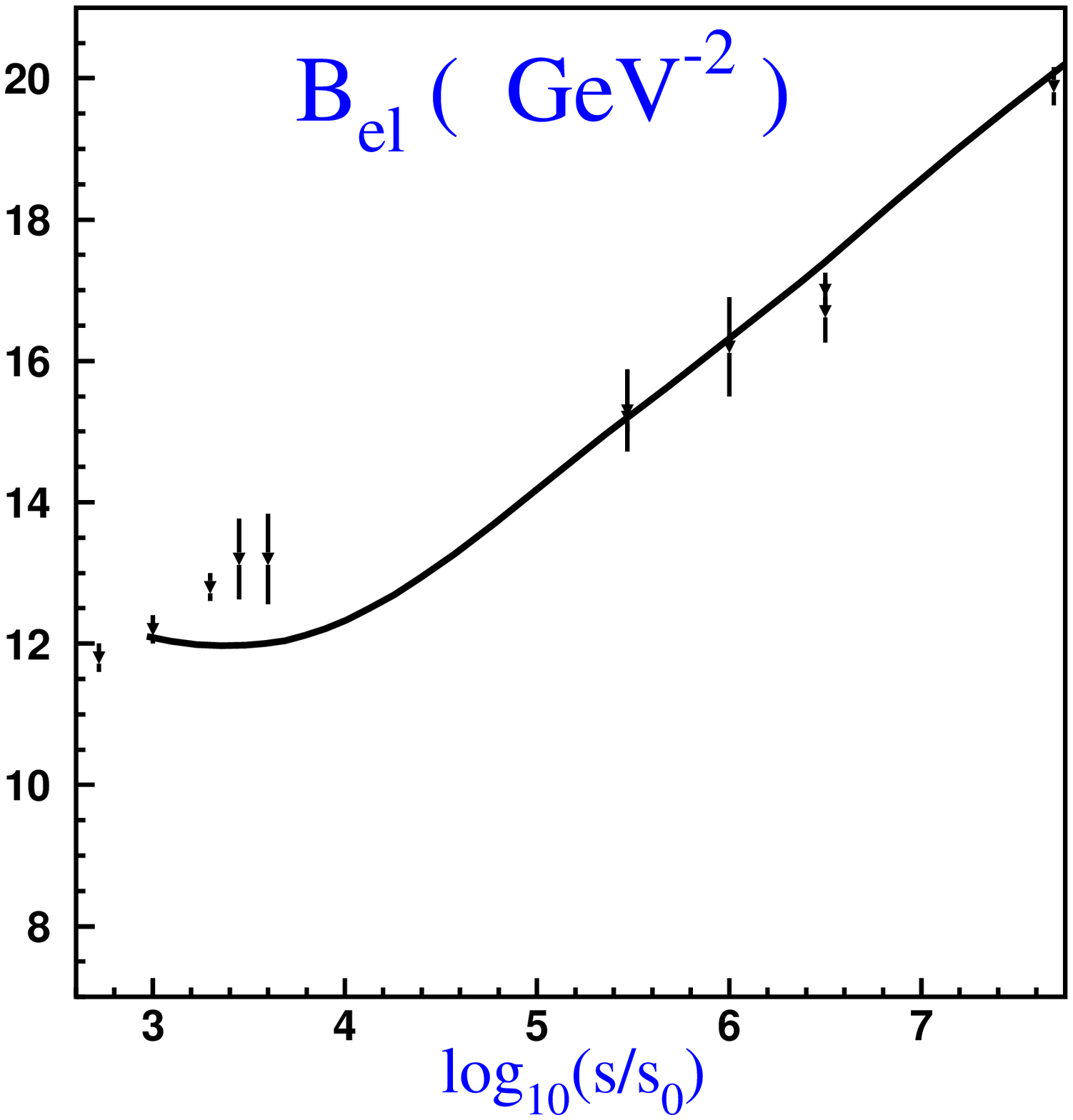}\\
\fig{fit}-a & \fig{fit}-b &\fig{fit}-c\\
 \includegraphics[width=0.3\textwidth]{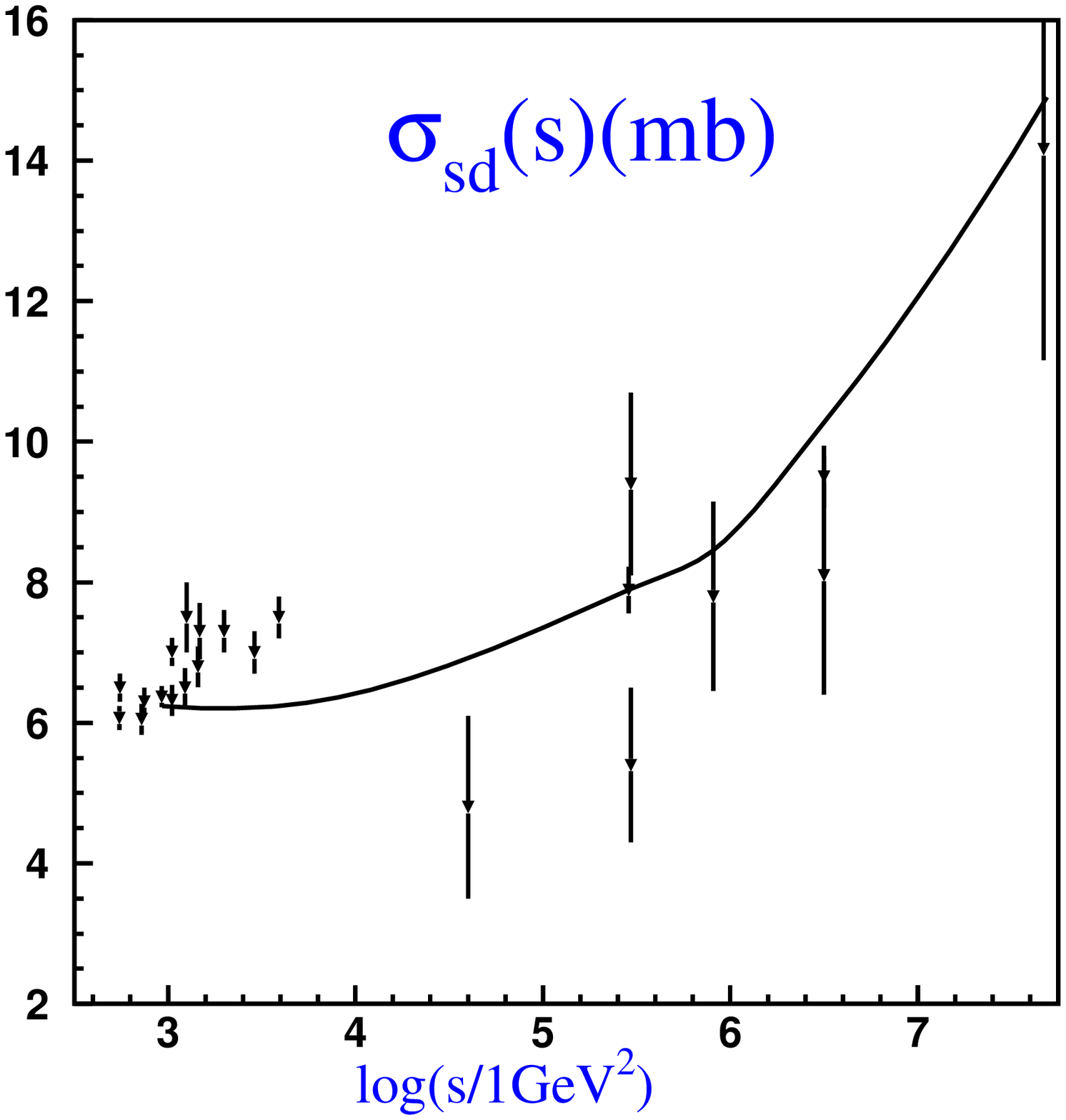}&
\includegraphics[width=0.3\textwidth]{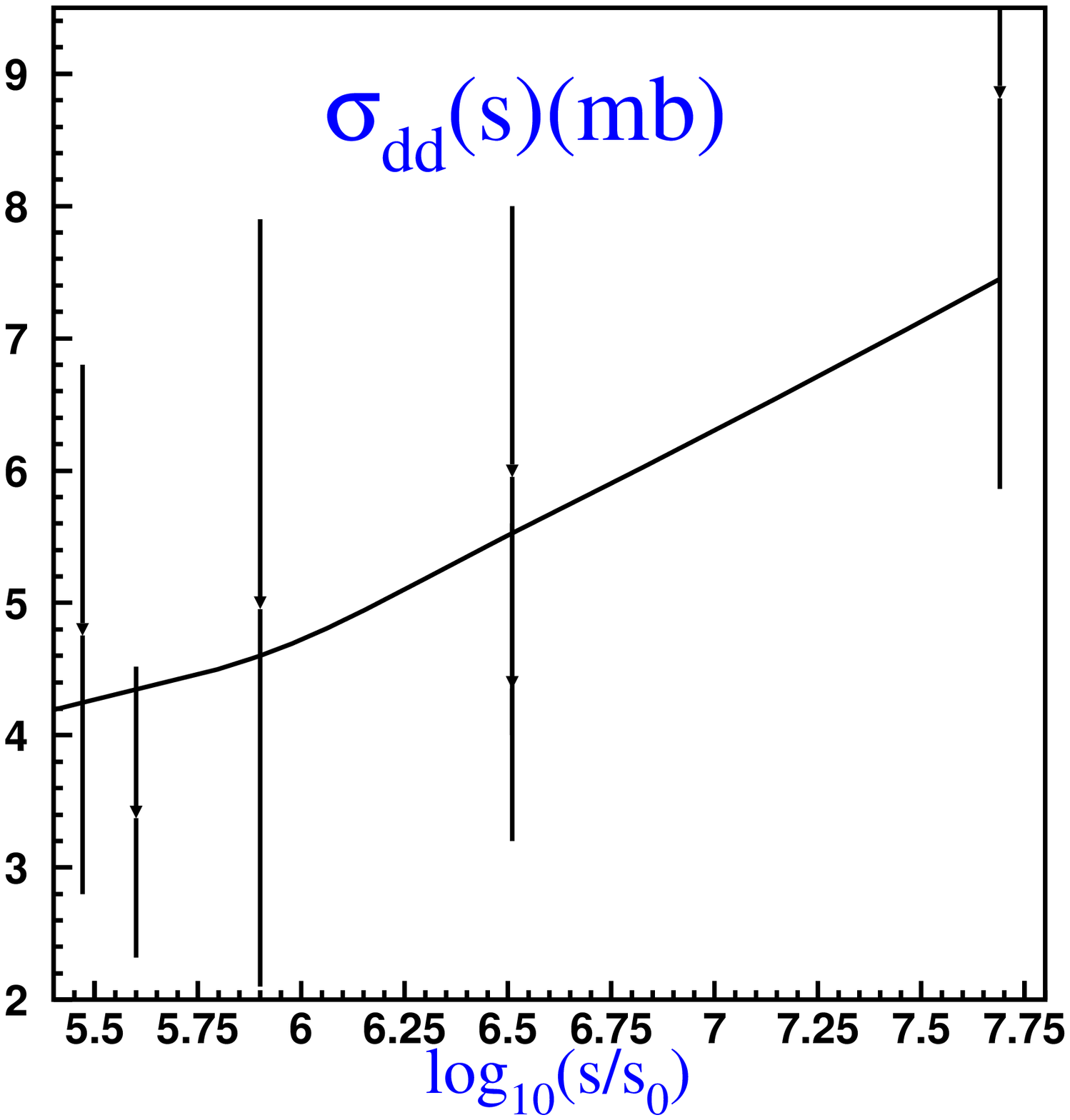}&
\includegraphics[width=0.3\textwidth]{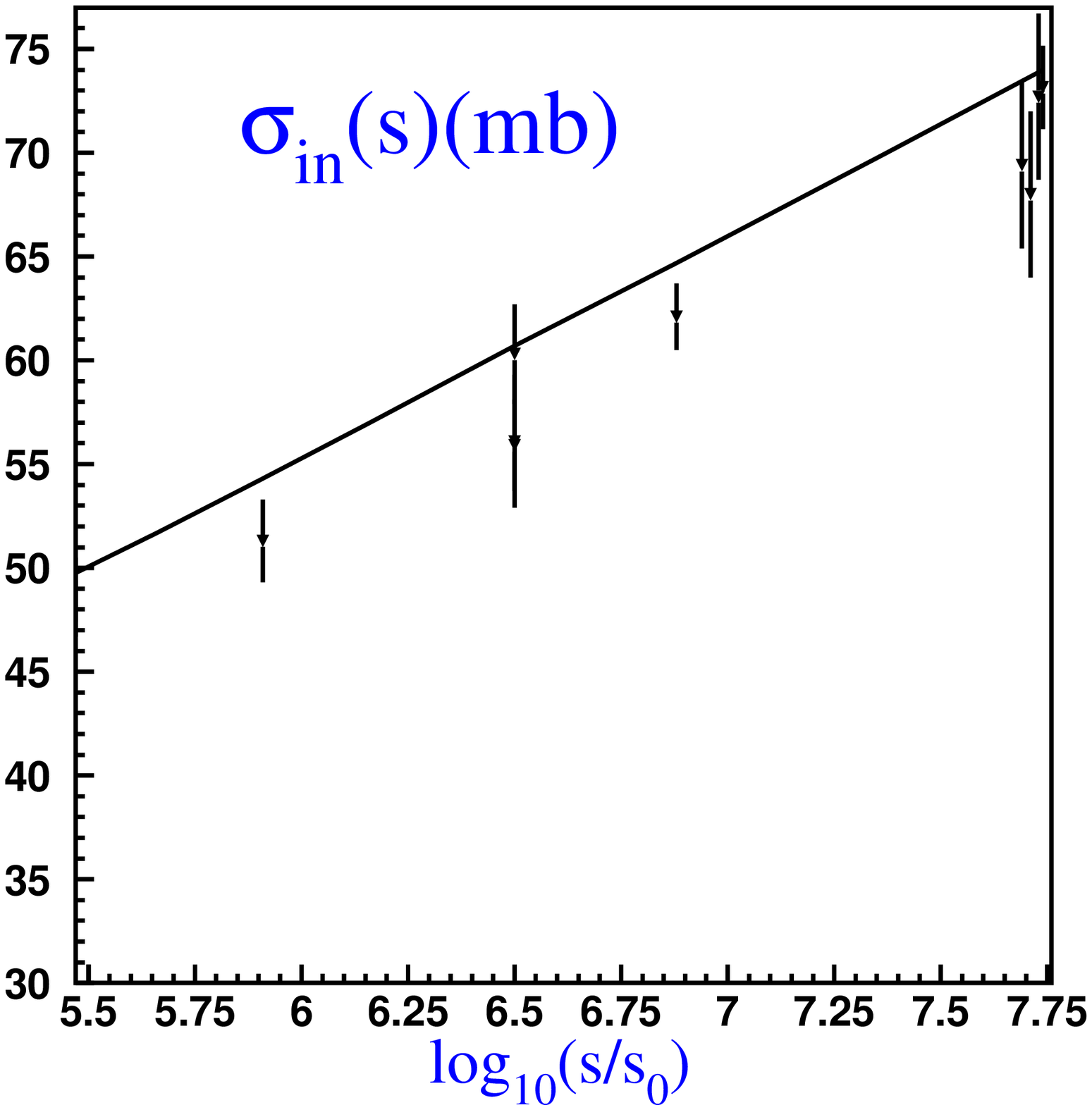}\\
\fig{fit}-d & \fig{fit}-e &\fig{fit}-f\\
\end{tabular}
\caption{Comparison with the experimental data: the energy behaviour
 of the total (\protect\fig{fit}-a), elastic  (\protect\fig{fit}-b),
 elastic slope ($B_{el}$) (protect\fig{fit}-c), single diffraction 
(protect\fig{fit}-d),
double diffraction (\protect\fig{fit}-e),  and inelastic 
(\protect\fig{fit}-f) cross sections.
The solid lines show our present fit. The data has been taken from            
Ref.\protect\cite{PDG}
for energies less than the LHC energy. At the LHC energy for total and
elastic cross
section we use TOTEM data\protect\cite{TOTEM} and for single and double
 diffraction
cross sections are taken from Ref.\protect\cite{ALICE}.}
\label{fit}
\end{figure}

\begin{figure}[ht]
\centerline{\epsfig{file=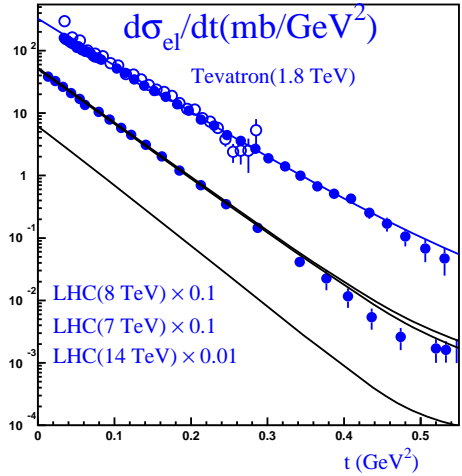,width=70mm}}
\caption{$d \sigma_{el}/dt$     versus $|t|$     at Tevatron (blue curve
and data)) and LHC ( black curve and data) energies ($W = 1.8 \,TeV
$ , $ 8 \, TeV$ and $7 \,TeV$ respectively) The solid line without data
shows our prediction for $W=14\,TeV$.}
\label{dsdt}
\end{figure}

The output of our current model is presented in \fig{fit}  , 
in which we display
our calculated cross sections and forward elastic slope. It is
interesting to compare the quality of our present results with
our previous output\cite{GLMLAST}. Recall that the two fits have
almost identical values for the
free parameters, with a single change of $\Delta_{\pom}$
from 0.21 to 0.23.\\
We list  the main features of our results:
\begin{itemize}  
\item
The main feature of our present calculation
is  the excellent
reproduction of TOTEM's values for $\sigma_{tot}$ and $\sigma_{el}$.
The quality of our good fit to $B_{el}$ is maintained.
As regards $\sigma_{inel}$, our results are in accord with the higher
alues obtained by ALICE\cite{ALICE} and TOTEM\cite{TOTEM};
 ATLAS\cite{ATLAS} and CMS \cite{CMS} 

\item
The quality of our output at lower energies,
when compared with ISR data,
is not as good as our
previous results\cite{KAP}, but still acceptable. Recall that Reggeon
exchange, which is included in our model, plays an
important role
at the low energy end of our data, and a negligible role at higher   
energies. As our main goal is to provide a good description of the LHC
data,  we have not tuned the Reggeon parameters, which could lead to an
improved characterization of the ISR measurements.
\item
An interesting observation is that our updated output strongly supports
the CDF total and elastic cross sections rather than the E710 values
\cite{E710}.
This is a common feature of other models\cite{KMR,DL}, that have succeeded
in reproducing the TOTEM results \cite{TOTEM,RISTO} by making a radical
change in their modelings.
\item
Note that our model is the one of the few which achieves a good
reproduction of ISR diffraction, and a reproduction
of the diffractive cross sections at higher energies (as shown in 
\fig{fit}-d
and \fig{fit}-e).
\item
 Our reproduction of SD and DD cross sections is complicated by the  
lack of common definitions of
signatures and mass bounds on the diffractive components.
All models on the market have introduced at least two different
mechanisms to describe diffraction production. In our model these
two mechanisms are: the Good-Walker production of the diffraction 
state with finite unspecified mass
(which is independent of energy and of the values of the parameters
in our model); and  the diffraction due to Pomeron
interactions, where the typical mass depends on $\Delta_\pom$.
In other models (see Refs.
\cite{KAP,KMR,OST}) the two different mechanisms are called:  "low mass"
diffraction and "high mass" diffraction. 
\item
Our calculated $d\sigma_{el}(t\leq\,0.55GeV^2)/dt$ is presented in \fig{dsdt}
together with the corresponding data. The quality of the
fit is very good in this region of t.
It shows that the impact parameter dependence of the model does
not change considerably and reproduces the experimental data
as well, as in our previous version of the model.
We refrain from trying to reproduce the
diffractive dip and higher $t$ cross sections since our model is confined
to the forward cone.
\item
Table 3 summarizes our calculated cross sections and $B_{el}$ at
1.8 - 57\,TeV. The Table provides the predicted values of the
cross sections and forward slope at 8, 14 and 57 TeV. 

Considering the energy behaviour of $\sigma_{inel}/\sigma_{tot}$,
the values of this ratio given by our model 
are $\sigma_{inel}/\sigma_{tot}$ = 0.77 at Tevatron energies
decreasing to 0.73 at 57 TeV. \\
The origin of such a slow approach to
the black disc limit of 0.5, turns out to be the same as
with our previous set of parameters (see Ref.\cite{GLMLAST}):
where one of the partial scattering
amplitude, $A_{1,1}\Lb s, b \Rb <\,1$
at all energies,
while $A_{1,2}\Lb s,b\Rb \approx A_{2,2}\Lb s,b \Rb =1$ for $b=0$ at
the same energies.
\end{itemize}

The comparison of our results with experimental data for
$\sigma_{tot}$, $\sigma_{el}$,  $B_{el}$, $\sigma_{sd}$,  $\sigma_{dd}$, 
and  $\sigma_{inel}$ are shown in \fig{fit}.


\begin{table}[h]
\tbl{Fitted parameters for our model. 
The quality of the fit is $\chi^2/d.o.f.$ = 0.86}
{\begin{tabular}{@{}cccccccccc@{}}\toprule
$\Delta_\pom $ & $\beta$ &  $\alpha^{\prime}_{\pom} $& $g_1$ &  $g_2$ &$m_1$ &
$m_2$ & $\gamma $ &$G_{3\pom}$& $\tilde{g}$\\
 &   &$GeV^{-2}$& $GeV^{-1}$ &$GeV^{-1}$ &$GeV$&$GeV$& &  $GeV^{-1}$& $GeV^{-1}$\\ \colrule
0.23& 0.46  & 0.028  & 1.89  &
61.99 &
  5.045  & 1.71  & 0.0045  & 0.03 & 14.6\\ \botrule  
 \end{tabular}
\label{t1}}
\end{table}
\begin{table}[h]
\tbl{Fitted parameters for the secondary Reggeon in our fit.
}
{\begin{tabular}{@{}ccccc@{}}\toprule 
 $ \Delta_{\reg} $  &  $\alpha^{\prime}_{\reg}\,$& $g^{\reg}_1$ &
$g^{\reg}_2$ & $ R^2_{0,1}\,$ \\
 & $GeV^{-2}$ & $GeV^{-1}$ &$GeV^{-1}$ &$GeV^{-1}$ \\ \colrule
-\,0.47  &  0.4 &  13.5  & 800  &
4.0  \\ \botrule
 \end{tabular}
\label{treg}}
\end{table}

\begin{table}[h]
\tbl{Experimental data versus GLM model.}
{\begin{tabular}{@{}ccccc@{}} \toprule
W & $\sigma^{model}_{tot}$(mb) &  $\sigma^{exp}_{tot}$(mb) &
$\sigma^{model}_{el}$(mb) &
 $\sigma^{exp}_{el}$(mb)\\ \colrule
7 TeV & 98.6 & TOTEM: 98.6 $\pm$2.2 & 24.6 & TOTEM: 25.4 $\pm$1.1\\ \botrule
 W & $\sigma^{model}_{in}(mb)$&
$\sigma^{exp}_{in}$(mb) &$B^{model}_{el}(GeV^{-2}) $&
$B^{exp}_{el}(GeV^{-2})$\\ \colrule
7 TeV & 74.0  & CMS: 68.0$\pm2^{syst}\pm 2.4^{lumi}\pm 4^{extrap}$ &
 20.2 &
TOTEM: 19.9$  \pm 0.3\,$\\
 &  &  ATLAS: 69.4$\pm 2.4^{exp}  \pm 6.9^{extrap}$  & &\\
& &  ALICE: 73.2 $(+2./-4.6)^{model}  \pm 2.6^{lumi}$  & & \\
& & TOTEM: 73.5  $\pm 0.6^{stat}  \pm 1.8^{syst}$ & & \\\botrule
W & $\sigma^{model}_{sd}(mb)$ &  $\sigma^{exp}_{sd}$(mb) &   
 $\sigma^{model}_{dd}$(mb)& $\sigma^{exp}_{dd}$(mb)\\  \colrule 
7 TeV & 10.7${}^{GW}$ +  4.18${}^{nGW}$  & ALICE : 14.9(+3.4/-5.9)
 & 6.21${}^{GW}$  + 1.24${}^{nGW}$  & ALICE: 9.0 $\pm$ 2.6 ~
\\ \botrule
W & $\sigma^{model}_{tot}$(mb)&  $\sigma^{exp}_{tot}$(mb)& $\sigma^{model}_{in}$(mb)& $\sigma^{exp}_{in})mb)$\\ \colrule
57 TeV & 130& AUGER: 133
$\pm 13^{stat} \pm 17^{sys} \pm 16^{Gl}$  
 & 95.2  & AUGER: 92 $\pm7^{stat}\pm11^{syst} \pm7^{Gl}$\\ \botrule
\end{tabular} \label{glm}}
\end{table}

\subsection{ Our partial   amplitudes 
and comparison with other models on the market}
\subsubsection{Amplitudes }

\begin{figure}
 \begin{tabular}{ c c}
\includegraphics[width=0.5\textwidth]{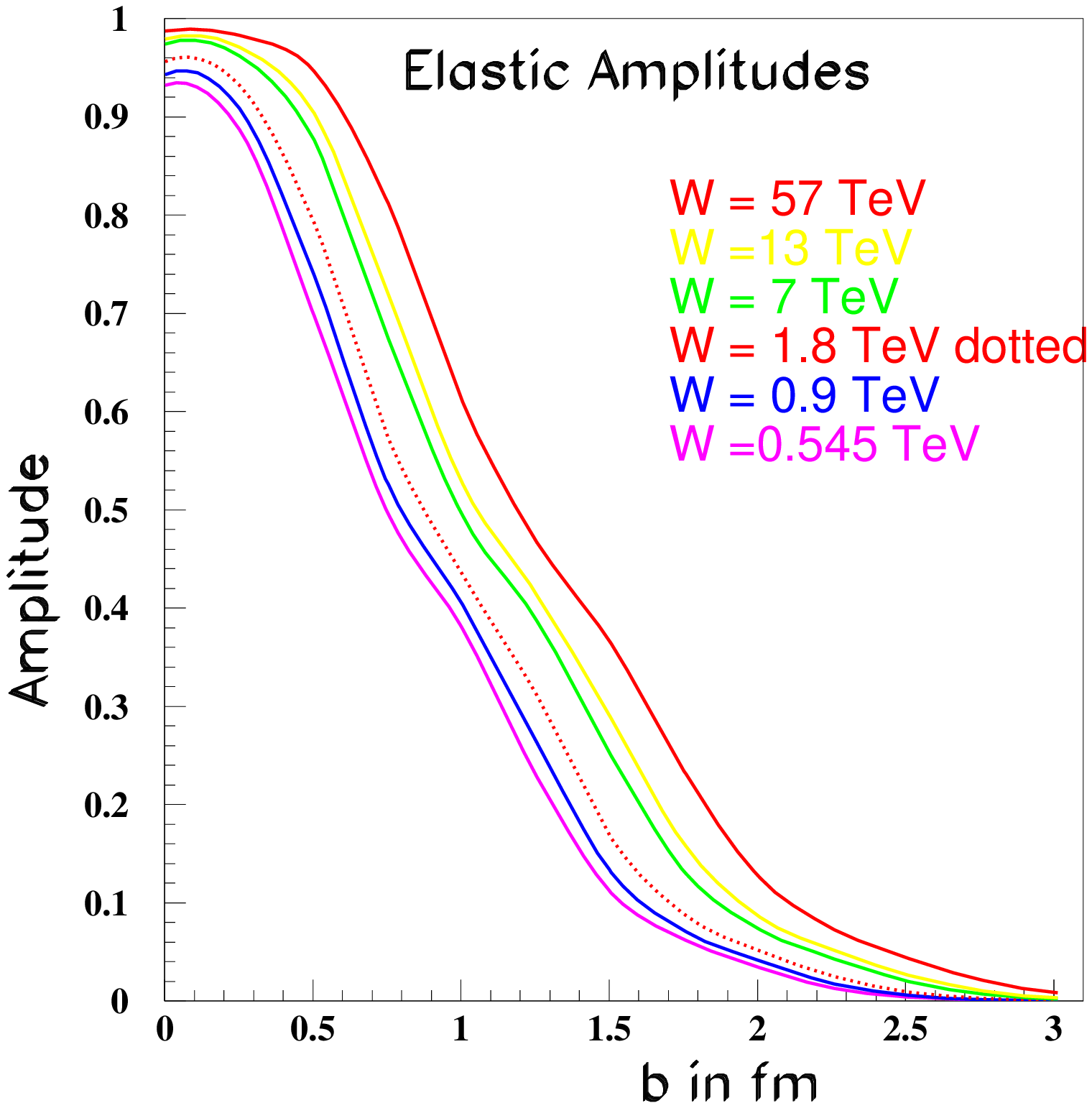}&
\includegraphics[width=0.5\textwidth]{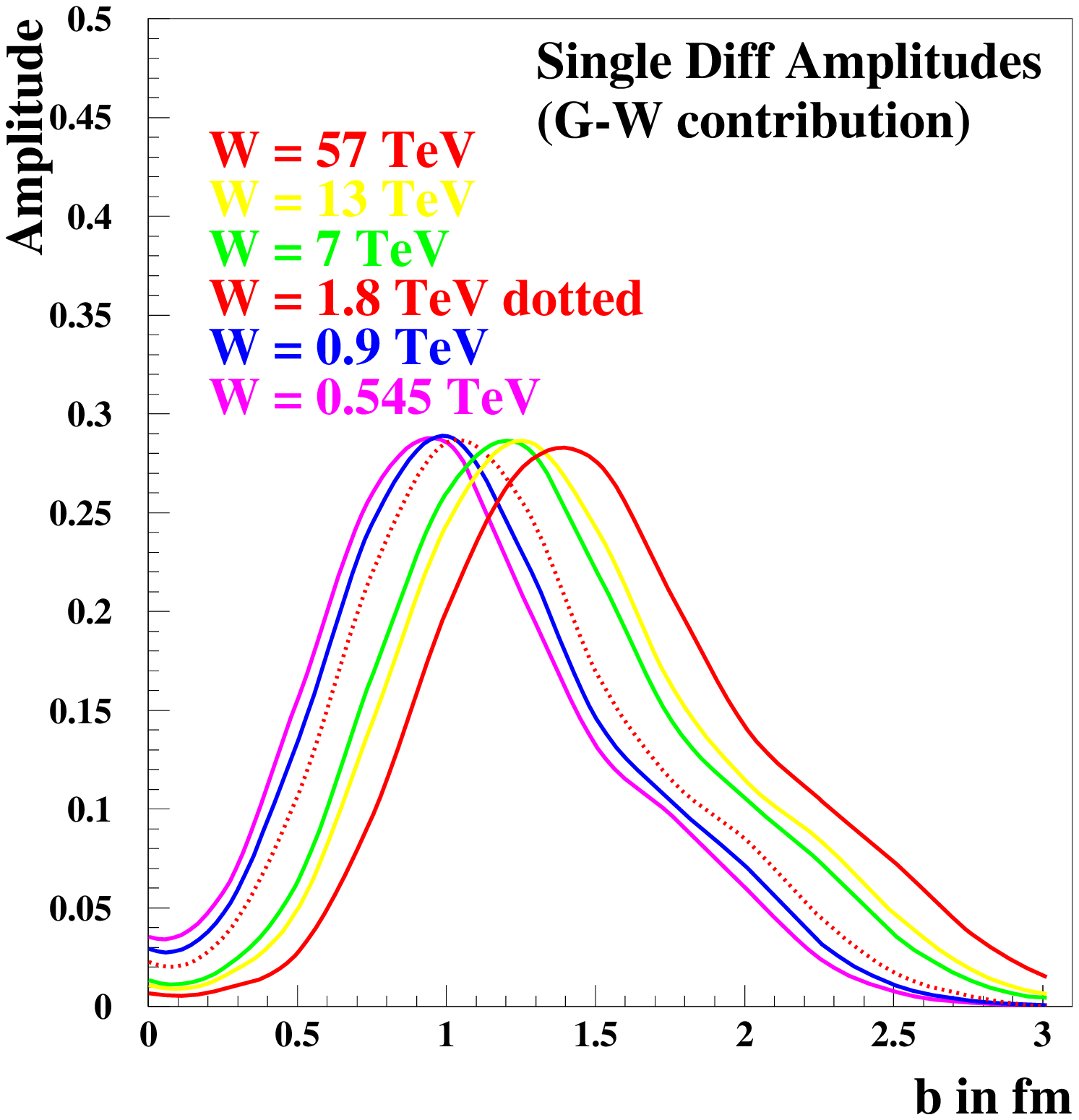}\\
\fig{amp}-a & \fig{amp}-b\\
\includegraphics[width=0.5\textwidth]{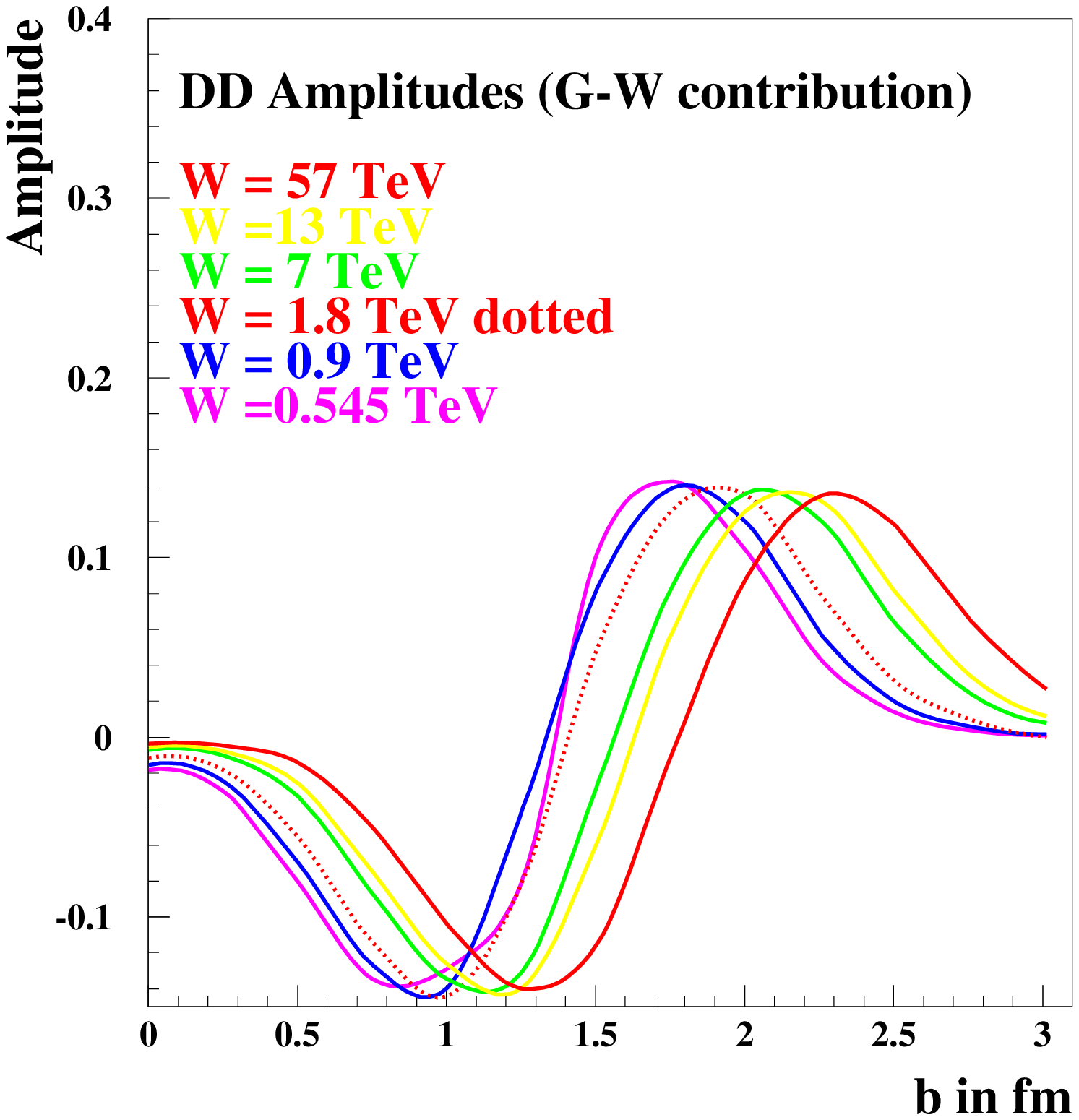} &\includegraphics[width=0.5\textwidth]{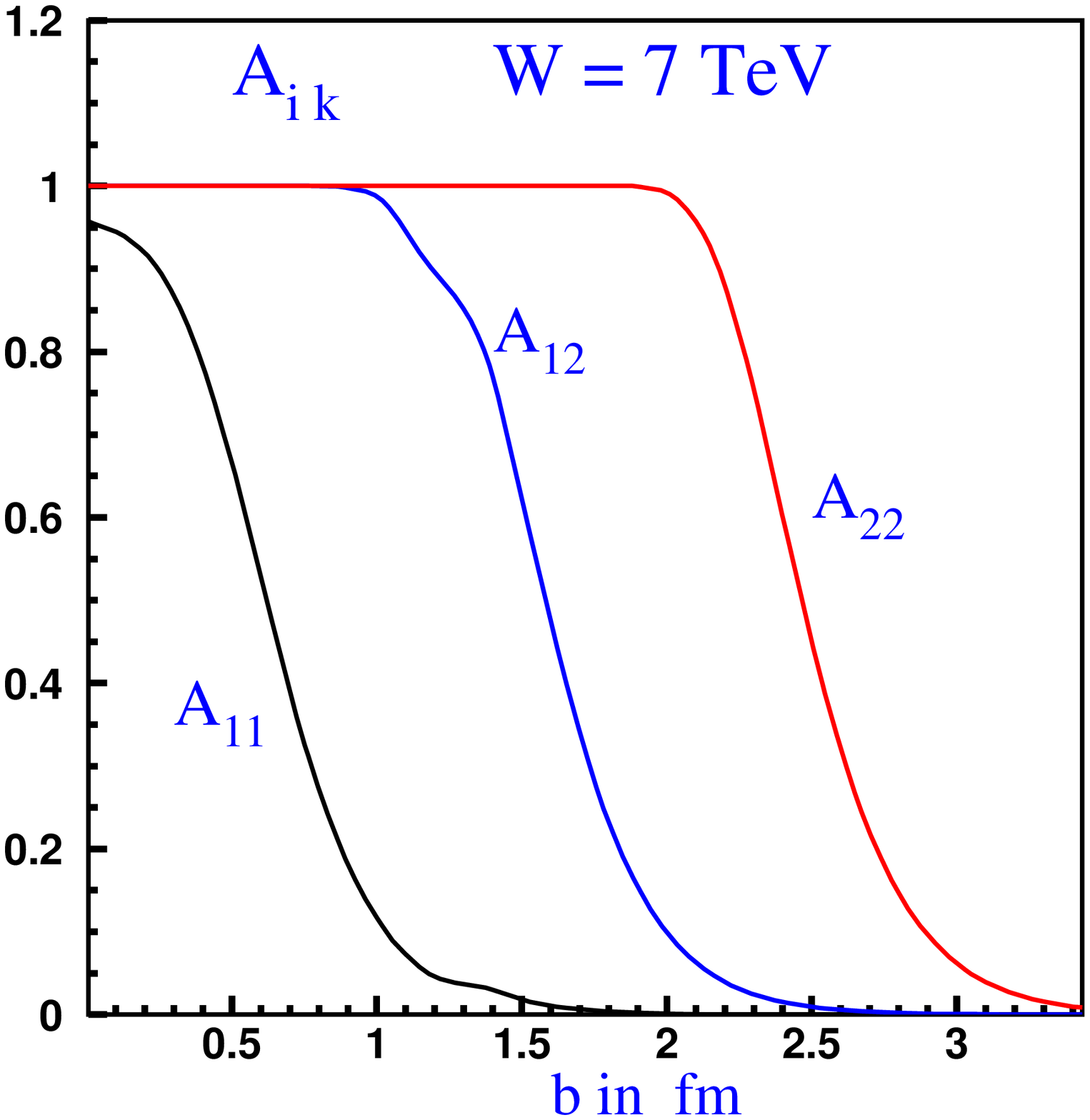}\\
\fig{amp}-c & \fig{amp}-d\\
\end{tabular}
\caption{\label{amp} \fig{amp}-a: GLM elastic amplitudes.  \fig{amp}-b: GLM-GW single diffraction amplitudes. \fig{amp}-cl: GLM
double diffractive amplitudes and \fig{amp}-d: the partial amplitude at the LHC energy (W = 7 TeV) in our model }
\end{figure}

 The Good-Walker formalism \cite{GW} provides an explicit form for the
various elastic and diffractive amplitudes.
 Until recently  most of the comparison of models has been confined
to the level of cross-sections (which are areas), and only reveal the
energy dependence, and do not display other features.
 Having the behaviour of the various
amplitudes as functions of impact parameter (momentum transfer) would be
more revealing.
  Unfortunately, there is a paucity of material available on amplitudes,  
and most refer only to the elastic amplitudes e.g. \cite{KMRL}.
In \fig{amp}-a we show elastic amplitudes emanating from the
GLM model for various energies. We note the overall gaussian shape of the
elastic amplitudes for all energies 0.545 $\leq\; W \; \leq$ 57 TeV,
with the width and height of the gaussian growing with increasing energy.
For
small values of b, the slope of the amplitudes decreases with increasing
energy. The elastic amplitude (as $ b \rightarrow $ 0) becomes almost flat
for W = 57 TeV, where it  is still below
 the Unitarity limit $ A_{el} = 1$ .

In \fig{amp}-b we show the energy behaviour of the GLM (G-W
contribution) of the single diffractive
amplitude as a function of impact parameter for different energies, and in   
\fig{amp}-c we display the behaviour of the double diffractive
amplitude. A common feature of both diffractive amplitudes is that with
increasing energy the peaks broaden and become more peripheral.
We can see in $\fig{amp}-d$ that the partial amplitudes have  
different b (impact parameter)
behaviour, $A_{1,2}$ approaches the black disc limit at b $\approx$ 1 fm,
while $A_{2,2}$ is already at the black disc limit at b $\approx$ 2.2 fm,  
while the third amplitude $A_{1,1}$  is always less than the black disc 
limit.
This is the reason that our elastic amplitude \fig{amp}-a does not reach 
the black 
disc limit.

\par
\begin{figure}
\begin{tabular}{c c}
\includegraphics[width=0.45\textwidth]{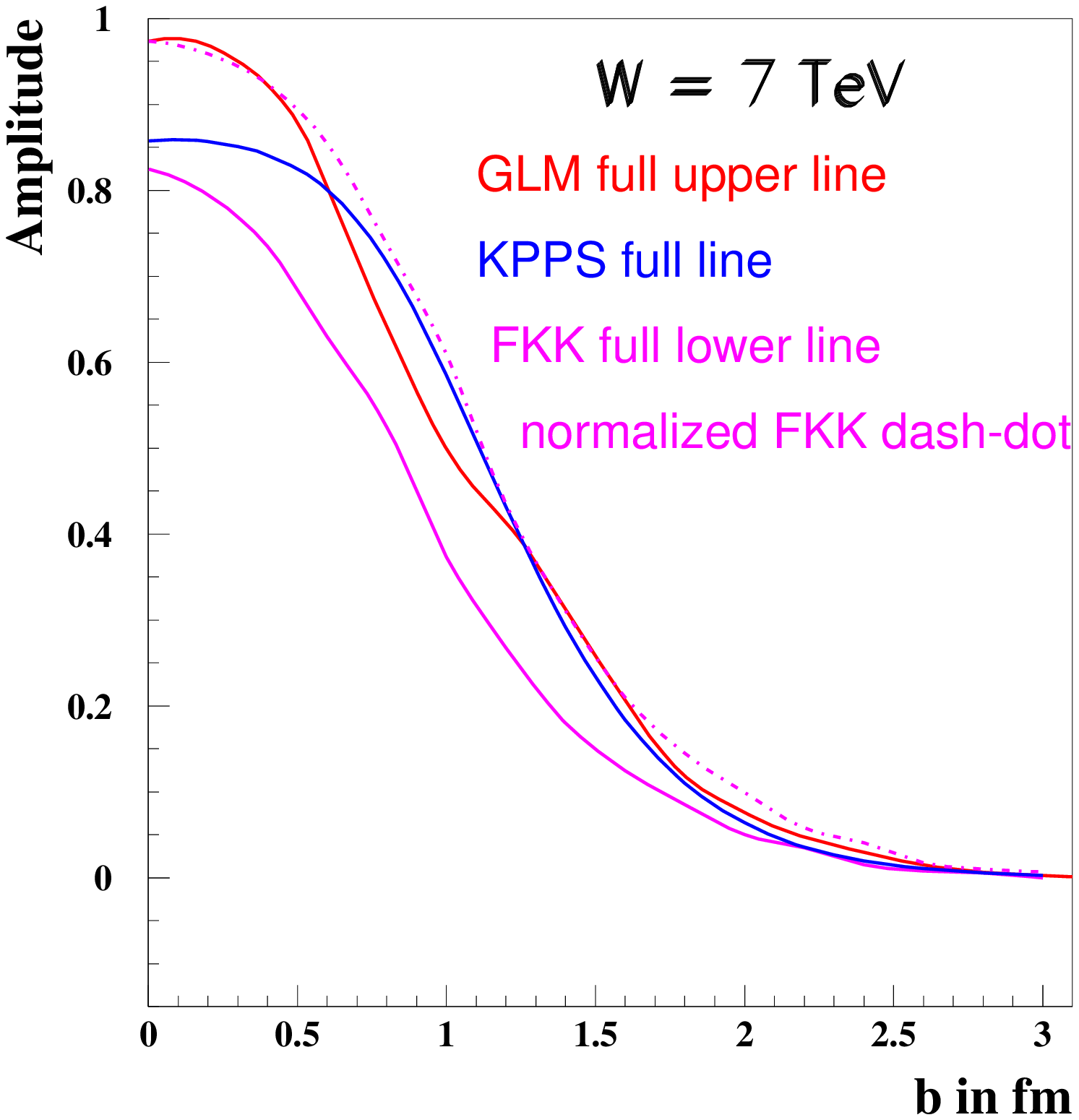} &
\includegraphics[width=0.45\textwidth]{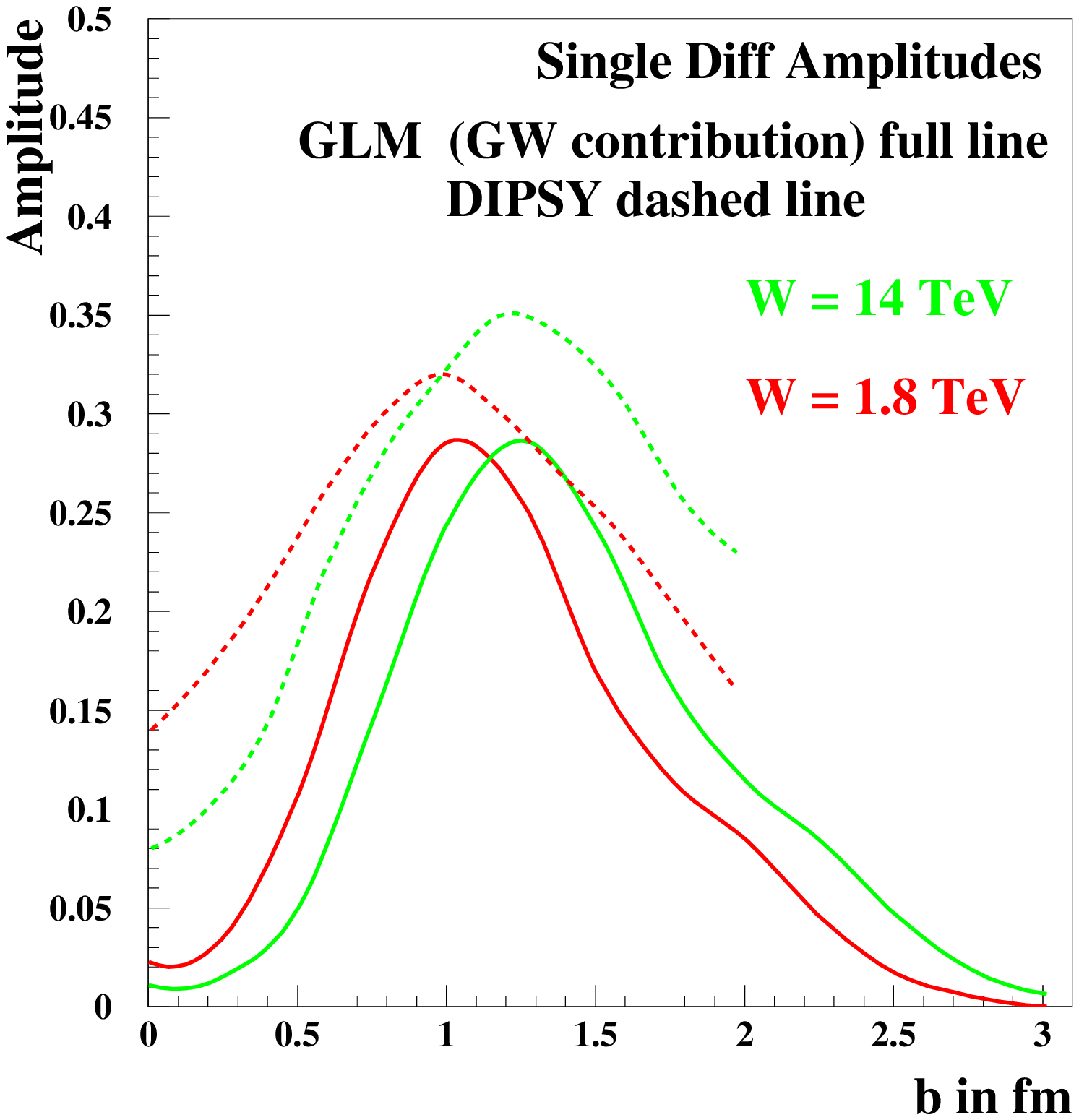}\\
\fig{am}-a & \fig{am}-b\\
\end{tabular}
\caption{\label{am}\fig{am}-a: Comparison of the elastic amplitude at
7 TeV
determined by
FKK \cite{FKK},  KPPS \cite{BKO} and  GLM.\fig{am}-b: Comparison of
single
diffraction amplitude, dashed line DIPSY \cite{FG} (which includes
enhanced and  
semi-enhanced contributions) and full line  GLM (only GW contribution) }
\end{figure}

 Kopeliovich, et al \cite{BKO} have calculated the proton-proton
elastic amplitude
within the framework of a two scale  dipole model. We show their result in
\fig{am}-a, as well as that of
Ferreira, Kodama and Kohara \cite{FKK} who  have recently made a detailed
study of the
proton-proton elastic amplitude for center of mass energy W = 7 TeV, based
on the Stochastic Vacuum Model.

\fig{am}-a we compare the GLM, KPPS and FKK elastic amplitudes
at W = 7 TeV as a function of the
impact parameter. Although the shapes are similar, the KPPS and FKK
amplitudes have  lower intercepts at   b = 0.
If we normalize the FKK amplitude to the GLM value at b
= 0, we note that the amplitudes which are gaussian in shape,  have very
similar behaviour as a function
of the impact parameter.
In \fig{am}-c we show the single diffraction amplitude as given
by the DIPSY Monte Carlo \cite{FG} (dashed line) at W = 1.8 and 14 TeV.   
This includes contributions both from the Good-Walker sector  and enhanced
and semi-enhanced sector. The full line is the GLM amplitude
which only contains the Good-Walker contribution.
Note that, although the amplitudes for the same value of W, peak
at the same value of b, the DIPSY amplitudes are broader and higher, due
to the additional enhanced contributions.
 For historical purposes we mention that the impact parameter behaviour
of
our diffractive amplitudes are in accord with the estimates of  Miettinen   
and Pumplin
\cite{MP} made over 35 years ago.
                                                                        
\par

\subsubsection{Other models on the market}

Ostapchenko~\cite{OS} [pre LHC] has made a comprehensive
calculation
 in the framework of Reggeon Field Theory,
 based on the resummation
 of both enhanced and semi-enhanced Pomeron diagrams.
To fit the total and diffractive cross sections he assumes two Pomerons:
(for his solution set C)
"Soft Pomeron"
 $\,\,\,\alpha^{Soft} = 1.14\, +\, 0.14t$ and a
"Hard Pomeron"
 $\,\,\,\alpha^{Hard} = 1.31\, +\, 0.085t$. His results are
quoted in Table~4, in the column Ostap(C).

Kaidalov and Poghosyan \cite{KP} have proposed a model which is based on 
Regge calculus, where they attempt to describe data on soft diffraction 
taking into account all possible non-enhanced absorptive corrections to 
three Reggeon vertices and loop diagrams. It is a single Pomeron model 
plus secondary Regge poles, they have a  Pomeron intercept $\Delta_{\pom}$ 
=0.12 and a Pomeron slope $\alpha'_{\pom}$ = 0.22 $GeV^{-2}$.
The KP model forms the basis of the Monte
Carlo program used by the ALICE collaboration to analyze their soft
scattering data.
The  predictions of  Kaidalov-Poghsyan \cite{KP} appear in Table 4, in the
column KP.

Ciesielski and Goulianos have proposed an "event generator"~\cite{MBR}
which
 is based on the MBR-enhanced PYTHIA8 simulation. In
Table 4 their results are denoted by MBR.

The Durham group's approach for describing soft hadron-hadron
scattering~\cite{RMK1} is similar  to the GLM~\cite{GLM} approach, they
include both  enhanced and semi-enhanced diagrams. The two
groups utilize different
techniques for summing the multi-Pomeron diagrams.
 The Durham Group \cite{KMR3} to be consistent with the TOTEM result
\cite{TOTEM}, have a model, based on a three channel eikonal description,
with three diffractive eigenstates of different sizes, but with only one
Pomeron. $\Delta_{\pom} \,= \,0.14$; and $\alpha'_{\pom} \,=\,0.1$
GeV$^{-2}$. Which we will refer to as KMR3C.

Recently KMR \cite{KMR4} suggested a two channel eikonal model where the
Pomeron couplings to the diffractive eigenstates depend on the collider
energy. They have four versions of the model. The parameters of the
Pomeron of their "favoured version" Model 4 are:
$\Delta_{\pom}\,=\,0.11$; and $\alpha'_{\pom}\,=\,0.06$ GeV$^{-2}$.
We refer to this as KMR2C.

The predictions of the above models and GLM  are given in 
 Table 4.


\begin{table}[]
\tbl{Comparison of results of the different models for $W$ = 1.8, 7  and 14 TeV}
{\begin{tabular}{@{}ccccccc@{}} \toprule
 W = 1.8 TeV  & GLM & KMR3C & KMR2C &  Ostap(C)  &
 BMR$^{*}$ & KP\\ \colrule
$\sigma_{\rm tot}(mb)$ & 79.2 & 79.3 & 77.2 & 73.0 & 81.03 & 75.0
\\   
$\sigma_{\rm el}(mb)$ & 18.5 & 17.9 & 17.4 & 16.8 & 19.97 & 16.5   \\
$\sigma_{SD}(mb)$ & 11.27 & 5.9(LM) & 2.82(LM)  & 9.2 & 10.22 &10.1   \\
$\sigma_{DD}(mb)$ & 5.51 & 0.7(LM) & 0.14(LM) & 5.2 & 7.67 & 5.8     \\
$B_{el}(GeV^{-2})$ & 17.4 & 18.0 & 17.5  & 17.8 &  &    \\ \botrule
 W = 7 TeV  & GLM & KMR3C & KMR2C &  Ostap(C)  &
 BMR$^{*}$ & KP\\ \colrule
$\sigma_{\rm tot}(mb)$ & 98.6 & 97.4 & 96.4 & 93.3 & 98.3 & 96.4
\\ 
$\sigma_{\rm el}(mb)$ & 24.6 & 23.8 & 24.0 & 23.6 & 27.2 & 24.8   \\
$\sigma_{SD}(mb)$ & 14.88 & 7.3(LM) & 3.05(LM) & 10.3 & 10.91 &12.9   \\
$\sigma_{DD}(mb)$ & 7.45 & 0.9(LM) & 0.14(LM) & 6.5 & 8.82 & 6.1     \\
$B_{el}(GeV^{-2})$ & 20.2 & 20.3 & 19.8 & 19.0 &  &19.0    \\ \botrule
 W = 14 TeV  & GLM & KMR3C & KMR2C & Ostap(C)  &
BMR
& KP  
\\ \colrule
$\sigma_{\rm tot}(mb)$ & 109.0 & 107.5 & 108. & 105. & 109.5 &108.\\
$\sigma_{\rm el}(mb)$ & 27.9 & 27.2& 27.9  & 28.2 & 32.1 & 29.5  \\ 
$\sigma_{SD}(mb)$ & 17.41 & 8.1(LM)& 3.15(LM)  & 11.0 & 11.26 & 14.3   \\
$\sigma_{DD}(mb)$ & 8.38 & 1.1(LM) & 0.14(LM) & 7.1 & 9.47 & 6.4   \\
$B_{el}(GeV^{-2})$ & 21.6 & 21.6 & 21.1 & 21.4 &  & 20.5    \\ \botrule
\end{tabular}
\label{comp}}
\end{table}

\subsection{ Survival probability of the large rapidity gaps (Ref.\cite{GLMSP})}
\par
\begin{figure}[ht]
\centerline{\epsfig{file=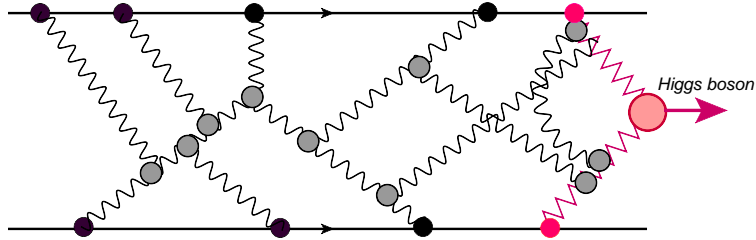,width=100mm}}
\caption{MPSI approximation: diagrams contributing to the 
survival probability. Wavy lines denote the BFKL Pomerons,
the blob stands for the triple BFKL Pomerons vertex. Zigzag line denote 
the hard amplitude}
\label{mpsisp}
\end{figure}

The calculation of the survival probability for the hard processes 
is an excellent example to show that the cross section of  hard 
processes cannot be found without  knowledge of  `soft' physics.  
Indeed,  hard processes  originate from the one parton shower 
interaction. 
In order to select this process, 
it is necessary to apply an additional suppression, so that the 
interaction of more than one parton shower does not take place. 
Only in a restricted numbers of cases, when we have inclusive production, 
do these contributions  cancel, and  not lead to a suppression.  
This suppression factor (survival probability) 
was introduced in Refs.\cite{BJSP,DKSSP} and its general form is:
\beq \label{SP1}
\langle\mid S^2 \mid \rangle = 
\frac{\int \,\,d^2\,b_1\,d^2\,b_2\left\{\sum_{i,k} 
\,<p|i>^2 <p|k>^2\,e^{-\frac{1}{2} \Omega_{i,k}
\Lb(s,(\mathbf{b}_1+\mathbf{b}_2)^2\Rb}\,\,S^H_{i,k}(b,b_1,b_2)\right\}^2}
{\int\,d^2\,b_1\,d^2\,b_2
\left\{\sum_{i,k} <p|i>^2 <p|k>^2\,A^i_H(s,b_1)\,A^k_H(s,b_2)\right\}^2},
\eeq
where,
$<p|i>$ is  equal to $\langle \Psi_{proton}\mid \Psi_i \rangle$ and , 
therefore, $<p|1>=\alpha$ and $ <p|2>=\beta$. 
The factors $\exp[- \Omega_{i,k}/2$ are responsible for the 
Good-Walker mechanism contribution to the survival probability, 
while the factors $S^H_{i,k}(b,b_1,b_2)$ take into account an additional 
suppression due to the structure of $\Omega_{i,k}$ (see \fig{mpsisp} 
for different contributions to $S^H_{i,k}$). Recall that the amplitudes 
$A_{i,k}$ of \eq{SCL2} are equal to 
$A_{i,k} = i \Lb 1 - \exp\Lb - \h\Omega_{i,k}\Rb\Rb$.
$A^i_H$ is the hard process amplitude. Its contribution is 
critical in the production 
of the Higgs boson, in a one parton shower.
\par
$S^H_{i,k}$ can be expressed through the functions $\tilde{G}$, 
in the following way:
\bea \label{SP2}
S^H\Lb Y,b\Rb\,&=&\,\int d^2 b_1 \,
\frac{A^H_i(\vec{b}-\vec{b_1})}{\Lb 1+
\tilde{g}_k(\vec{b}-\vec{b}_1)\,T\Lb Y\Rb \,+
\,\tilde{g}_i(b_1)\,T\Lb Y/2 - y_h/2\Rb\Rb}\nn\\
&\times& \frac{A^H_k( b_1)} {\,\Lb 1 + \tilde{g}_i(\vec{b} - 
\vec{b}_1)\,T\Lb Y\Rb \,+\,\tilde{g}_k(b_1)\,T\Lb Y/2 +y_h/2\Rb\Rb}.
\eea
$y_h \,=\,\ln (M^2_{Higgs}/s_0)$ with $s_0 = 1 \,GeV^2$.

\begin{figure}[ht]
\centerline{\epsfig{file=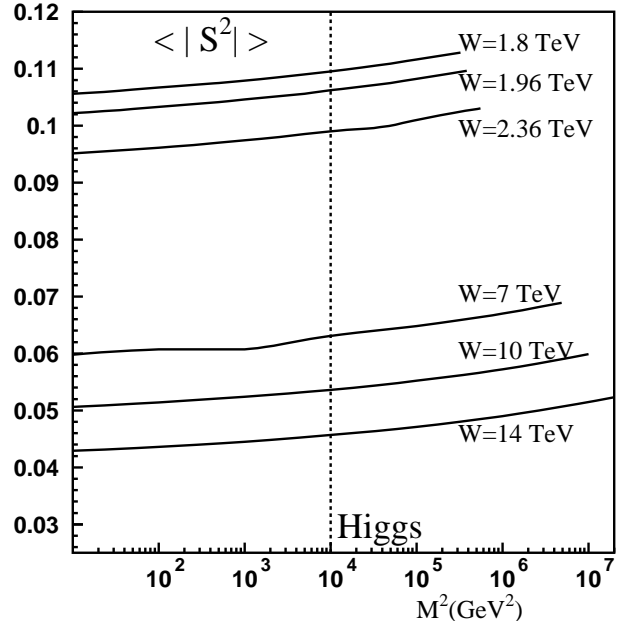,width=90mm}}
\caption{The estimates for the value of SP  for  dijets with mass $M$.   
 The scale on the y axis denotes the value of $<|S^{2}|>$ .
 The dotted line corresponds
 to the Higgs boson production with $M_{Higgs} = 100 \,GeV$.
 We estimate a  $60 \div 70 \% $      margin of error in our results.
  }
\label{sp}
\end{figure}

Substituting in \eq{SP1} for specific values of M at the relevant
energy, we obtain values of the SP which are plotted in \fig{sp}.

 Based on the errors assigned to our fitted parameters, we estimate the
 margin of errors of our results to range from 60\%  at W=2 TeV to 70\% at 
W=
14 TeV. The central values obtained are shown in \fig{sp}.
 The margin of error in our results were estimated by the uncertainty
present in our parametrization of the hard amplitude  (see \cite{GLMM} and 
\eq{SP1}), e.g.
the slope and coupling of the elastic and inelastic hard amplitudes were
taken from the HERA data for $J/ \Psi$ production \cite {KT}, where
 $ 4\,\leq\,B^{H}_{el}\,=\,R^2_H/2\leq \, \, 6 \,GeV^{-2}$, and
$ 0.5\,\leq\,B^{H}_{inel}\,\leq \, \, 1.5 \,GeV^{-2}$.
Our
estimates are 
 compatible with the CDF Tevatron data \cite{CDFt} as well as with the 
theoretical 
 estimates \cite{KMRCDF, ROYON}.

   Our   values of  SP and its  mild dependence of 
the
 mass of produced di-jets have the same origin, the small values of     
 $G_{3\pom}$.

\begin{table} 
 \tbl{Comparison of values obtained by GLM \cite{ GLMSP} and KMR \cite{KMR4}. 
KMR present four models, the numbers quoted are those of their "favoured" 
model 4.}
{\begin{tabular}{@{}ccccc@{}} \toprule
W (TeV) & & GLM ($\%$) & & KMR2 ($\%$) \\  \colrule
1.8  & & 7.02  & & 2.8 \\ 
7 & & 2.98 & &1.5 \\
14 & &1.75 & &1.0 \\ \botrule
\end{tabular}
\label{tsp}}
\end{table}

  Our main conclusions are  that
 the value of the SP, as well as its mass dependence are very
 sensitive to both  the particular form adopted for the Pomeron 
interaction, and
to the values of the fitted parameters that determine the strength of the   
contribution 
 of the
 enhanced diagrams.

\subsubsection{Inclusive cross sections(Ref.\cite{GLMINC})}
\par
\begin{figure}[ht]
\includegraphics[width=120mm]{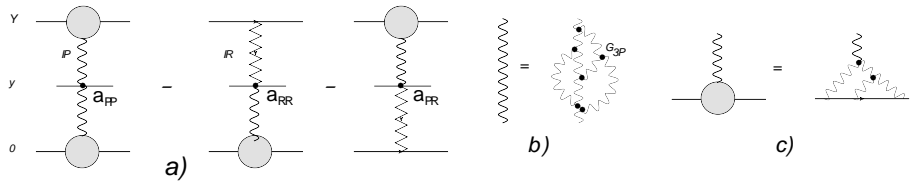}
\caption{Mueller diagrams\cite{MUDI} for a single inclusive cross section. 
A bold  wavy line presents the exact Pomeron Green function of \eq{MPSI6},  
which is the sum of the enhanced diagrams of \fig{mpsiinc}-b. 
A zig-zag line corresponds to the exchange of a Reggeon.
\fig{mpsiinc}-c is a graphic form presenting the vertex that includes the 
Pomeron interaction (see \eq{INC2}.}
\label{mpsiinc}
\end{figure}

In the framework of Pomeron calculus, single inclusive 
cross sections can be calculated 
using the Mueller diagrams \cite{MUDI} shown in \fig{mpsiinc}-a.  
They lead to 
\bea \label{INC1}
&&\frac{1}{\sigma_{in}}\frac{d \sigma}{dy}
=\frac{1}{\sigma_{in}(Y)}\left\{a_{PP}(\alpha^2 g_1 
+\beta^2 g_2)^2 G_{enh}\Lb T\Lb Y/2-y\Rb\Rb\times G\Lb T\Lb Y/2 
+y\Rb\Rb \right.\nn\\
&&-\left.a_{RP} (\alpha^2 g^R_1 
+\beta^2 g^R_2)(\alpha^2 g_1+\beta^2 g_2)\right. \\
&&\left.\left[e^{(\Delta_R (Y/2-y)}\times G\Lb T\Lb Y/2+y\Rb\Rb
+e^{(\Delta_R(Y/2-y)}\times G\Lb T\Lb Y/2+y\Rb\Rb\right]\right\},
\nonumber
\eea
where, $G(T)$ is given by \eq{MPSI6} and sums all enhanced diagrams of 
\fig{mpsiinc}-b.
We need to replace 
\beq \label{INC2}
g_i\Lb b\Rb\,\,\rightarrow\,\,\Gamma_i\Lb y,b \Rb \,=
\,\frac{g_i\,S_i(b)}{1 + g_iG(y)\,S_i(b)},
\eeq
so as to take into account the fan diagrams shown in \fig{mpsiinc}-c.

\begin{figure}
\begin{tabular}{c c}
\epsfig{file=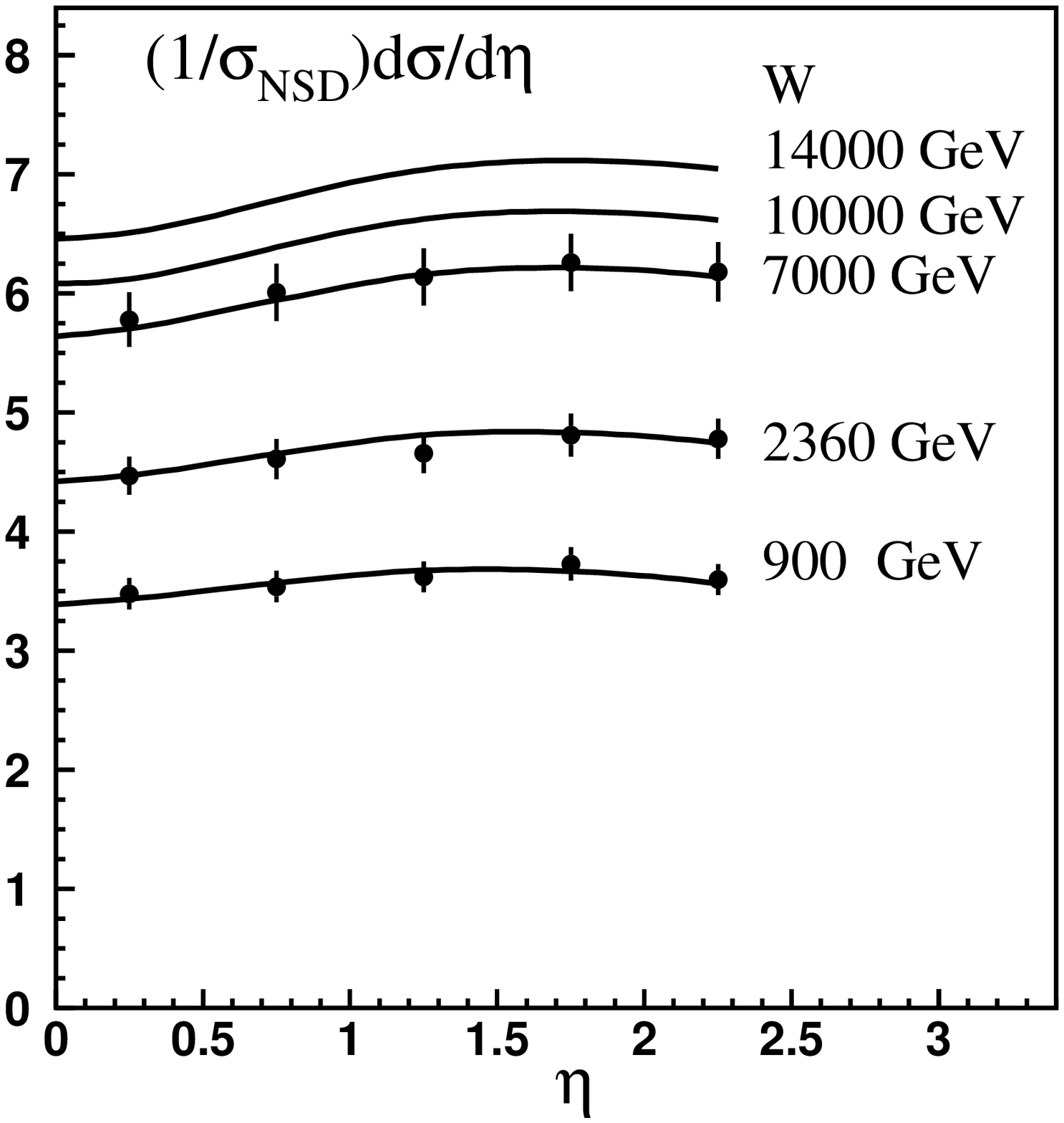,width=60mm} & 
\epsfig{file=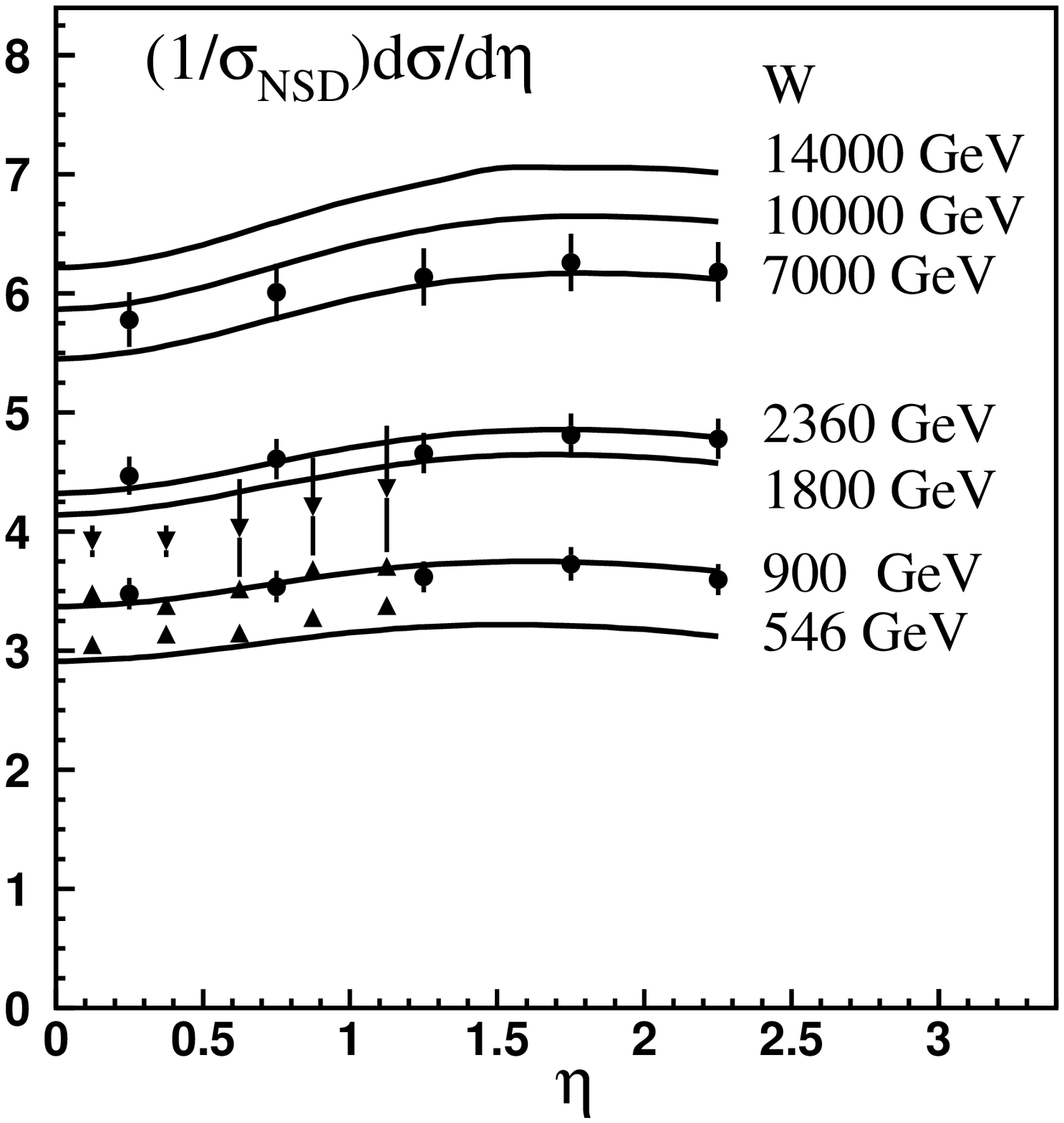,width=60mm}\\
\fig{incl}-a & \fig{incl}-b\\
\end{tabular}
\caption{The single inclusive density versus energy.
The data were taken from Refs.\cite{ALICE,CMS,ATLAS}
and from Ref.\cite{PDG}.
The fit to the CMS data is plotted in \fig{incl}-a,
while \fig{incl}-b presents the description of all inclusive spectra
with $W \geq 546 GeV$.}
\label{incl}  
\end{figure}

This calculation entails three additional
parameters.
The determination of these parameters from existing data \cite{PDG} is
not trivial. Comparing the numbers corresponding to the data shown in
 \fig{incl}, it is evident that a conventional overall
 $\chi^2$ analysis is impractical, owing to the quoted error bars of the  
 546 GeV data points, which are considerably smaller than the error bars
quoted for the other energies.
The full lines in \fig{incl}  are the  results derived from a $\chi^2$
fit 
to the  200-1800 GeV data, excluding the 546 GeV points. This fit
yields a seemingly poor $\chi^2/d.o.f = 3.2$. Despite this,
we consider this fit to be acceptable, as the data points
"oscillate" about a uniform  line with error bars
which are much smaller than their deviation from a smooth average.
In our procedure, the line for 546 GeV  in \fig{incl} is calculated
with the model parameters and is visually compatible with the experimental
data points. Note that both the axes of
 \fig{incl} are linear,
and that  our calculation
coincides with the  LHC experimental results \cite{ALICE} and \cite{CMS}.
We have
also
made predictions for the higher energies at which the LHC is expected to
run, see \fig{incl}.
 The contributions of the secondary Regge trajectories are minimal.
The experimental values for
$\sigma_{in} = \sigma_{tot} - \sigma_{el} - \sigma_{diff}$
were taken from Refs.\cite{PDG,ALICE,CMS}. For our predictions we have
used
the values of $\sigma_{in}$ calculated in our GLMM model.
Our output over-estimates the few data points with
$\eta > 4$ data at 546 and 900 GeV by up to 20$\%$.
This is to be expected, as we have not taken into account the
parton correlations due to energy conservation, which are important in the
fragmentation region, but  difficult to include
in the framework of  Pomeron calculus.


\begin{table}
\tbl{Values of parameters used for the two fits.   }
{\begin{tabular}{@{}cccc@{}} \toprule
Data & $a_{\pom \pom} $ &   $a_{\pom \reg}$  & $Q_0/Q$ \\  \colrule
CMS  & 0.39  & 0.186 & 0.427\\ 
All & 0.413 & 0.194 & 0.356\\ \botrule

\end{tabular}
\label{tincl} 
}
\end{table}                                     

We extract the three new parameters: $a_{\pom \pom},
a_{\pom \reg}$ and $Q_0/Q$ from the experimental inclusive data.
$a_{\pom \pom}$ and
 $a_{\reg \pom}$
   describe the emission of  hadrons from the
 Pomeron and  the Reggeon.
 As well as two dimensional
 parameters
 $Q$ and $Q_{0}$,
 $Q$ is  the average transverse momentum
of produced minijets, and $\frac{Q_{0}}{2}$ denotes the mass of the
slowest hadron produced in the decay of the minijet(see Ref.\cite{KHL}
 for more details).

We made two separate fits:
(a) fitting only the CMS data at different LHC
energies (see \fig{incl}-a); and (b) fitting all inclusive
data for $W \geq 546 \,GeV$ (see \fig{incl}-b).
We choose only data in the central region of rapidity,
as we have not included energy conservation, and therefore
our model is inadequate
to describe the data behavior in the fragmentation region.
\fig{incl} shows that the soft model based on the Pomeron approach is able 
to
describe the behavior and the value of the inclusive production
observed experimentally.
Our predictions are shown in the same figure.
We note that the final version of our approach which includes
the contributions of enhanced, semi-enhanced and net diagrams
(see Ref.\cite{GLMLAST}) provides a much better description of
the data than we obtained
in our previous attempt\cite{GLMINC},
where only enhanced diagrams were summed.
\subsubsection{Double inclusive production: 
two particle correlations)(Ref.\cite{GLMCOR})}

In this section we  take the next step (following the
successful description of  inclusive hadron production) in   describing
 the structure of the
bias events without the aid of  Monte Carlo codes.
 Two new results are presented :(i)  a method for calculating
  the two particle correlation functions in the BFKL Pomeron calculus
 in zero transverse dimension;
 and (ii)  an estimation of the   values of these correlations in a model
of  soft interactions.
 
\par
\begin{figure}[h]
\begin{center}
\includegraphics[width=0.7\textwidth]{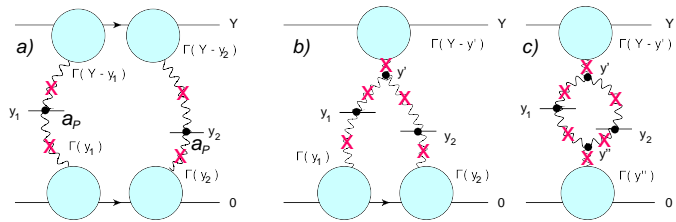}
\end{center}
\caption{Mueller diagrams for double inclusive 
production . 
Crosses mark the cut Pomerons. $G\Lb y \Rb$ is given by \protect\eq{MPSI6} and
$\Gamma_i$ by \eq{INC2}. 
All rapidities are in the laboratory reference frame.}
\label{mpsicor}
\end{figure}

The Mueller diagrams\cite{MUDI} that contribute to double inclusive cross section 
are shown in \fig{mpsicor}.
\par
All ingredients of \fig{mpsicor} have been discussed above: 
the Pomeron propagator is given by \eq{MPSI6} and 
$\Gamma_i\Lb Y,b \Rb$ should be taken from \eq{INC2}.

  In Pomeron calculus the long range correlations in rapidity stem from 
the
production of two hadrons from two different Pomerons
(two different parton showers), as shown in (l.h.) figure.
i.e. two hadrons in the central rapidity region can be produced
in an event with more than two parton showers.
\beq
R\Lb y_1, y_2\Rb\,\,\,=\,\,\,\frac{\rho_2(y_1,y_2)\,\,-\rho(y_1)
 \rho(y_2)}{\rho(y_1) \rho(y_2)}\,\,=\,\,\,
\frac{\frac{1}{\sigma_{in}}\,
\frac{d^2 \sigma}{d y_1\,d y_2}}{\frac{1}{\sigma_{in}}\,
\frac{d \sigma}{d y_1}\,
\frac{1}{\sigma_{in}}\,\frac{d \sigma}{d y_2}}\,\,\,-1. \nonumber
\eeq
 $\sigma_{in}$  denotes the inelastic cross section 
   $ \rho_2(y_1,y_2)\,\,=\,\,\frac{1}{\sigma_{in}}\,\frac{d^2 \sigma}{d
y_1\,d y_2}$ and
$\rho (y)\,\,=\,\,\frac{1}{\sigma_{in}}\frac{d \sigma}{dy}$.

If particles were emitted independently   then
$
\rho_2(y_1,y_2)\,\,=\,\,\rho(y_1) \rho(y_2).
$

\begin{figure}
\centerline{\includegraphics[width=80mm,height=80mmm]{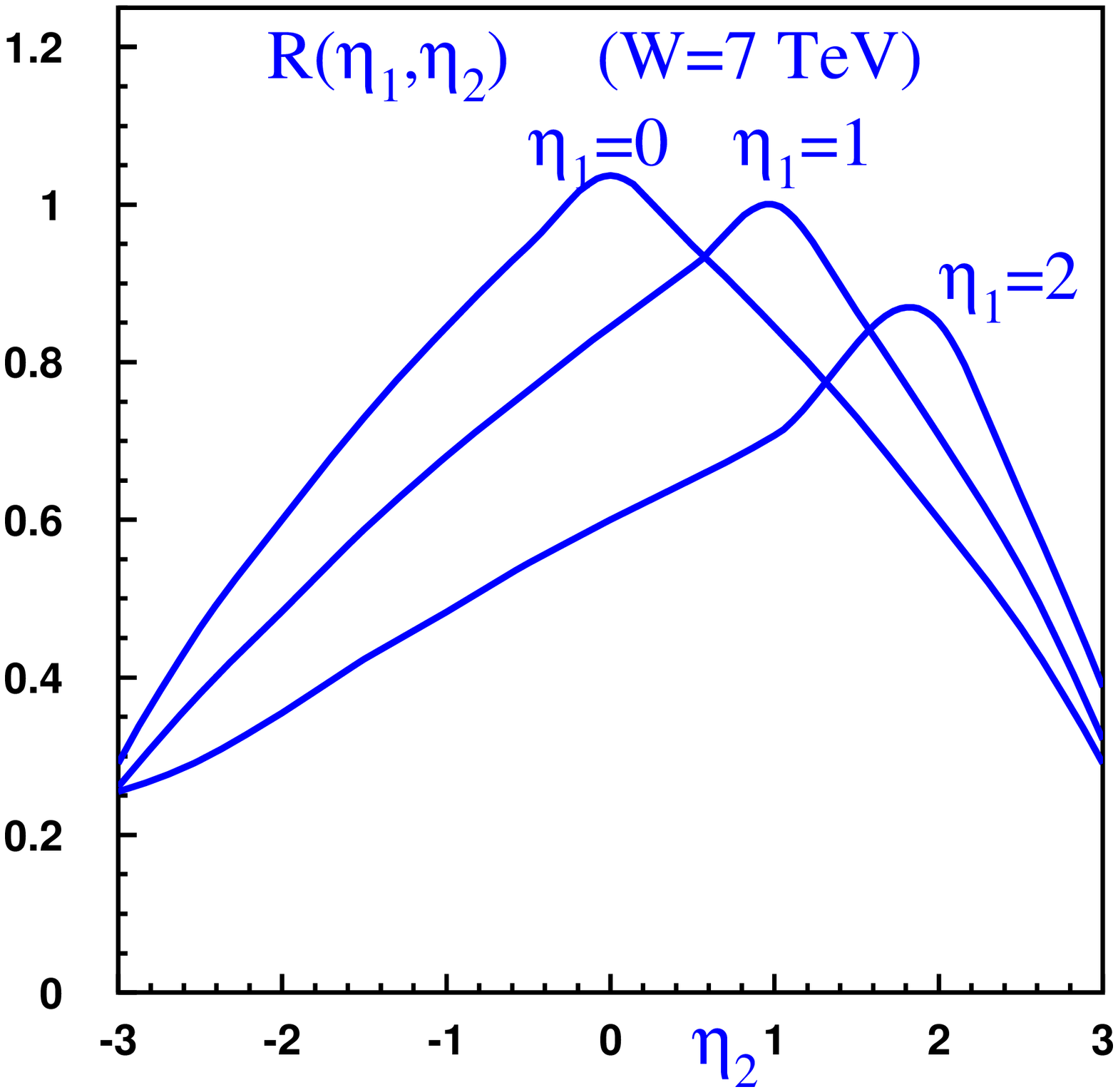}  }
\caption{Multiplicity distribution measured by the CMS collaboration 
\cite{CMSMULT} and our parameters discussed in \cite{GLMCOR}}
\label{cor1}
\end{figure}

Our results for $R\Lb \eta_1, \eta_2\Rb$ versus $\eta_2$
at different values of $\eta_1$ at $W = 7 \,TeV$ is shown in \fig{cor1}.

Surprizingly this function falls
steeply  at large $\eta_2$. This is not expected
 since all  three diagrams  which contribute
generate long range rapidity correlations.
The main contribution comes from the enhanced diagram \fig{mpsicor}-c.
 The eikonal-type diagram of \fig{mpsicor}-a leads 
 to long range rapidity correlations which do not depend on the
values
of $\eta_1$ and $\eta_2$.
 The diagram of \fig{mpsicor}-b gives
a negligible contribution.

 \begin{table}[th]
\tbl{Normalized Moments $C_q\,=\,\langle n^q\rangle/\langle n\rangle^q$ 
 for W = 7 TeV  and $|\eta| \leq 0.5$ .}
{\begin{tabular}{@{}cccc@{}} \toprule 
q  & 3 & 4 &  5 \\ \colrule
CMS & 5.8$\pm$0.6   & 22 $\pm$2   &
90 $\pm$ 18 \\
 GLM &
5.65  & 21.18 & 98.2 \\  \botrule
\end{tabular}\label{moments}}
\end{table}

Using parameters determined in our fit to the Single Inclusive Cross
Section for W = 7 TeV and $|\eta| \leq 0.5$, i.e. $\langle n \rangle
\,=\,$5.8 and $r = 1.25 $\cite{GLMLAST}.

Then
\beq
\frac{\sigma_n}{\sigma_{in}}\,\,=\,\,\Lb \frac{r}{r +
\langle n \rangle}\Rb^r
\frac{\Gamma\Lb n + r\Rb}{n!\, \Gamma\Lb r \Rb}\Lb
\frac{\langle n \rangle}{r\,+\,\langle n \rangle}\Rb^n. \nonumber
\eeq


\begin{table}[h]
\tbl{$R(y_1=0, y_2=0)$ for different energies.}
{\begin{tabular}{@{}ccccc@{}}\toprule
 W(TeV) & 0.9 & 1.8 & 2.36 & 7 \\ \colrule
$R(y_1 = 0,y_2 = 0)$& 1.0 & 1.12 & 1.026 & 1.034\\  \botrule
\end{tabular} \label{cor11}}
\end{table}   

This result is in a good agreement with the CMS data on
multiplicity distribution \cite{CMSMULT}. Indeed, experimentally,
$C_2 \,=\,\langle n^2 \rangle/\langle n \rangle^2$
was measured for the rapidity window $|\eta|\, < \,0.5$
in the energy range W = 0.9 to  7 TeV
(see \fig{cor1} in Ref.\cite{CMSMULT}) and $C_2 \approx 2$.
For this small range of rapidity,
we can consider that $C_2 = R\Lb 0, 0\Rb + 1$. 
It is worthwhile mentioning that using our calculation of $R\Lb0,0\Rb$,
we can calculate the parameters of the negative binomial distribution.   
Using this distribution we calculate
$C_q\,=\,\langle n^q\rangle/\langle n \rangle^q$ given in 
Table 7 .

The attraction of the Pomeron approach reveals itself in the possibility
to discuss not only the forward scattering data but, also,
to make predictions relating to multiparticle production processes
using the AGK cutting rules \cite{AGK}.

\subsubsection{ Proton-Air scattering (Ref.\cite{GLMpAir})}

As we have discussed above,the LHC data \cite{ALICE,ATLAS,CMS,TOTEM} provided two important lessons for our
 understanding of soft interactions at high energy. The first, regrettably, none
 of the phenomenological models based on the Reggeon 
approach
 were able to predict the data, in spite of having a large number of 
fitting
 parameters.  
The  second, a more 
encouraging one, the LHC data could be fitted by choosing a  new set of the 
parameters
 without changes in the theoretical scheme of the models. The natural question 
that arises, is
 whether the new set of parameters has any predictable power,
 or its life time is
 only until  new measurements at higher energies appear. In particular, we ask 
this question
 in relation to our model, and to  our final set of  fitting parameters (see Table 1)
 .  Our model has passed the first check:  as one can see from Table 3 it is
 able to describe the proton-proton inelastic and total cross section at
 $W = 57\,TeV$ that has been extracted from the Pierre Auger Collaboration
 data on proton-Air interactions\cite{PACOL}. 
 Our goal in this section is to compare our model directly with the cosmic 
ray data
 on the  proton-Air interactions.
We re-visit the problem of hadron-nucleus interactions at high 
energies. It is
well known that  the Glauber-Gribov approach
\cite{GLAUB,GRIBA}, 
where
the total cross section of the hadron-nucleus interaction is expressed 
through the inelastic cross section of hadron-proton scattering, can only 
be
 justified  at rather low energies, where  corrections due to Pomeron
 interactions may  be neglected. A more general approach has been
 developed\cite{SCHW,KAID,BGLM,GLMpAir} in which the Pomeron interactions
 have been taken into account in the energy range for :
\bea \label{KRHA}
&g\,S_{A}(b)\,G_{3\pom}\,e^{\Delta_{\pom} Y} \,\,\propto\,\,g\,
G_{3\pom} A^{1/3}\,e^{\Delta_{\pom} Y} \,\,\approx \,\,1;&\nn\\\,\,\,\,\,\,\,\,\,\,\,\,
&g\,G_{3\pom}\,e^{\Delta_{\pom} Y} \,\,< \,\,1;\,\,\,\,\,\,\,\,\,\,\,\,\,\,\,\,\,\,\,g\,
G^2_{3\pom}\,e^{\Delta_{\pom} Y}\,e^{\Delta_{\pom} Y} \,\,\ll\,\,1.& 
\eea

 For the nuclear profile $S_A(b)$ we use the general 
expression
\beq \label{SA}
S_A(b)\,\,=\,\,\int^{+ \infty}_{- \infty}\,d z\,\rho\Lb z,b\Rb\,\,\,\,\,\,\,\,
\,\int d^2 b\,S_A\Lb b \Rb\,\,=\,\,A\,;
\eeq
where $\rho(z, b)$ is the density of nucleons in a nucleus.

\begin{figure}
\centerline{\epsfig{file=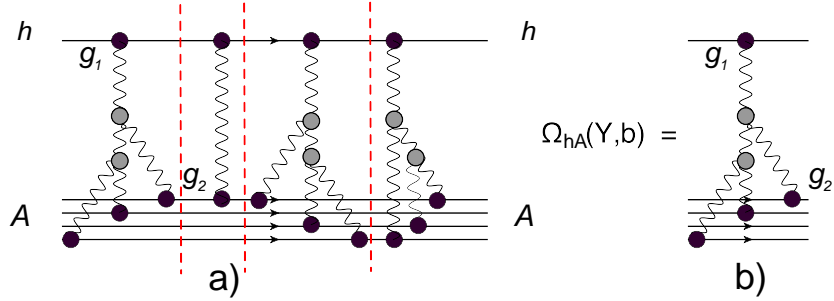,width=110mm}}
\caption{The set of diagrams that contribute to the scattering amplitude of 
hadron-nucleus scattering in the kinematic region given by \eq{KRHA}.
\fig{hAset}-b shows the hadron-nucleus irreducible diagrams while the
general case is shown in \fig{hAset}-a. . The vertical dashed lines indicate 
the hadron-nucleus states. The wavy lines denote the soft Pomerons.}
\label{hAset}
\end{figure}
\par
In this region we need to sum the diagrams of \fig{hAset}.
The final formula that includes both the Good-Walker mechanism of low
mass diffraction and the enhanced Pomeron diagrams is:
\bea \label{FHA}
\sigma_{in}\Lb p+A; Y\Rb\,\,&=&\,\int d^2 b \Bigg(1 - \exp \Bigg( - 
\Big\{ 2\, \mbox{Im}\, a^{el}_{\mbox{\tiny pp}}\Lb Y,\vec{b}\,- \,\vec{b}^{\,'}
\Rb\,{\cal S}_A\Lb Y;\, \vec{b}^{\,'}\Rb\\
&-& \Big[ \,\sigma_{el}\Lb Y,\vec{b}\,-\,\vec{b}^{\,'}\Rb
+\sigma_{diff}\Lb Y,\vec{b} \,-\,\vec{b}^{\,'}\Rb \Big]
\frac{{\cal S}^2_A\Lb Y;\, \vec{b}^{\,'}\Rb}{S_A \Lb \vec{b}^{\,'}\Rb}
\Big\} \Bigg) \Bigg);
\nonumber\\
&&\sigma_{el}\Lb Y,\vec{b}\,-\,\vec{b}^{\,'}\Rb\,
+\,\sigma_{diff}\Lb Y,\vec{b} \,-\,\vec{b}^{\,'}\Rb\,\,=\nn\\
&&\,\,|a^{el}_{\mbox{\tiny pp}}\Lb Y,\vec{b} \,-\,\vec{b}^{\,'}\Rb|^2\,+
\,|a^{sd}_{\mbox{\tiny pp}}\Lb Y,\vec{b} \,-\,\vec{b}^{\,'}\Rb|^2\,+
\,|a^{dd}_{\mbox{\tiny pp}}\Lb Y,\vec{b} \,-\,\vec{b}^{\,'}\Rb|^2.
\nonumber
\eea

   Before comparing with the experimental results, we would like to draw
the reader's attention to the fact that some of the experimental results 
shown might be overestimated, 
due to the possibility of the airshowers being created by helium nuclei, 
as well as protons.
 The importance of this phenomena has been investigated by Block\cite{BLO} 
and the Auger 
colloboration\cite{PACOL}. We refer the reader to these references for 
further details.
In the paper of the Pierre Auger collaboration, a possible contamination of 25\% of
helium was assumed  which produces an 
uncertainty of about
~ 30 mb (which is less than 10\% of their final result), and is included 
in their systematic error.

\begin{figure}
\centerline{\epsfig{file=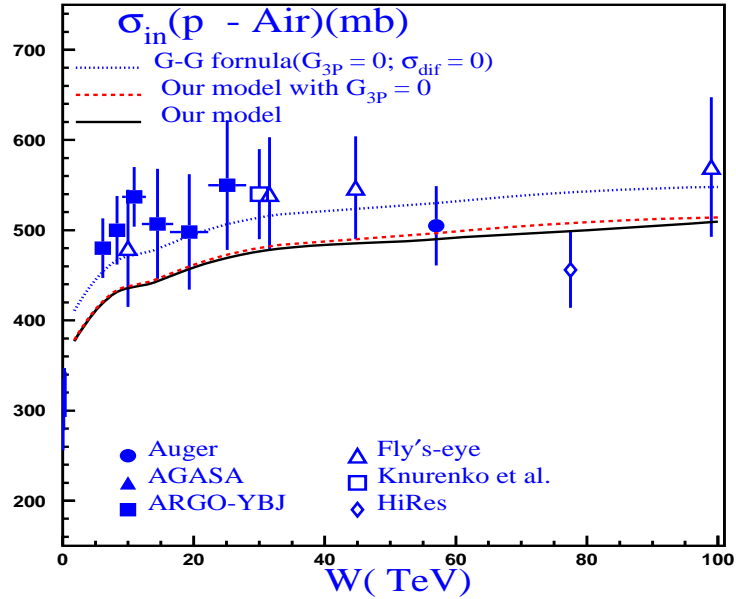,width=110mm,height=90mm}}
\caption{ Comparison the energy dependence of the total cross
 section for proton-Air interaction with the high energy experimental
 data. Data are taken from Refs.\cite{HRE,KNU,AGASA,EAS,ARGO}.
  $p_{lab} = W^2/(2 m) $ where m denotes the proton mass.
    }
\label{sig}
\end{figure}

The results of our calculations are shown in \fig{sig}.  For our calculations
  we used the parameters of our model, presented in Table 1 and Table 2.  For the
 scattering with air we use $S_{\mbox{\tiny Air}}\Lb b\Rb\,\,=\,\,0.78
 S_{\mbox{\tiny Ng}}\Lb b \Rb
+ 0.22 S_{\mbox{\tiny O}}\Lb b \Rb$ where $Ng$ and $O$ denote nitrogen
 and oxygen,
 respectively. For both these nuclei we used the harmonic oscillator
 parametrization, following  Ref. \cite{WS}.

Two conclusions follow from the results of our calculations. First, all 
formulae
  including the Glauber-Gribov one, give good agreement with the experimental
 data. This agreement improves (at least does not deteriorate) at
 ultra high energies beyond  the accelerator region ($W\, > \,8 TeV$).

Second, the inelastic Gribov corrections ( \fig{sig} the curve
 with $G_{3 \pom} = 0$),  decrease the value of the inelastic
 p-Air cross section by  7 - 10\% which are within the experimental
 errors. The corrections  due to the  Pomeron interaction turns out to be
 negligibly small (see our model curve in  \fig{sig} ).
 We would like to stress that our model \cite{GLM} gives a
 smaller contribution for Pomeron interactions, when compared
 to other attempts to describe the LHC data\cite{KMR,OST}. 
 
 We also calculate 
 the total and inelastic cross sections for proton-lead interaction
 at high energy, to check whether the corrections due to Pomeron 
interactions
 are visible in the collisions with heavy nuclei.  For a heavy nucleus
 such as lead
we can use \eq{SEA1}, and our prediction will  not depend on the 
details
 of $b$ distribution for the proton-proton scattering. We employ
the Wood-Saxon
 parametrization for $S_A\Lb b \Rb$ 
\beq \label{WS}
S_A\Lb b \Rb\,\,=\,\ \int^\infty_{- \infty} d z \,\frac{\rho_0}{1 \,+\,
 \exp\Big(  \frac{\sqrt{z^2 + b^2} - R_A}{h}\Big)}.
\eeq

\begin{figure}
\centerline{\epsfig{file=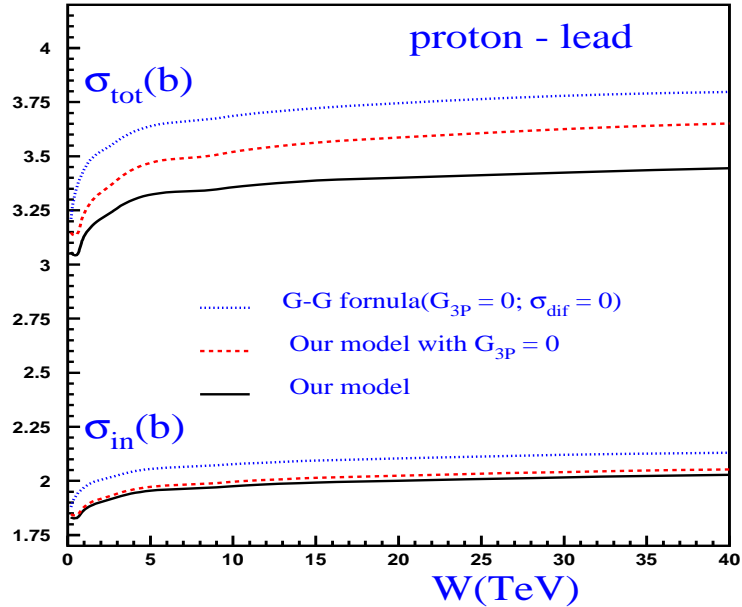,width=110mm,height=90mm}}
\caption{   Our model predictions for proton-lead cross sections.
 In this figure $\sigma$ in barns while $W$ in TeV.  }
\label{sigpb}
\end{figure}

 \fig{sigpb} shows our predictions for the total and inelastic
 cross sections for the proton-lead interaction at high energy.
   For heavy nuclei the difference between our
 approach and Glauber-Gribov formula is not
 large, reaching about 11\% for 
 the total 
and 5\% for the  inelastic cross sections. It is instructive to note
 that the inelastic cross section for heavy nuclei is not sensitive
 to  Pomeron interactions, and the major difference from Glauber-Gribov
 formula stems from Good-Walker mechanism for low mass diffraction in
 proton-proton collisions. However, all three contributions influence
  the value and energy behaviour of the total cross sections.

\section{Conclusions}

 Our model, reviewed here, gives an example of a self 
consistent approach, that incorporates our theoretical understanding
 of long distance physics, based both on N=4 SYM for strong coupling 
 and on  the matching with the perturtbative QCD approach.
 We hope that we have demonstrated how important and decisive   the LHC 
data are
 on strong interactions which led  us to  the set of the phenomenological
 parameters that 
fully confirmed our theoretical expectations.  We also showed how far we have
 come towards creating the framework for the description of the minimal 
bias
 events for high energy scattering without need to generate  Monte Carlo 
codes.

A result emerging from  our rather complex calculations and 
phenomenology is a simple
 picture of a hadron, in which a fast hadron is a quantum mixture 
of
 two state: the large component leads to gray disc regime with the
 transparency which is sizeable even at   energies as high as $W = 57 
TeV$;
 and the small black component which reaches the unitarity bound at rather 
low
 energies. The sizes of these components do not depend on energy while the
 blackness increases at high energies.

  On the qualitative level, this picture is very similar to the structure
 of the interaction that arises from the saturation/CGC approach 
 (see \cite{KOLE} and references therein).  Unfortunately, our
 model does not  include a matching with this approach in spite of providing
a qualitatively close picture. Some attempts have been made\cite{GLMMASS,
 KMRSAT} to include the main features of the saturation/CGC approach in
 the soft interaction models, but they are  far from being self 
consistent
 approaches, and so we feel  it is premature to discuss them.

 Lev Landau taught his students:`` A model is a theory which has not yet 
been 
proven to be 
correct". We hope that our model is just an illustration of these words.

\section{Acknowledgements}
We would like to thank Boris Kopeliovich and Gosta Gustafson for supplying 
the numerical values of their amplitudes, and Christophe Royon for 
suggesting that we write this review.
The research  was supported by the BSF grant 2012124 and by  
 the  Fondecyt (Chile) grants  1140842 and 1130549.

\end{document}